\documentclass[12pt,preprint]{aastex}

\def\imaps{{\it IMAPS\/}}
\def\fuse{{\it FUSE\/}}                 

\def\hst{{\it HST\/}}
\def\lya{Ly${\alpha}$}

\def\EE#1{\times 10^{#1}}

\def\teff{$T_{\rm{eff}}$}
\def\logg{log~$g$}

\def\hmol{H$_{\rm 2}$\/}

\def\kms{km s$^{-1}$}

\def\mfarcs{\hbox{$~\!\!^{\prime\prime}$}}

\shorttitle{D/H toward 3 stars}
\shortauthors{Oliveira et al.}

\begin{document}

\title{Variations in D/H and D/O from new {\it FUSE} observations\altaffilmark{1}}

\author{Cristina M. Oliveira\altaffilmark{2}, H.~Warren~Moos\altaffilmark{2}, Pierre Chayer\altaffilmark{2,3}, and Jeffrey W. Kruk\altaffilmark{2}}
  
\altaffiltext{1}{Based on observations made with the NASA-CNES-CSA {\it Far Ultraviolet Spectroscopic Explorer}. \fuse~is operated for NASA by The Johns Hopkins University under NASA contract NAS5-32985.}
\altaffiltext{2}{Department of Physics and Astronomy, The Johns Hopkins University, 3400 N. Charles St., Baltimore MD 21218}
\altaffiltext{3}{Primary affiliation: Department of Physics and Astronomy, University of Victoria, P.O. Box 3055, Victoria, BC V8W 3P6, Canada.}

\begin{abstract} 

We use data obtained with the {\it Far Ultraviolet Spectroscopic Explorer} ({\it FUSE}) to determine the interstellar abundances of D\,I, N\,I, O\,I, Fe\,II, and \hmol~along the sightlines to WD\,1034$+$001, BD$+$39\,3226, and TD1\,32709. Our main focus is on determining the D/H, N/H, O/H, and D/O ratios along these sightlines, with log $N$(H) $>$ 20.0, that probe gas well outside of the Local Bubble. {\it Hubble Space Telescope} ({\it HST}) and {\it International Ultraviolet Explorer} ({\it IUE}) archival data are used to determine the H\,I column densities along the WD\,1034+001 and TD1\,32709 sightlines, respectively. For BD$+$39\,3226, a previously published $N$(H\,I) is used. We find (D/H)$\times10^5$ = 2.14 $\pm~^{0.53}_{0.45}$, 1.17 $\pm~^{0.31}_{0.25}$, and 1.86 $\pm~^{0.53}_{0.43}$, and (D/O)$\times10^2$ = 6.31 $\pm~^{1.79}_{1.38}$, 5.62 $\pm~^{1.61}_{1.31}$, and 7.59 $\pm~^{2.17}_{1.76}$, for the WD\,1034$+$001, BD$+$39\,3226, and TD1\,32709 sightlines, respectively (all 1$\sigma$). The scatter in these three D/H ratios exemplifies the scatter that has been found by other authors for sightlines with column densities in the range 19.2 $<$ log $N$(H) $<$ 20.7. The D/H ratio toward WD\,1034$+$001 and all the D/O ratios derived here are inconsistent with the Local Bubble value and are 
some of the highest in the literature. We discuss the implications of our measurements for the determination of the present-epoch abundance of deuterium, and for the different scenarios that try to explain the D/H variations. We present a study of D/H as a function of the average sightline gas density, using the ratios derived in this work as well as ratios from the literature, which suggests that D/H decreases with increasing gas volume density. Similar behaviors by other elements such Fe and Si have been interpreted as the result of depletion into dust grains.

\end{abstract}


\section{INTRODUCTION} 

The present day abundance ratio of deuterium to hydrogen places important constraints on Big Bang nucleosynthesis (BBN) and the chemical evolution of galaxies. Because deuterium is only produced in appreciable amounts in primordial BBN and destroyed in stellar interiors, the measurement of D I/H I in the interstellar medium (ISM) places a lower limit on the primordial abundance of deuterium. In addition, by comparing the ISM abundance of deuterium to its abundance in high-redshift intergalactic gas we should be able to better understand the effects of astration and chemical evolution of galaxies.
Measurements of the D/H ratio in intervening clouds of gas seen toward distant quasars have yielded a range of values D/H = (1.65 -- 4.0)$\EE{-5}$ \citep[and references therein]{2001ApJ...552..718O,2001ApJ...560...41P,2002ApJ...565..696L}. \citet{2003ApJS..149....1K} measured D/H = (2.42 $\pm~^{0.35}_{0.25}$)$\times10^{-5}$ toward Q1234$+$3047 ($z$ = 2.526). These authors calculate the value they believe is the best estimate of the primordial D/H abundance, D/H$_{\rm prim}$ = (2.78 $\pm~^{0.44}_{0.38}$)$\times10^{-5}$ (1$\sigma$ errors in the mean), by taking the weighted mean of five D/H measurements toward QSOs ranging from $z$ = 2.079 -- 3.572. This value is in good agreement with that determined from the cosmic microwave background measurements performed with $WMAP$~and previous missions \citep{2003ApJS..148..175S}. \citet{2004ApJS..150..387S} found D/H = (2.2 $\pm$ 0.7)$\times10^{-5}$ for Complex C, a high velocity cloud falling into our galaxy, which has low metallicity and has presumably experienced more stellar processing than the gas seen toward QSOs.
 
Measurements of D/H in the local ISM have been made with {\it Copernicus} \citep[e.g.][]{1973ApJ...186L..95R}, \hst, \citep[e.g.][]{1995ApJ...451..335L},~\imaps,~\citep{1999ApJ...520..182J,2000ApJ...545..277S}, and more recently~\fuse~\citep[e. g.][and references therein]{2002ApJS..140....3M}.
A nearly constant ratio of D/H = $(1.5~\pm~0.1)\EE{-5}$ (1$\sigma$~on the mean) has been obtained in the Local Interstellar Cloud (LIC) by \citet{1998SSRv...84..285L}; recent measurements inside the Local Bubble \citep[LB, log $N$(H\,I) $\le$ 19.2][]{1999A&A...346..785S} appear to be consistent with a single value for D/H in the LB \citep{2002ApJS..140....3M,2003ApJ...587..235O}. From a compilation of measurements from the literature \citet{2004ApJ...609..838W} derive (D/H)$_{\rm LB}$ = (1.56 $\pm$ 0.04)$\times10^{-5}$. Using (D/O)$_{\rm LB}$ = (3.84 $\pm$ 0.16)$\times10^{-2}$ from \citet{2003ApJ...599..297H} and (O/H)$_{\rm LB}$ = (3.45 $\pm$ 0.19)$\times10^{-4}$ from \citet{2005ApJ...625..232O} one can derive indirectly (D/H)$_{\rm LB}$ = (1.33 $\pm$ 0.09)$\times10^{-5}$. The direct and indirect determinations of (D/H)$_{\rm LB}$ are only consistent when one considers the 2$\sigma$ uncertainties in both quantities. A detailed discussion of the possible causes for this discrepancy can be found in \citet{2003ApJ...599..297H}.
 
Outside the Local Bubble, however, there still is not a consistent picture of the D/H behavior. Measurements performed along sightlines probing gas outside the LB 
suggest variations of the interstellar D/H ratio beyond the Local Bubble, at the distance of a few hundred parsecs. D/H varies for 19.2 $\le$ log $N$(H\,I) $\le$ 20.7 \citep[e.g.][]{1999ApJ...520..182J,2000ApJ...545..277S,2002ApJS..140...37F} and apparently remains constant for log $N$(H\,I) $\ge$ 20.7, albeit with a lower value. The possibility of a trend towards low D abundance at high H column densities was noted in the D/O survey by \citet{2003ApJ...599..297H}. \citet{2004ApJ...609..838W}, using additional measurements and values in the literature, confirmed the trend for D/H. So far, only five measurements for sightlines with log $N$(H\,I) $\ge$ 20.7 have been published. \citet{2003ApJ...586.1094H}, \citet{2004ApJ...609..838W} and \citet{2005guillaume} measured D/H along five extended lines of sight ($d \ge$ 500 pc). The weighted mean of these five measurements yields D/H = (0.87 $\pm$ 0.08)$\times10^{-5}$, in clear disagreement with the LB value. 

The value of the present-epoch Milky Way abundance of deuterium, (D/H)$_{\rm PE}$, is still an issue subject to debate. Two opposite explanations have emerged. \citep{2003ApJ...599..297H} argue for a value of the current-epoch D/H ratio lower than D/H$_{\rm LB}$, on the basis of the generally low values found for more distant sightlines. On the other hand, Linsky et al. (2005, in prep) argue that an important fraction of deuterium along many sightlines is trapped in a population of grain material and hence the current-epoch deuterium abundance (gas plus grains) is (D/H)$_{\rm PE}$ $\ge$ (2.19 $\pm$ 0.27)$\times10^{-5}$. The astration factors, (D/H)$_{\rm prim}$/(D/H)$_{\rm PE}$, derived from both of these scenarios ($\sim$4 and $\sim$1.2, respectively) challenge current models of galactic chemical evolution.

Because the O/H variations in the diffuse ISM from the LB out to 1 Kpc, are small, consistent with the measurement uncertainties \citep{1998ApJ...493..222M,2003ApJ...591.1000A,2004ApJ...613.1037C,2005ApJ...625..232O}, O\,I has been used as a proxy for H\,I \citep[for a more detailed discussion see][and references therein]{2003ApJ...599..297H}. Thus D/O and D/H are expected to trace each other in the diffuse ISM. However, how well this works at high column densities is a subject of discussion \citep[][and $\S$\ref{litvalues} of this paper]{2006guillaume}.

At the present time, definitive conclusions are prevented by the limited number of D/H and D/O measurements along high column density sightlines. In this work we present D/H and D/O measurements along three sightlines (WD\,1034$+$001, BD$+$39\,3226, and TD1\,32709) with log $N$(H\,I) $\ge$ 20.0. Using data obtained with $FUSE$, $IUE$, and the Goddard High Resolution Spectrograph (GHRS, onboard {\it HST}) we derive column densities of atomic and molecular species which are then used to determine important ratios (D/H, N/H, O/H, etc.) that can be compared to values in the literature. 

This paper is organized as follows. The three targets used for the analyses are described in \S\ref{targets}. The observations and data processing are presented in \S\ref{obs}. In \S\ref{ana} we determine the column densities of H\,I, D\,I, N\,I, O\,I, Fe\,II, and \hmol~along the WD\,1034$+$001, BD$+$39\,3226, and TD1\,32709 sightlines. Lower limits to the column densities of other elements are also reported. The D/H, D/N, D/O, N/H, O/H, and O/N ratios for the three sightlines are presented in $\S$\ref{results} and discussed in \S\ref{discussion}, where evidence for above average D abundance is presented for the WD\,1034$+$001 and TD1\,32709 sightlines. The D/O ratios for the three sightlines are substantially higher than the mean LB value, and thus they considerably increase the scatter in the published D/O ratios. Implications of the derived ratios for hypothetical mechanisms, such as dust depletion and infall of metal poor gas, that try to explain the D/H variations in the Galactic disk are considered in \S\ref{implica}. Our derived D/H, D/O, and O/H ratios are compared with previously published ratios in \S\ref{litvalues}. In $\S$\ref{iron} we consider the relationship between D/O and Fe/O. In \S\ref{nh_disc} we perform a study of D/H as a function of the average sightline gas density ($N$(H)/$d$). Our study indicates that D/H is constant up to densities of 0.10 cm$^{-3}$ and decreases with increasing density after this point, similarly to what has been observed for other elements such as Fe and Si. We summarize our findings in \S\ref{summary}. All uncertainties are quoted at the 1$\sigma$ level unless stated otherwise.

\section{THE TARGETS}
\label{targets}

The properties of the three stars are listed in Table \ref{star_properties}. Below we discuss each star in detail.

\subsection{WD\,1034$+$001}
\label{wd1034_target}

WD\,1034$+$001 is a hot DO white dwarf with \teff~=100,000 $\pm~^{15,000}_{10,000}$ K, \logg~=~7.5 $\pm$ 0.3, at a distance of 155 $\pm~^{58}_{43}$ pc \citep[photometric,][]{1995A&A...298..567W} in the direction $l$ = 247.55$^\circ$ and $b$ = $+$47.75$^\circ$. This star was first observed in the ultraviolet with {\it IUE} by \citet{1985ApJ...292..477S}, who identified several photospheric features (C\,IV, N\,V, and O\,V) as well as interstellar features (N\,I, C\,II, Si\,II, Mg\,II, Si\,II, and Si\,III). These authors derived $v_{\rm PH} - v_{\rm ISM} \sim$50 km s$^{-1}$ ($v_{\rm PH}$ and $v_{\rm ISM}$ correspond to stellar and interstellar absorptions, respectively). Non-stellar absorption by C\,IV was also detected along this sightline by \citet{1985ApJ...292..477S}, however the resolution of the data did not allow them to decide whether this absorption had an interstellar or circumstellar origin.

\citet{1995A&A...298..567W} used GHRS onboard {\it HST} to observe this target at a resolution of $\sim$20,000. In addition to stellar absorption by He\,II, C\,IV, N\,V, O\,V, Si\,IV, Fe\,VI, and Fe\,VII, they also identified absorption by highly ionized species of C\,IV, N\,V, and Si\,IV. With resolution similar to that of \citet{1985ApJ...292..477S}, \citet{1995A&A...298..567W} could not determine whether the absorption had an interstellar or circumstellar origin. Using H\,I Ly$\alpha$ observations obtained with GHRS in conjunction with NLTE model atmospheres, \citet{1995A&A...298..567W} determined log $N$(H\,I) = 20.05 $\pm$ 0.20. In this work we reanalyze these H\,I Ly$\alpha$ observations together with $FUSE$~data to place tighter constraints on $N$(H\,I) (see \S\ref{hi-wd1034}).

Using Sloan Digital Sky Survey data, \citet{2003ApJ...599L..37H} discovered a region of ionized gas around WD\,1034$+$001, seen in H$_{\rm \alpha}$, H$_{\rm \beta}$, [N\,II], [O\,III], and [S\,II] emission, with a diameter greater than 2$^\circ$ and identified it with a planetary nebula, Hewett 1. This is one of the largest planetary nebula in the sky, and also the first to be unambiguously associated with a DO white dwarf. Using data obtained with the Southern H-Alpha Sky Survey Atlas, \citet{2004A&A...417..647R} found two extended emission structures surrounding WD\,1034$+$001, in addition to the planetary nebula discovered by \citet{2003ApJ...599L..37H}. The inner halo has a linear diameter of $16.2~\pm~^{6.1}_{4.5} \times 24.3 \pm~^{9.1}_{6.8}$ pc while the wider shell has a linear diameter of 27$\times$43 pc \citep{2004A&A...417..647R}, at the quoted distance for WD\,1034$+$001 (see Table \ref{star_properties}). During its evolution, WD\,1034$+$001 is likely to have passed through a much hotter phase with \teff = 150,000K, roughly 30,000 years ago \citep{2004A&A...417..647R}. These authors estimate that WD\,1034$+$001 is most likely the exciting star of the nebula and the halo given that the recombination time scale in such a low-density ($n_e$ = 3 cm$^{-3}$) nebula is longer than 30,000 years.

\subsection{BD$+$39\,3226}

BD$+$39\,3226, first identified as a sdO by \citet{1978A&A....64L...9B}, lies at a distance of 290 $\pm~^{140}_{70}$ pc ($Hipparcos$) in the direction $l$ = 65.00$^\circ$, $b$ = $+$28.77$^\circ$. \citet{1999A&A...352..287B} determined the abundances of the atomic (H\,I, D\,I, etc.) and molecular species (\hmol) along the line of sight using data obtained with {\it IUE} and {\it ORFEUS II}, and determined $v_{\rm PH} - v_{\rm ISM} \sim$ $-$255 \kms. However, the spectral resolution of only $\sim$30 \kms~for both the {\it IUE} and {\it ORFEUS~II} datasets led to fairly large uncertainties on the column densities determined with those data. The new \fuse~data with a resolution of $\sim$20 \kms~and better S/N ratio allows us to determine more accurate column densities and in addition, offers us the possibility of comparing column densities determined from data obtained with different instruments. 

In this work we use the \fuse~data to revisit the column densities along the BD$+$39\,3226 sightline for the species that have transitions in the \fuse~bandpass. Since the \lya~transition of H\,I is not covered by \fuse~we use the value reported by \citet{1999A&A...352..287B}. These authors used \lya~observations from $IUE$ and $ORFEUS~II$, in conjunction with a stellar model with \teff~=~45,000 K and log$g$ = 5.5, to determine log $N$(H\,I) = 20.08 $\pm$ 0.09.

\subsection{TD1\,32709}

TD1\,32709 (UVO 0904$-$02) was first identified as a sdO by \citet{1980A&A....85..367B} using data obtained with the Ultraviolet Sky Survey Telescope S\,2/68. Using optical data in conjunction with NLTE stellar atmosphere models, \citet{1993A&A...273..212D} determined \teff~=~46,500 $\pm$ 1,000 K and log $g$ = 5.55 $\pm$ 0.15. The corresponding photometric distance is $d$~= 520 $\pm$ 90 pc ($z$ = 245 $\pm$ 40 pc). $IUE$~observations covering the Ly$\alpha$~region in combination with $FUSE$~data and stellar models allow us to determine $N$(H\,I) along this sightline (see \S\ref{hi-uv0904}). From our work we determine $v_{\rm PH}-v_{\rm ISM}$ = $-$13 \kms~(see $\S$\ref{uv0904atomic}).

\section{OBSERVATIONS AND DATA PROCESSING} 
\label{obs}

\subsection{$FUSE$~Observations}

The {\it FUSE} observatory consists of four coaligned prime-focus telescopes and Rowland-circle spectrographs that produce spectra over the wavelength range 905 -- 1187 \AA, with a spectral resolution of $\sim$15 -- 20 \kms~(wavelength dependent) for point sources. Details about the $FUSE$~mission, its planning, and on-orbit performance can be found in \citet{2000ApJ...538L...1M} and \citet{2000ApJ...538L...7S}.

Table \ref{fuse_obs} summarizes the $FUSE$~observations of the three targets. The data were obtained through the medium size aperture (MDRS, $4\mfarcs\times20\mfarcs$), in histogram (H) or ttag modes (T). The two-dimensional \fuse~spectra are reduced using the CalFUSE pipeline v2.4.1 or v2.4.2\footnote{The CalFUSE pipeline reference guide is available at http://fuse.pha.jhu.edu/analysis/pipeline\_reference.html}. The processing includes data screening for low quality or unreliable data, thermal drift correction, geometric distortion correction, heliocentric velocity correction, dead time correction, wavelength calibration, detection and removal of event bursts, background subtraction, and astigmatism correction.
The spectra are aligned by cross-correlating the individual exposures over a short wavelength range that contains prominent spectral features and then coadded by weighting each exposure by its exposure time, using the CORRCAL software developed by S. Friedman. All the spectra are binned to three pixel samples, or $\sim$20 m\AA, for analysis (the line spread function, LSF, is about 11 pixels or $\sim$70 m\AA~wide). For each target, all the observations were coadded in order to increase the S/N of the dataset. The S/N per unbinned pixel in the SiC 1B channel around 940 \AA~is 8, 26, and 14, for WD\,1034$+$001, BD$+$39\,3226, and TD1\,32709, respectively.

\begin{figure}
\begin{center}
\epsscale{0.87}
\plotone{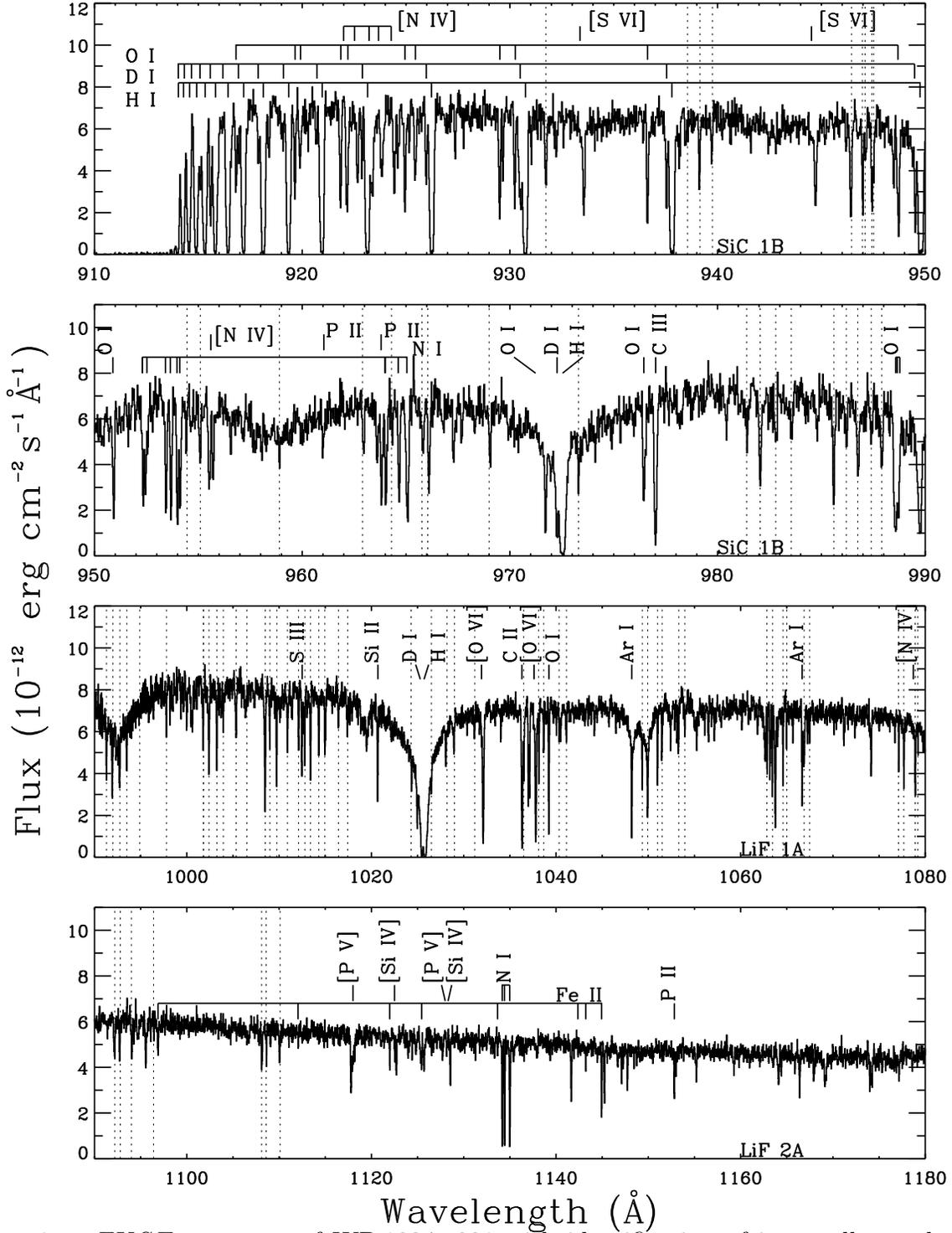}
\caption{$FUSE$~spectrum of WD\,1034$+$001 with identification of interstellar and stellar absorptions. Labels in [\,] refer to stellar lines; dashed vertical lines mark the position of the strongest \hmol~lines along this sightline. Stellar H\,I absorption is displaced from the ISM one by the velocity shift specified in Table \ref{star_properties}. Note the isolated broad He\,II stellar lines around 992 and 958 \AA. Other stellar He\,II lines fall close to the positions of the Ly$\alpha$ series and are not resolved. The {\it FUSE} channel used for each panel is indicated at the bottom right. The spectrum was binned by 4 for display purposes only.\label{wdspectra}}
\end{center}
\end{figure}

\begin{figure}
\begin{center}
\epsscale{0.87}
\plotone{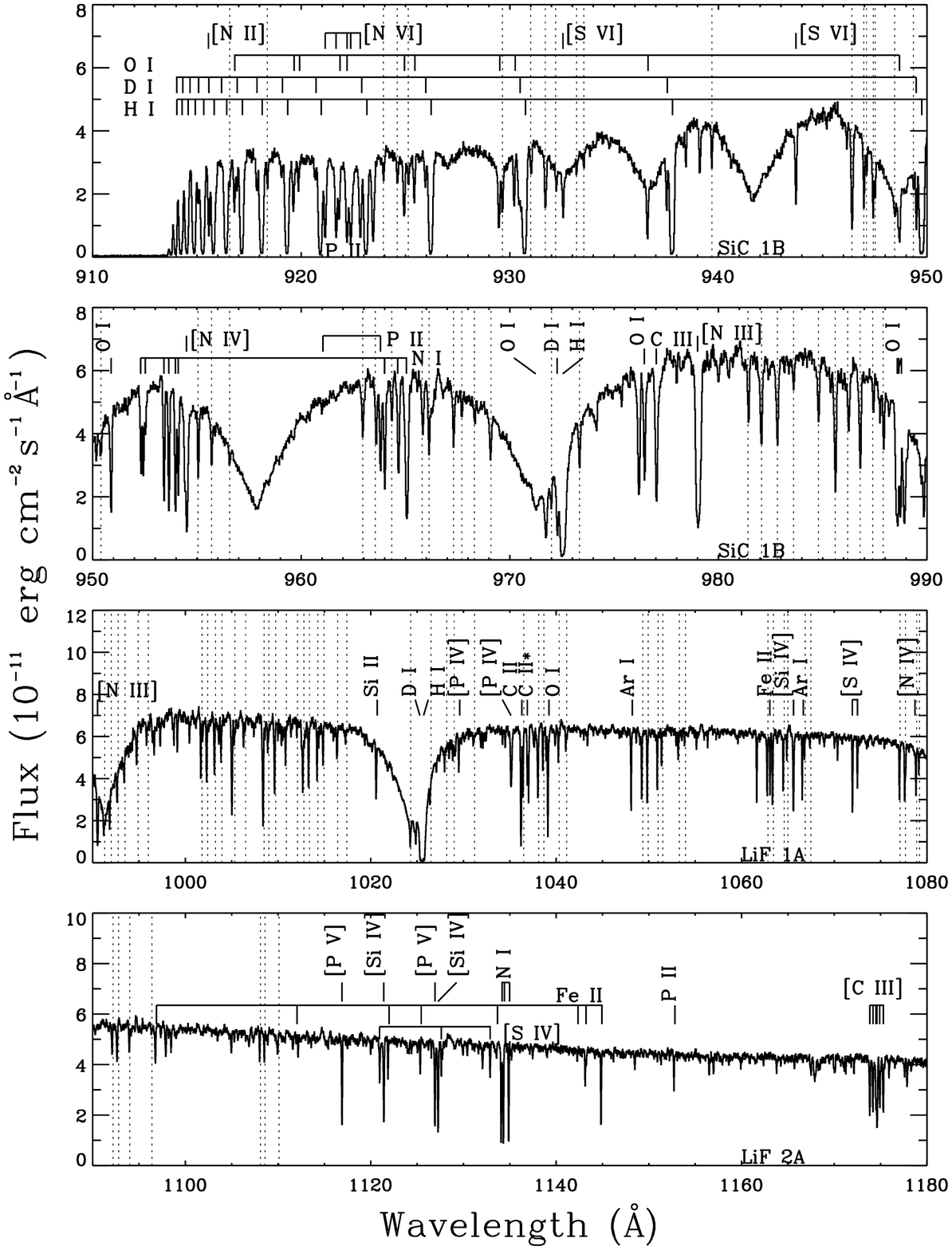}
\caption{Same as Fig. \ref{wdspectra} but for the BD$+$39\,3226 sightline.\label{bdspectra}}
\end{center}
\end{figure}

\begin{figure}
\begin{center}
\epsscale{0.87}
\plotone{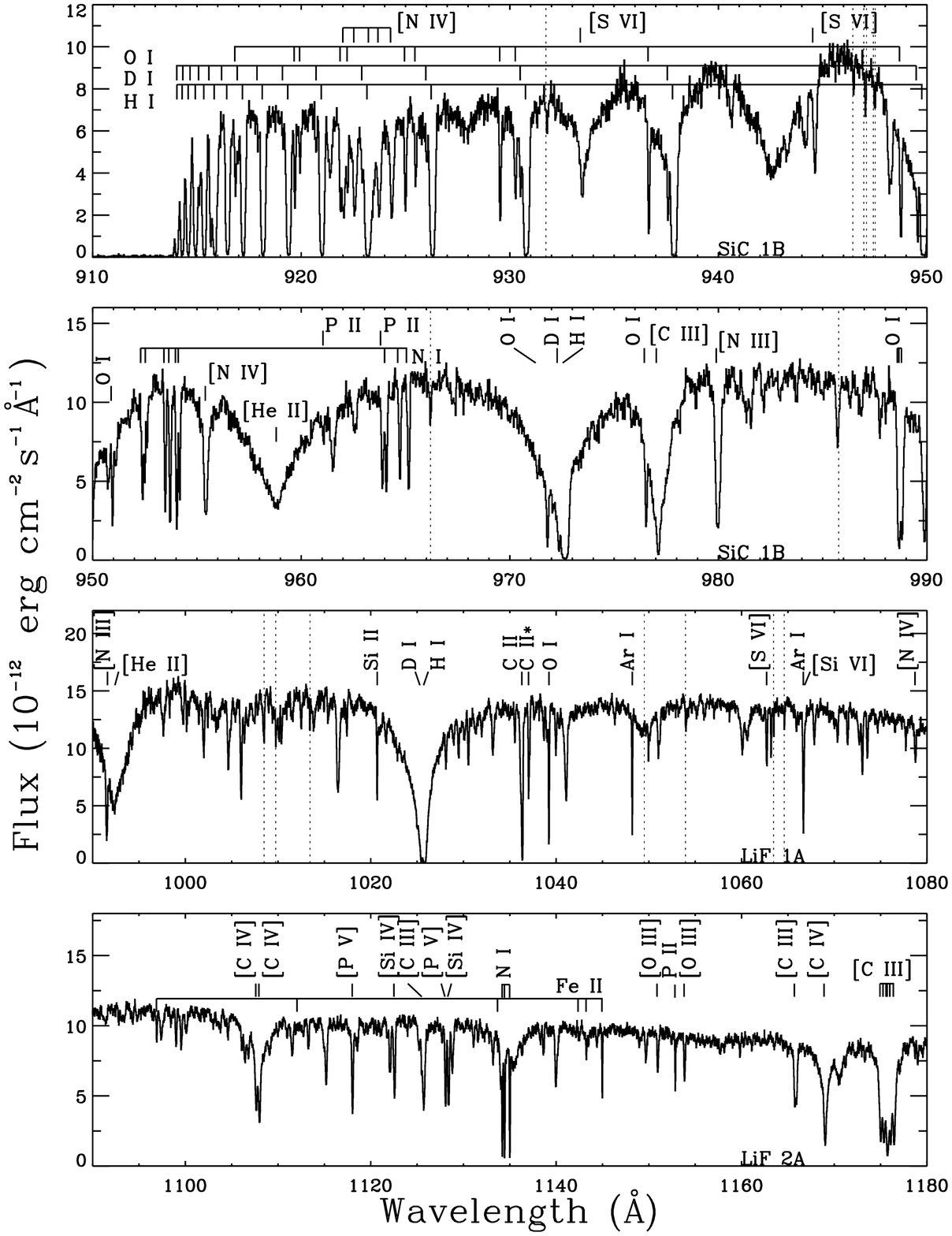}
\caption{Same as Fig. \ref{wdspectra} but for the TD1\,32709 sightline. \label{uvspectra}}
\end{center}
\end{figure}

Figures \ref{wdspectra}, \ref{bdspectra}, and \ref{uvspectra}, present the \fuse~spectra of the three targets. The most prominent interstellar and photospheric lines are labeled; photospheric lines are indicated by [\,], and \hmol~absorption by dashed vertical lines.

\section{ANALYSIS}
\label{ana}

Whenever possible we use apparent optical depth, curve of growth and profile fitting methods (hereafter AOD, COG, and PF, respectively) to determine column densities \citep[see e.g.][and references therein for a further discussion of these methods]{2003ApJ...587..235O}. AOD is used on weak transitions to determine column densities of  species that have multiple transitions in the {\it FUSE} bandpass, which are then compared to the values derived with the other methods. For species that only have saturated transitions in the \fuse~bandpass we use the AOD technique to place lower limits on the column densities of those species. 

We use the COG method to determine the column densities of species that have several transitions in the {\it FUSE} bandpass (one or more of these transitions must be unsaturated). We measure the equivalent widths of all the non-blended transitions, in all the {\it FUSE} channels where those transitions are covered, and compare them to check for inconsistencies (due to fixed pattern noise, for instance). A single component Gaussian curve-of-growth is then fit to the measured equivalent widths of the atomic species, allowing us to determine $N$ and $b$, for each species. For \hmol~we fit a single-component Gaussian COG to the equivalent widths of all the $J$~levels, simultaneously. We determine then $N$($J$) and $b$, common to all $J$~levels.

With PF we fit a single absorption component to one or more non-saturated transitions of the species for which $N$~is being sought. We use the profile fitting code {\it Owens}, developed by Martin Lemoine and the French {\it FUSE} Team. Details of how the PF method is used for the analysis of each sightline studied in this paper are given below. More information about the fitting code can be found in \citet{2002ApJS..140...67L} and \citet{2002ApJS..140..103H}.

We follow the procedures outlined in \citet{2003ApJ...587..235O} to use these techniques and determine the uncertainties in $N$~associated with each method. We introduce a modification from \citet{2003ApJ...587..235O} regarding the PF technique. A single-Gaussian with a FWHM of 10.5 pixels (constant across all wavelengths and channels and during the fitting procedure) is used here to describe the \fuse~line spread function (LSF), whereas previously the LSF was a free parameter of the fitting procedure during the initial stages of the fit \citep[see][for more details]{2003ApJ...587..235O}. However, since several studies have shown that column densities obtained with fits using variable and fixed LSFs are similar \citep[e.g.][]{2005astro.ph..1320W} we prefer to decrease the number of degrees of freedom of the profile fitting routine by using a fixed LSF.

All the atomic lines used in the AOD, COG, and PF analyses are shown in Table \ref{atomicdata} along with log~$f\lambda$~for each transition. We use the compilation of \citet{2003ApJS..149..205M} for the atomic data and the ones by \citet{1993A&AS..101..273A} and \citet{1993A&AS..101..323A} for the \hmol~data. For each star, A, C, and P denote transitions that are used with the apparent optical depth, curve of growth, and  profile fitting methods, respectively. Table \ref{eqw} presents the equivalent widths of the atomic lines used with the COG method for each sightline. After column densities and uncertainties are determined with these different techniques the results are examined to check the consistency between the different methods. The adopted results are a subjective compromise between the values obtained with the several methods used; the uncertainties adopted are such that they in general include the extreme values obtained with the different methods. No attempt was made to combine the different results in any statistical way. The results of the individual methods are also presented (see $\S$ \ref{atomicdis}).

In $\S$ \ref{h2disc} we determine the \hmol~column densities along the three lines of sight, atomic column densities are discussed in $\S$ \ref{atomicdis}. Determination of the H\,I column density along the WD\,1034$+$001 and TD1\,32709 sightlines is discussed in $\S$\ref{nhi}.

\subsection{Molecular Hydrogen}
\label{h2disc}

\subsubsection{WD\,1034$+$001}
\label{h2wd1034}

Along this sightline we detect absorption arising from the H$_{2}$ rotational levels $J$ = 0 -- 3. We use AOD, single-$b$ COG and PF to derive the column densities of the $J$ = 0~--~3 levels. From the single-$b$ COG fit we derive $b_{\rm COG}$ = 3.6 $\pm~^{0.4}_{0.3}$ km s$^{-1}$. Profile fitting of the different $J$~levels is done simultaneously with the atomic species along this sightline, under the assumption that the molecular gas does not trace precisely the atomic gas (i.e. atomic and molecular species are fit in different absorption components). Although the velocities of the atomic and molecular component are a free parameter of the fit, no significant velocity differences are found between the two components. The adopted column densities for each $J$-level are summarized in Table \ref{h2}. We adopt log~$N$(\hmol) = 15.72 $\pm~^{0.13}_{0.12}$. The excitation diagram for \hmol~along this sightline is presented in the top panel of Figure \ref{h2exc}. We derive $T_{02}$ = 341 $\pm$ 75 K.

\subsubsection{BD$+$39\,3226}
\label{h2bd39}

For this sightline we detect absorption arising from the \hmol~rotational levels $J$ = 0 -- 5. We use AOD, single-$b$ COG and PF to derive the column densities of the $J$ = 0 -- 3 levels. The $J$ = 4, 5 levels have very low column densities and we use only PF to determine the column densities associated with these two $J$~levels.

\begin{figure}
\begin{center}
\epsscale{0.75}
\plotone{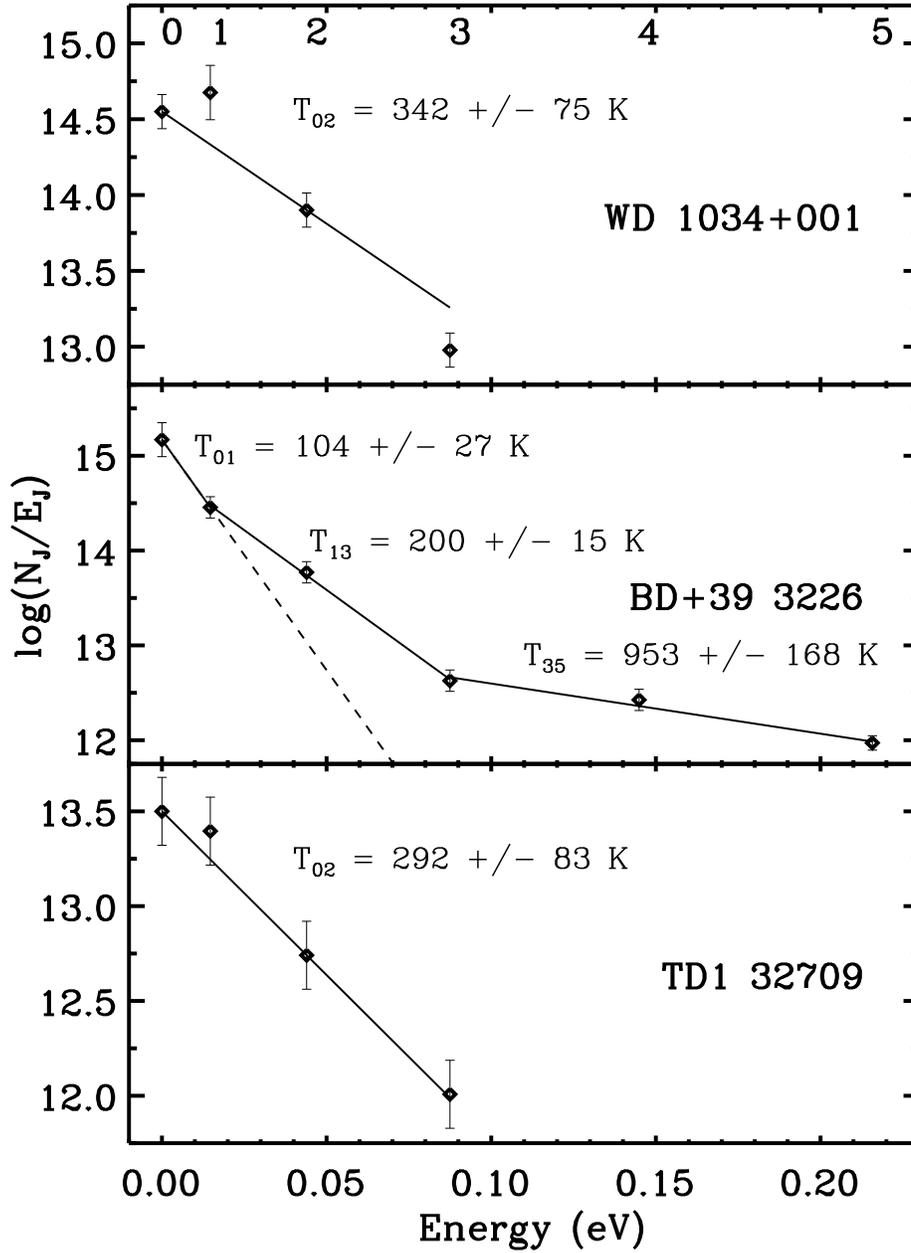}
\caption{\hmol~excitation diagram for the three sightlines. Each $J$~level is labeled at the top of the first panel. See $\S$ \ref{h2disc} for discussion.\label{h2exc}}
\end{center}
\end{figure}

From the single-$b$ COG fit to the $J$ = 0 -- 3 levels we derive $b_{\rm COG}$ = 4.3 $\pm~^{0.3}_{0.2}$ km s$^{-1}$. Similarly to WD\,1034$+$001, profile fitting of the $J$ = 0 -- 5 levels is done simultaneously with the atomic species, but in different absorption components. Table \ref{h2} presents the adopted column densities for this sightline. We adopt log $N$(\hmol) = 15.65 $\pm~^{0.06}_{0.07}$. The excitation diagram for \hmol~along this sightline is presented in the middle panel of Figure \ref{h2exc}. The populations of the levels with $J$ $\ge$ 2 are larger than expected if the different levels were populated according to a Boltzmann distribution with an excitation temperature corresponding to $T_{01}$ (indicated by a dashed line in Figure \ref{h2exc}). This nonthermal excitation is the result of pumping by UV photons followed by cascading transitions down through the various rotational levels \citep{1976ApJ...203..132B}. The high $J$ lines can typically be fit by a single excitation temperature. However, in the case of this sightline a single temperature fit to $J$ = 2 -- 5 clearly overestimates the column densities of the $J$ = 3 and $J$ = 4 levels. We find that the distribution of \hmol~through the different $J$ levels is better described by several temperatures. We derive then $T_{01}$ = 104 $\pm$ 27 K, $T_{13}$ = 200 $\pm$ 15 K, and $T_{35}$ = 953 $\pm$ 158 K.

\subsubsection{TD1\,32709}
\label{h2td1}

We use AOD and PF to determine the column densities of the H$_{2}$ rotational levels $J$ = 0 -- 3. For this sightline, the low column density of each $J$~level implies that the number of absorption lines that could be used with the COG method is small, and hence we do not perform a COG fit for \hmol~along this sightline. The consistency of PF results is checked against that of the AOD method. Table \ref{h2} presents the adopted $N$($J$) for this sightline. We derive $N$(\hmol) = 14.48 $\pm~^{0.12}_{0.11}$. The bottom panel of Figure \ref{h2exc} presents the \hmol~excitation diagram for this sightline. All the $J$ levels are consistent with a single temperature corresponding to $T_{02}$ = 292 $\pm$ 83 K.

No HD is detected along any of these sightlines. Typical HD/\hmol~ratios along these types of diffuse sightlines are of the order of 1$\times10^{-5}$ or less. Considering $N$(\hmol) quoted in Table \ref{h2} we expect log $N$(HD) $<$ 11.

\subsection{Atomic Species}
\label{atomicdis}

In this section we discuss the determination of the atomic column densities along the three sightlines. However, determination of $N$(H\,I) along the WD\,1034$+$001 and TD1\,32709 sightlines will be discussed separately in \S\ref{nhi}. Tables \ref{nwd1034}, \ref{nbd39}, and \ref{ntd1} present the column densities for D\,I, N\,I, O\,I, and Fe\,II obtained with the different methods discussed above. The adopted column densities for the atomic species along the three sightlines are summarized in Table \ref{natomic}.

\subsubsection{WD\,1034$+$001}
\label{wd1034atomicdis}

We use the AOD method to derive lower limits to the column densities of C\,II$^*$, C\,III, N\,II, Si\,II, P\,II, and Ar\,I, because all the transitions of these species in the {\it FUSE} bandpass are saturated. The AOD method is also used for weak transitions of D\,I, N\,I, O\,I, and Fe\,II, to determine column densities which are then compared to the column densities determined with the other methods.

\begin{figure}
\begin{center}
\epsscale{0.55}
\rotatebox{90}{
\plotone{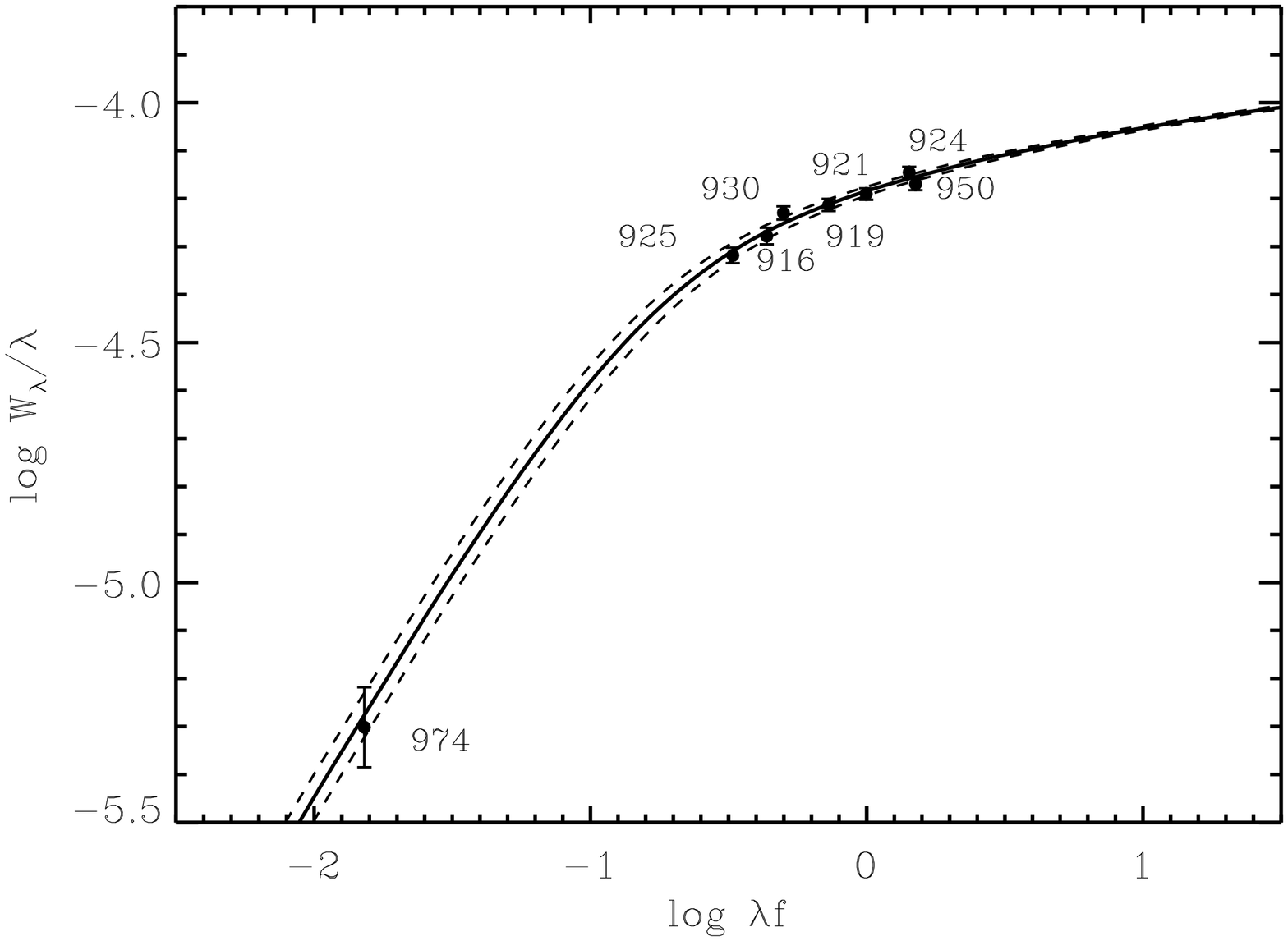}
}
\caption{Curve of growth for O\,I along the WD\,1034$+$001 sightline. We derive log $N$(O\,I) = 16.62 $\pm$ 0.05. Dashed lines indicate the 1 $\sigma$ uncertainty in $N$.\label{oicogwd1034}}
\end{center}
\end{figure}

We fit a single-Gaussian COG to the measured equivalent widths of D\,I, N\,I, O\,I, and Fe\,II. The four COGs yield $b$-values in the range $b$ = 5 -- 6 \kms. Figure \ref{oicogwd1034} presents the curve of growth for O\,I along this sightline.

With PF we fit a single absorption component to non-saturated lines of D\,I, N\,I, O\,I, and Fe\,II. In order to model the continuum in the vicinity of the D\,I lines an extra absorption component with H\,I only, is also included in the fit. This component is not used to derive $N$(H\,I) along this sightline. No stellar model is used in the fitting process because the photospheric spectrum is generally flat in the regions of interest. As discussed above, \hmol~is also included in the fit, in a different absorption component. The D\,I lines used in this analysis have a minor blending with \hmol, hence $N$(D\,I) is not affected by the assumed $N$(\hmol).

Along this sightline we also detect absorption by O\,VI ($\lambda$1032, 1038), S\,III ($\lambda$1012), and S\,IV ($\lambda$1063). O\,VI absorption has two components, separated by $\sim$ 50 km s$^{-1}$. The strongest component falls at the stellar velocity derived by \citet{1985ApJ...292..477S} and is likely of stellar origin. Absorption by S\,III, S\,IV, and the weaker O\,VI absorption component fall at the velocity of the atomic gas (D\,I, N\,I, O\,I, and Fe\,II) along this sightline. However, we cannot conclude with certainty whether these features have an interstellar or circumstellar origin. Note that as discussed in $\S$\ref{wd1034_target}, two large ionized halos have been discovered in the vicinity of WD\,1034$+$001. The O\,VI lines are weak and blended with stellar O\,VI absorption; we derive $N$(O\,VI) with PF, following the method described in \S\ref{ana}. The S\,III absorption line is strong and likely saturated. We derive only a lower limit to its column density using the AOD technique. The S\,IV line is blended with an unidentified feature on the blue side, and is likely saturated. We derive a lower limit to its column density using the AOD technique.

\begin{figure}
\begin{center}
\epsscale{0.85}
\plotone{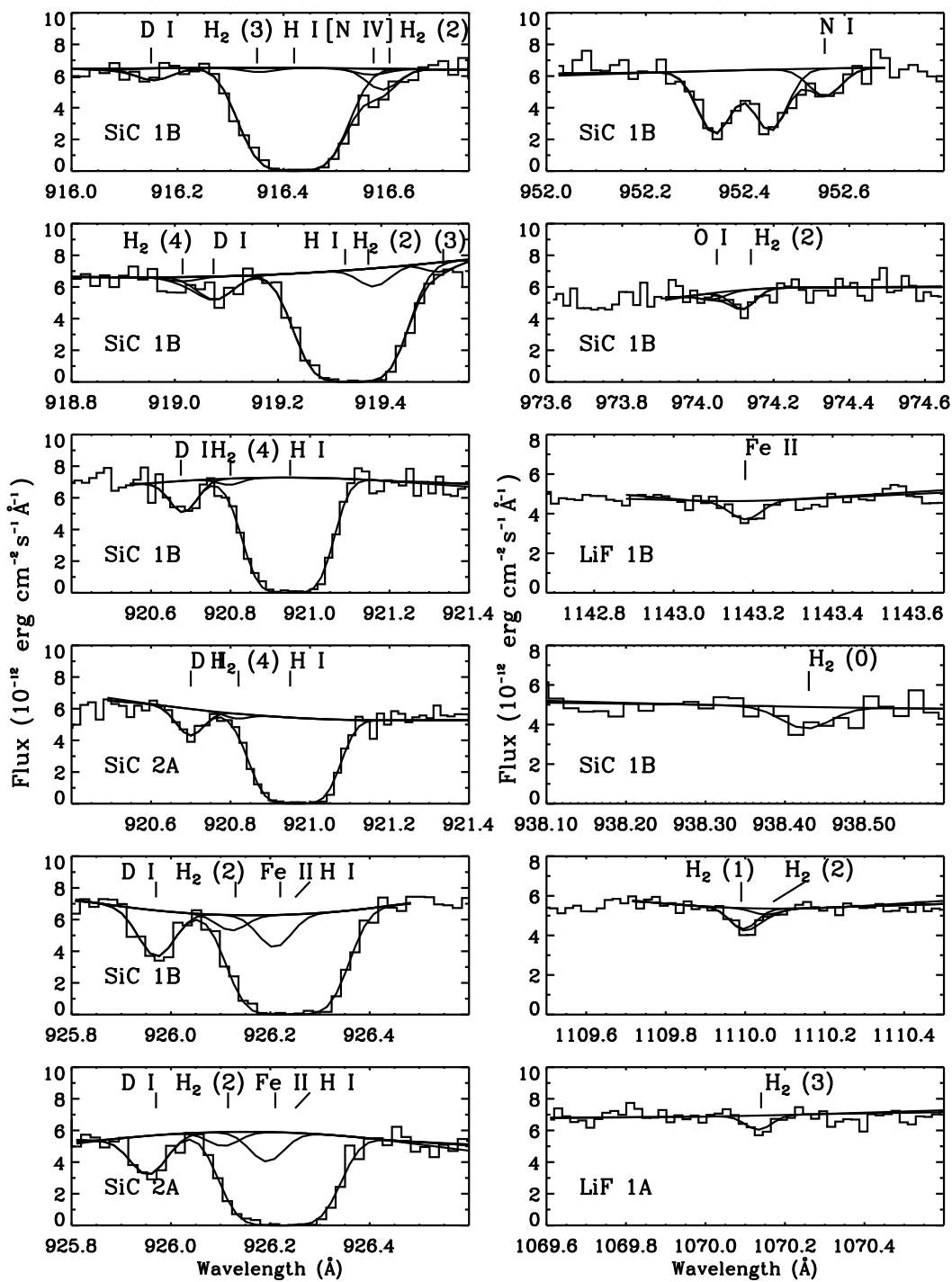}
\caption{Fits to some of the lines used in the analysis of the WD\,1034$+$001 sightline. The {\it FUSE} channels corresponding to the plotted data are indicated in the bottom left of each panel. For \hmol, the corresponding $J$~level is indicated in parenthesis. [] are used to indicate lines of photospheric origin.\label{wd1034fits}}
\end{center}
\end{figure}

Figure \ref{wd1034fits} presents fits to some of the lines used in the analysis of the WD\,1034$+$001 sightline. The line inside [] is of photospheric origin. Table \ref{nwd1034} presents the column densities for D\,I, N\,I, O\,I, and Fe\,II along this sightline, obtained with the three different techniques.

\subsubsection{BD$+$39\,3226}
\label{bd+39atomicdisc}

\begin{figure}
\begin{center}
\includegraphics[scale=.5,angle=90]{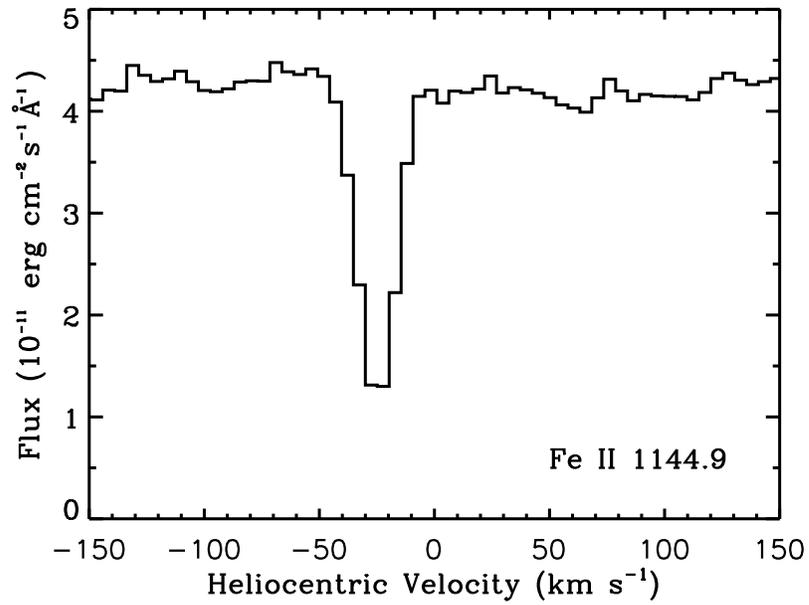}
\caption{Fe\,II $\lambda$1144.9 interstellar absorption toward BD$+$39\,3226 as seen with \fuse. Only one absorption component, at $v_{\odot}$ = $-$25 \kms, is present. We see no indication of a weaker absorption component at $v_{\odot}$ = $-$75 \kms, as reported by \citet{1999A&A...352..287B}. \label{feii}}
\end{center}
\end{figure}

\citet{1999A&A...352..287B} report the detection of a weak interstellar absorption component at $v_{\odot} = -$75 \kms~(Component B in their Fig. 1), seen in O\,I ($\lambda$1039), Si\,II ($\lambda$1526), and Fe\,II ($\lambda$2374). We see no evidence for such a component in the higher resolution and higher S/N \fuse~data. Figure \ref{feii} displays the absorption profile of Fe\,II 1144.9 as seen with \fuse. This transition is $\sim$1.6 times stronger than the Fe\,II $\lambda$2374 transition in which \citet{1999A&A...352..287B} claim to see a second component. The \fuse~data shows only one component at $v_{\odot}$ = $-$25 \kms, consistent with their Component A at $v_{\odot}$ = $-$25 \kms. Close inspection of the O\,I $\lambda$ 971.7 transition (stronger than O\,I $\lambda$1039) in our data also shows only one absorption component at the same velocity of Fe\,II $\lambda$1144.9. There is however an absorption feature detected in the {\it FUSE} data, shifted by $\sim -50$ km s$^{-1}$ from O\,I $\lambda$1039 (consistent with absorption at $v_{\odot} = -$75 \kms). This feature is due to absorption by stellar N\,III $\lambda$1038.988 and $\lambda$1039.0, at the radial velocity of the star. It is possible that the lower S/N {\it IUE} data used by \citet{1999A&A...352..287B} have lead them to mistakenly identify coincident noise features with extra absorption components of Fe\,II and Si\,II.

Although scattered flux is generally low in the \fuse~channels \citep{2000ApJ...538L...1M,2000ApJ...538L...7S}, there is some apparent residual flux, $\le$4\% of the continuum, in the core of the H\,I Lyman lines (see Fig. \ref{bdspectra}). Apparent residual flux is also present, albeit at a smaller level, below the Lyman break. This is probably due to the fact that this target was observed in histogram mode. For this type of observation the {\it FUSE} pipeline takes into account a background model which is scaled only by the exposure time. For instance, WD\,1034$+$001 was observed in time-tag mode. In this case the background model used by the {\it FUSE} pipeline is scaled to match the observed counts in the unilluminated regions of the detector, hence doing a better job of removing the scattered flux (note how the cores of the H\,I lines in the spectrum of WD\,1034$+$001 in Fig. \ref{wdspectra} have no residual flux). To properly determine column densities it is important to account for this residual flux, which is approximately constant within each channel, but varies between channels. Consequently, we removed this residual flux from all the channels, before performing any analysis of the data \citep[a similar procedure has been used in other D/H analyses, see][for an example]{2002ApJS..140..103H}.

The large separation between stellar and interstellar absorption \citep[$\sim$ 255 km s$^{-1}$][]{1999A&A...352..287B} along this sightline is high enough to remove the need for using a stellar model to account for the shape of the stellar continuum in the vicinity of the D\,I lines due to H\,I and He\,II stellar absorption. Hence, no stellar model is used in this analysis.

We use the AOD technique to determine column densities for D\,I, N\,I, O\,I, and Fe\,II, and place lower limits on the column densities of C\,II, C\,II$^*$, C\,III, Si\,II, P\,II, and Ar\,I.

With the COG method we fit single-component Gaussian COGs to the measured equivalent widths of D\,I, N\,I, O\,I, and Fe\,II. No other atomic species detected along this sightline have enough transitions to perform COGs. From the four COGs we derive $b$-values in the range $b$ = 5 -- 6 km s$^{-1}$.

With the PF technique we fit a single component to unsaturated lines of D\,I, N\,I, O\,I, and Fe\,II. \hmol~is also included in the fit, but in a separate absorption component. The D\,I lines used in determining $N$(D\,I) (see Table \ref{atomicdata}) with the PF method are not blended with \hmol~or stellar absorption (to the best of our knowledge), hence $N$(D\,I) is not affected by the \hmol~column density.

Weak absorption by O\,VI $\lambda$1032 is detected at the same velocity as the one of the low ionization interstellar metals. O\,VI $\lambda$1038 is blended with stellar absorption by O\,VI$^{**}$ and is not used in our analysis. We use AOD and PF to determine $N$(O\,VI) along this sightline. 

\begin{figure}
\begin{center}
\epsscale{0.85}
\plotone{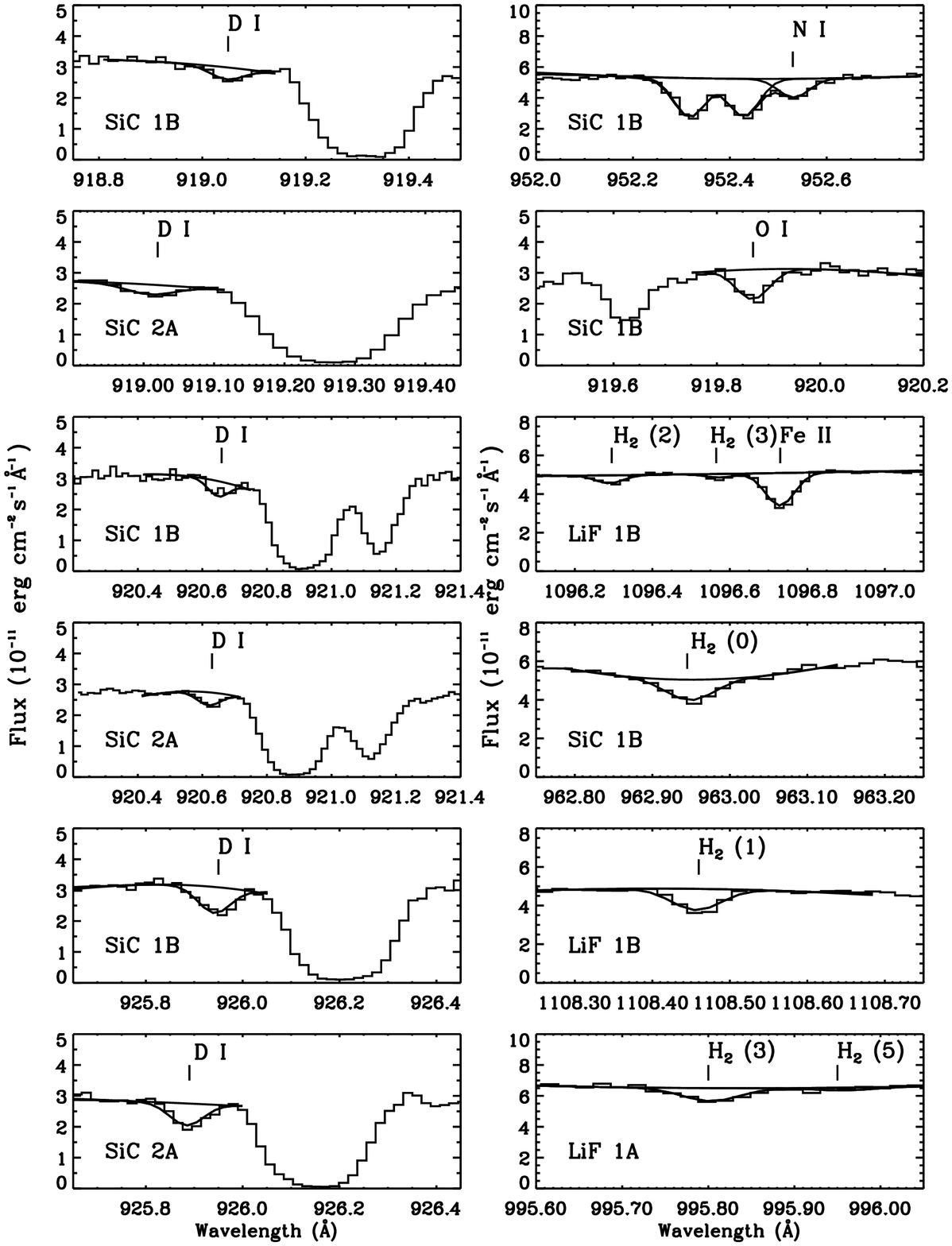}
\caption{Same as Fig. \ref{wd1034fits} but for the BD$+$39\,3226 sightline. \label{bd39fits}}
\end{center}
\end{figure}

Fits to some of the lines used in the analysis of this sightline are presented in Figure \ref{bd39fits}. Table \ref{nbd39} presents the column densities for D\,I, N\,I, O\,I, and Fe\,II along this sightline, obtained with the three different techniques.

\subsubsubsection{O\,I Column Density}
\label{oibd39text}



\begin{figure}
\begin{center}
\includegraphics[scale=.65,angle=90]{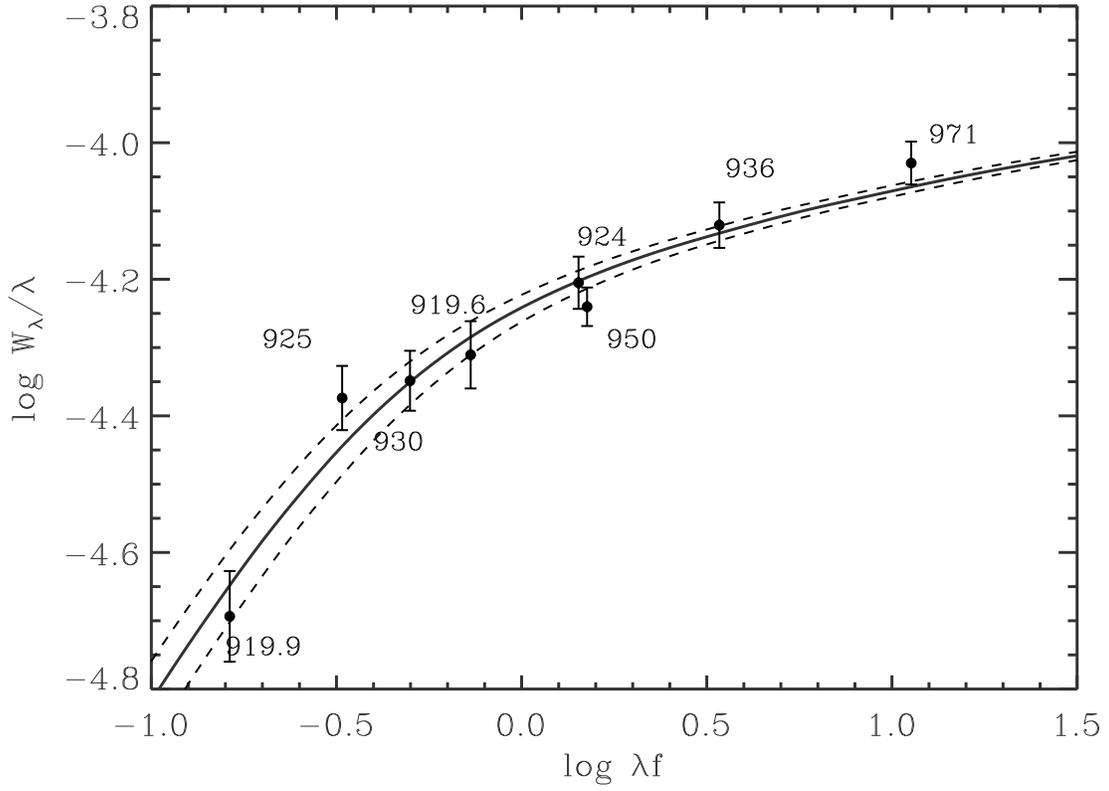}
\caption{Curve-of-growth for O\,I along the BD$+$39\,3226 sightline with the 1$\sigma$ uncertainties in $N$ plotted as dashed lines. The fit yields log $N$(O\,I) = 16.31 $\pm~^{0.07}_{0.06}$.\label{oicogbd39}}
\end{center}
\end{figure}

\begin{figure}
\begin{center}
\includegraphics[scale=.65,angle=90]{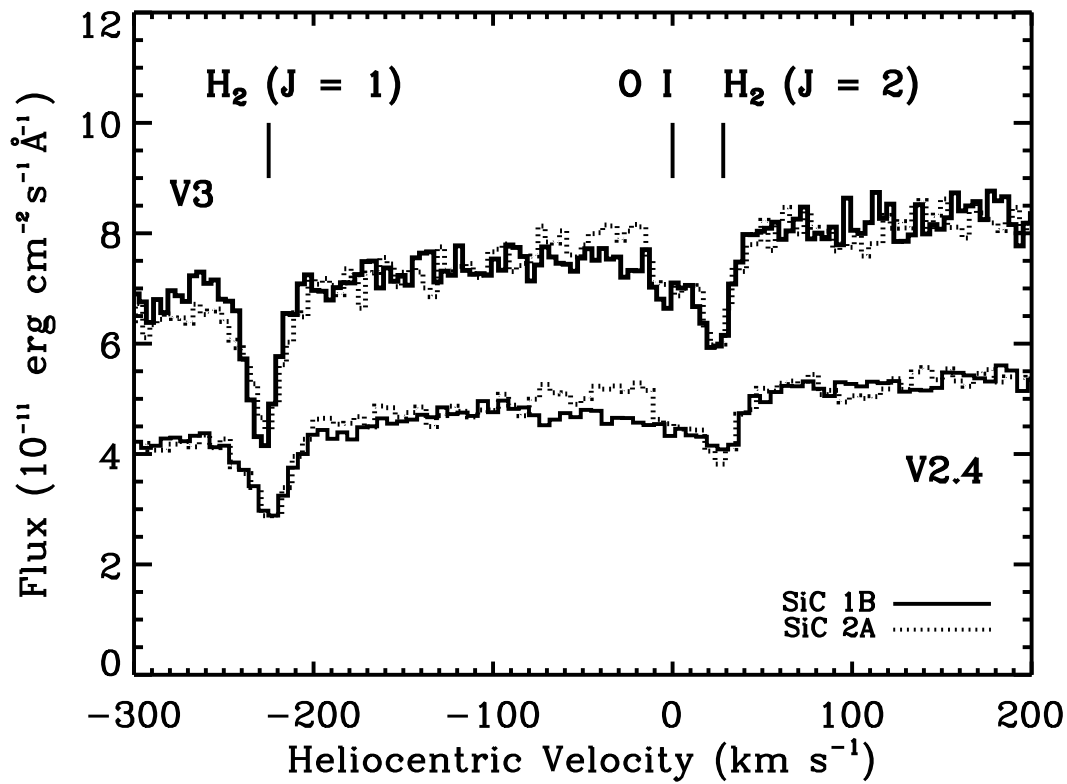}
\caption{Comparison between the SiC 1B ({\it solid}) and SiC 2A ({\it dotted}) data for the BD$+$39\,3226 sightline in the region of the O\,I $\lambda$974 transition. Data at the top were calibrated with version 3 of CalFUSE (V3), data at the bottom were calibrated with version 2.4 (V2.4) (data shifted vertically for clarity). The apparent discrepancy between the profiles around O\,I $\lambda$974 in the V2.4 SiC 1B and SiC 2A data disappears when V3 is used. See $\S$\ref{oibd39text} for discussion.\label{bd39channels}}
\end{center}
\end{figure}

To determine $N$(O\,I) along the BD$+$39\,3226 sightline we analyze the 919.9 (log ($f\lambda$) = $-$0.79), 925.4 (log ($f\lambda$) = $-$0.49), and 974 \AA~(log ($f\lambda$) = $-$1.82) O\,I lines. We derive log $N$(O\,I) = 16.33 $\pm$ 0.08 with the AOD method for the 919.9 \AA~transition; a similar value is obtained for O\,I $\lambda$925.4. We did not use the AOD technique on the weak $\lambda$974 transition, or measured its equivalent width, because this line is blended on the red side with \hmol~($J$ = 2). Figure \ref{oicogbd39} presents the curve-of-growth for O\,I along this sightline. From the COG we derive log $N$(O\,I) = 16.31 $\pm~^{0.07}_{0.06}$, also in good agreement with the AOD-derived $N$(O\,I) values from the 919.9 and 925 \AA~transitions (see Table \ref{nbd39}). Figure \ref{bd39channels} compares two sets of SiC 1B and SiC 2A data, in the vicinity of the O\,I $\lambda$974 transition (data shifted vertically for clarity). Data plotted at the top were calibrated with version 3 of CalFUSE (V3), data at the bottom were calibrated with version 2.4 (V2.4), which was used throughout this work. The discrepancy between V2.4 SiC 1B and SiC 2A data around O\,I $\lambda$974 disappears when CalFUSE V3 is used. Hence, for the O\,I $\lambda$974 transition, we use the V3 data (no other discrepancies between lines were found when the two datasets were compared). With the PF technique we determine $N$(O\,I) (as described in $\S$\ref{bd+39atomicdisc}) by assuming that D\,I, N\,I, O\,I, and Fe\,II trace the same absorption component. Hence, the $b$-value of each species is constrained by the $b$-values of the other species in the absorbing gas. Determining $N$(O\,I) with O\,I $\lambda$974 only leads to log $N$(O\,I) = 16.82 $\pm~^{0.06}_{0.07}$, while a similar fit with only O\,I $\lambda$919.9 and $\lambda$925.4 leads to log $N$(O\,I) = 16.51 $\pm$ 0.04. Fits using either $\lambda$919.9 or $\lambda$925.4 alone lead to $N$(O\,I) consistent with $N$(O\,I) quoted above when these two lines are fit together.

The discrepancy between the $N$(O\,I) values derived with or without the $\lambda$974 O\,I line ($>$ 0.30 dex in the log) could in principle indicate that there was a problem with saturation with some of the lines used in the analysis. However, the AOD determined $N$(O\,I) for $\lambda$919.9 is in good agreement with the one determined with the same method from the $\lambda$925 O\,I line, which is $\sim$2 times stronger. This indicates that there is little or no unresolved saturation in these lines, and hence that $N$(O\,I) derived from them is reliable. In addition, the COG-derived O\,I along this sightline (see Figure \ref{oicogbd39}) is in good agreement with the AOD results quoted above.

Overplotting log $N$(OI) = 16.82 on the profiles of the $\lambda$919.9 and $\lambda$925.4 O\,I lines clearly indicates that this column density is overestimated, unless one assumes that some unknown effect causes simultaneously the 925.4 \AA~and 919.9 \AA~lines, in both channels, to be narrower and shallower than they should be. This seems unlikely. Note that the $b$-values obtained in the fits of the three lines ($\lambda$974 alone, or $\lambda$919.9 together with $\lambda$925.4, or all three transitions simultaneously) is similar.

Three possibilities can be explored to try to explain the disagreement between N(O\,I) derived from $\lambda$974 and the one derived from the $\lambda$925 and $\lambda$919.9 O\,I lines: 1) there could be a problem with the $f$-value of the $\lambda$974 transition, 2) there could be a problem with the $f$-values of $\lambda$919.9 or $\lambda$925, and 3) there could be a stellar line blended with the $\lambda$974 O\,I transition and/or fixed pattern noise in one of the SiC channels (see below). A problem with the $f$-values of the $\lambda$974 or $\lambda$919.9 transitions seems unlikely as \citet{2003ApJ...599..297H} have reported that no inconsistencies were found between $\lambda$974 and $\lambda$919.9 in the analysis of the BD$+$28\,4211 sightline. In their O\,I analysis of the WD\,2211$-$495 sightline \citet{2002ApJS..140..103H} performed tests in which N(O\,I) derived independently from each individual O\,I transition (including amongst others the $\lambda$919.9 and $\lambda$925.4 transitions) was compared to N(O\,I) derived from fitting all the O\,I lines simultaneously. They found that N(O\,I) obtained from fits to the individual lines varied at most by 20\% from the value obtained when all the lines were fit simultaneously and concluded that the $f$-values of the O\,I transitions used in their study were consistent with each other. The study by \citet{2002ApJS..140..103H} indicates then that there are no large discrepancies between the $f$-values of the $\lambda$919.9 and $\lambda$925.4 O\,I transitions (at most 20\%). The discrepancy between $N$(O\,I) derived from $\lambda$974 and the one derived from the $\lambda$925 and $\lambda$919.9 O\,I lines is much larger than 20\%. Figure \ref{bd39channels} compares the SiC 1B and SiC 2A data for the BD$+$39\,3226 sightline in the O\,I $\lambda$974 region. The positions of two neighboring \hmol~lines are also marked. Even though O\,I is blended on the red side with \hmol~($J$ = 2), this blend should have little effect on the determination of $N$(O\,I) since the two lines are resolved and $N$(\hmol), $J$ = 2, is well constrained by other weak unblended $J$ = 2 lines.  The most likely possibility is then that part or all of the weak feature falling at the position of O\,I $\lambda$974 along the BD$+$39\,3226 sightline is of stellar origin. The spectra of the sdO BD$+$39\,3226 contains many absorption lines, likely of stellar origin, that we are not able to identify, and so it is not surprising that one of them would fall at the expected position of the O\,I $\lambda$974 line. We note that because stellar absorption is shifted by $\sim -$255 km s$^{-1}$ from its rest frame wavelength, the possible blend of O\,I $\lambda$974 with a stellar line does not affect the $N$(O\,I) determination along other sightlines when the $\lambda$974 transition is used. Finally, there is also the possibility, although unlikely, that $N$(O\,I) derived from the $\lambda$974 transition is correct and some unknown random effect causes $N$(O\,I) derived from $\lambda$919.9 and $\lambda$925.4 to agree even though these lines might suffer from saturation effects. Taking into account $N$(O\,I) obtained with the different methods and disregarding the value derived when the $\lambda$ 974 transition is used, we adopt log $N$(O\,I) = 16.40 $\pm$ 0.10.

\citet{2005guillaume} and \citet{2005friedman} have cautioned against not using the $\lambda$974 O\,I line when deriving $N$(O\,I) along high column density sightlines. In particular, \citet{2005guillaume} has determined a new $N$(O\,I) along the Feige\,110 sightline using the transition above and found it to be a factor of $\sim$2 larger than the original $N$(O\,I) reported by \citet{2002ApJS..140...37F}, which did not use the O\,I $\lambda$974 line. There are however differences between our work and that of \citet{2005guillaume} and \citet{2005friedman} as neither of these authors used the apparent optical method in their analyses and in particular to compare $N$(O\,I) derived from the $\lambda$919.9 and $\lambda$925.4 lines. As three of the authors of this paper also participated in the original analysis of the Feige\,110 sightline by \citet{2002ApJS..140...37F} we have access to the same dataset used by \citet{2005guillaume} in the new analysis. Using this dataset we performed AOD measurements of the $\lambda$919.9, $\lambda$925.4, and $\lambda$974 O\,I lines and found that $N$(O\,I) derived from $\lambda$925.4 (log $N$(O\,I) = 16.43 $\pm$ 0.03) is $\sim$5$\sigma$ smaller than $N$(O\,I) derived from the $\lambda$919.9 line (log $N$(O\,I) = 16.58 $\pm$ 0.03). From the $\lambda$974 line we derive log $N$(O\,I) = 16.98 $\pm$ 0.08 (1$\sigma$), in perfect agreement with the new log $N$(O\,I) = 17.06 $\pm$ 0.15 (2$\sigma$) quoted by \citet{2005guillaume}. The different results we obtained from the $\lambda$919.9 and $\lambda$925.4 lines for the Feige\,110 sightline are a strong indication of saturation of one or both of these two lines, and unless one compares the AOD derived $N$(OI) from $\lambda$919.9 with $N$(OI) from a weaker line one can not decide whether $\lambda$919.9 suffers from saturation or not. As mentioned before, in our case the $N$(OI) results from $\lambda$919.9 and $\lambda$925.4 lines are in excellent agreement, which gives us confidence that our results are not biased due to not using the $\lambda$974 transition.

\subsubsection{TD1\,32709}
\label{uv0904atomic}

Along this sightline the separation between stellar and interstellar absorption is $\sim$13 km s$^{-1}$. We use the stellar model computed to determine $N$(H\,I) (see \S\ref{nhi} below) to ascertain which ISM lines might suffer from stellar blendings. The stellar model indicates that the profiles of the D\,I lines used in this analysis (see Table \ref{atomicdata}) are not affected by the shape of the stellar continuum in the vicinity of these lines, hence no stellar model is needed to determine the column densities discussed below. In addition, the small molecular hydrogen content along this sightline (log $N$(\hmol) = 14.48 $\pm~^{0.12}_{0.11}$, see \S\ref{h2td1}) implies that the deuterium column density derived below is not sensitive to $N$(\hmol).

We use the AOD technique on weak lines to determine the column densities of D\,I, N\,I, O\,I, and Fe\,II, which are then compared to $N$ derived with the other methods. C\,II, C\,II$^*$, Si\,II, P\,II, and Ar\,I only have saturated transitions in the {\it FUSE} bandpass. We use the AOD method to determine lower limits to their column densities. We use the COG technique to determine the column densities of D\,I, N\,I, O\,I, and Fe\,II. The $b$-values derived from the four COGs fall in the range 3 -- 6 \kms. With PF we fit a single absorption component to unsaturated lines of D\,I, N\,I, O\,I, and Fe\,II. \hmol~is also included in the fit, but in a separate absorption component. In order to provide a smooth continuum for the D\,I lines we also fit H\,I as an independent component. This fit, however, is not used to determine $N$(H\,I). Absorption by O\,VI ($\lambda$1032, 1037) is also detected along this sightline but, the small separation between stellar and interstellar absorption along this sightline ($\sim$ 13 km s$^{-1}$) does not allow us to determine if this absorption is of stellar or interstellar absorption origin (or both), as stellar and interstellar absorptions are blended.

\begin{figure}
\begin{center}
\includegraphics[scale=.65,angle=90]{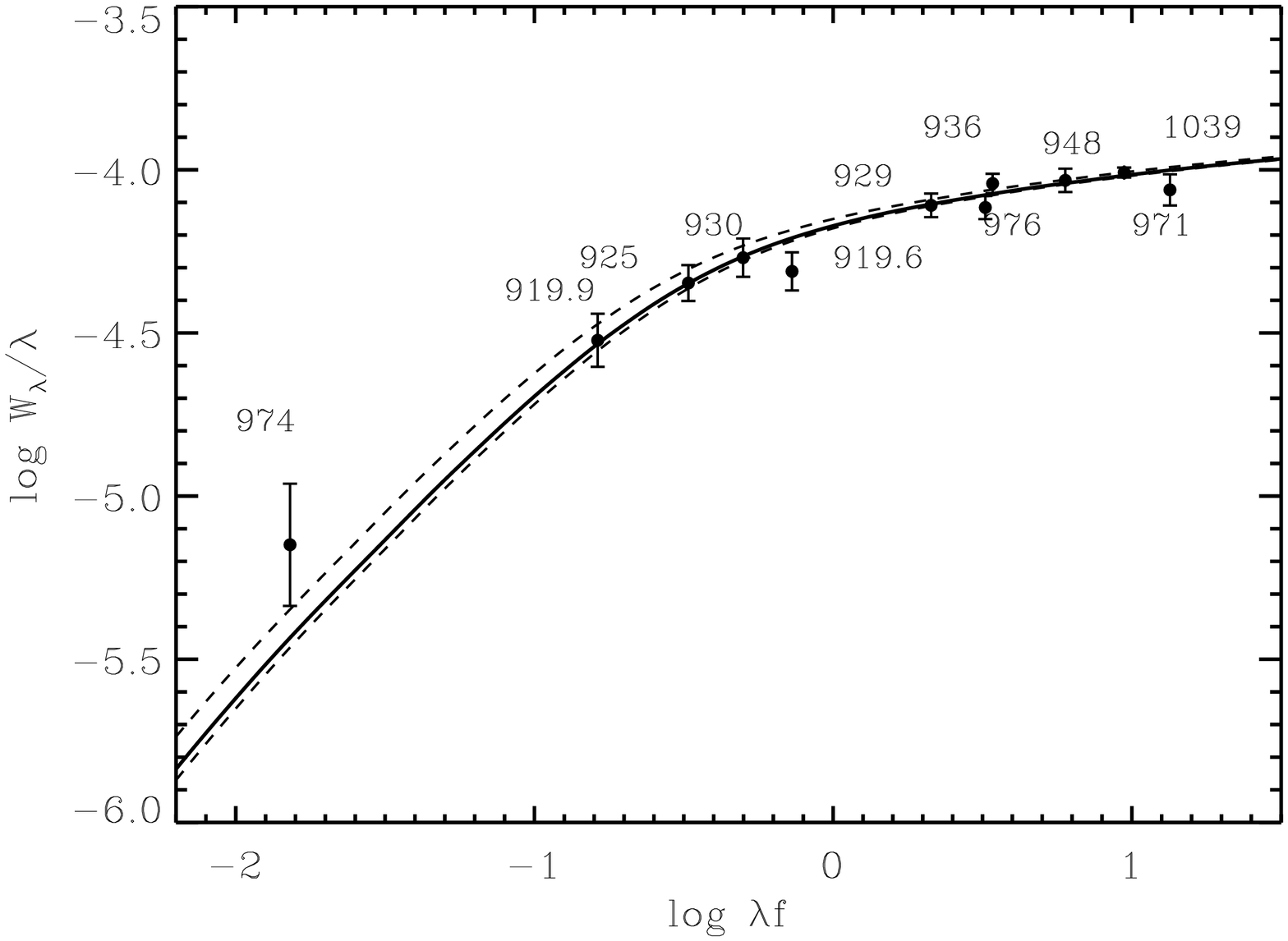}
\caption{Curve of growth for O\,I along the TD1\,32709 sightline. We derive log $N$(O\,I) = 16.45 $\pm~^{0.09}_{0.03}$. Dashed lines indicate the 1 $\sigma$ uncertainty in $N$.\label{uv0904oicog}}
\end{center}
\end{figure}

\begin{figure}
\begin{center}
\epsscale{0.85}
\plotone{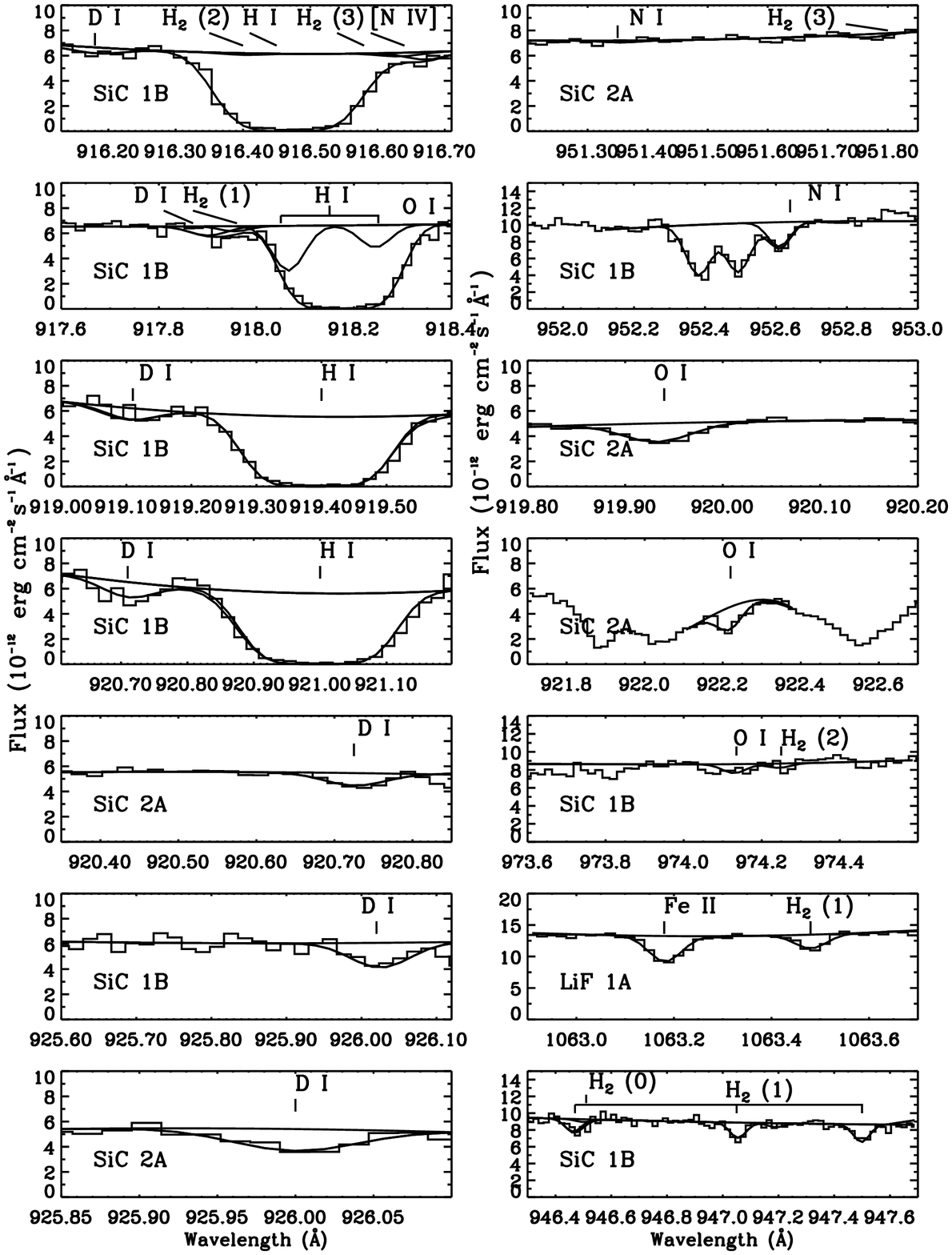}
\caption{Same as Fig. \ref{wd1034fits} but for the TD1\,32709 sightline. \label{uv0904fits}}
\end{center}
\end{figure}

The curve of growth for O\,I along this sightline is presented in Figure \ref{uv0904oicog}. We derive log $N$(O\,I) = 16.45 $\pm~^{0.09}_{0.03}$. Figure \ref{uv0904fits} presents fits to some of the lines used in the analysis of the TD1\,32709 sightline. Table \ref{ntd1} presents the column densities for D\,I, N\,I, O\,I, and Fe\,II along this sightline, obtained with the three different techniques.

\subsection{H\,I Column Densities}
\label{nhi}

\subsubsection{WD\,1034$+$001}
\label{hi-wd1034}

To determine $N$(H\,I) along this sightline we use data obtained by the GHRS onboard {\it HST}. GHRS observations were obtained in June of 1992, with the G160M grating and the LSA aperture (R $\sim$ 20,000) for a total exposure time of 288 s (rootname: z0ye0c08t). Unfortunately these observations cover only the wavelength range 1185 -- 1221 \AA, which does not include the complete red wing of \lya. However, the existing data still allow us to determine $N$(H\,I).

The calibrated data were retrieved from the Multimission Archive at the Space Telescope Science Institute and no further processing was applied. The error-weighted data, obtained at four different FP-split positions, were combined to increase the signal to noise ratio of the final dataset. We consider the influence on the \lya~profile of several stellar models with different effective temperatures and gravities. These are described in detail below.

\subsubsubsection{Stellar Model}

\begin{figure}
\begin{center}
\includegraphics[scale=.75,angle=90]{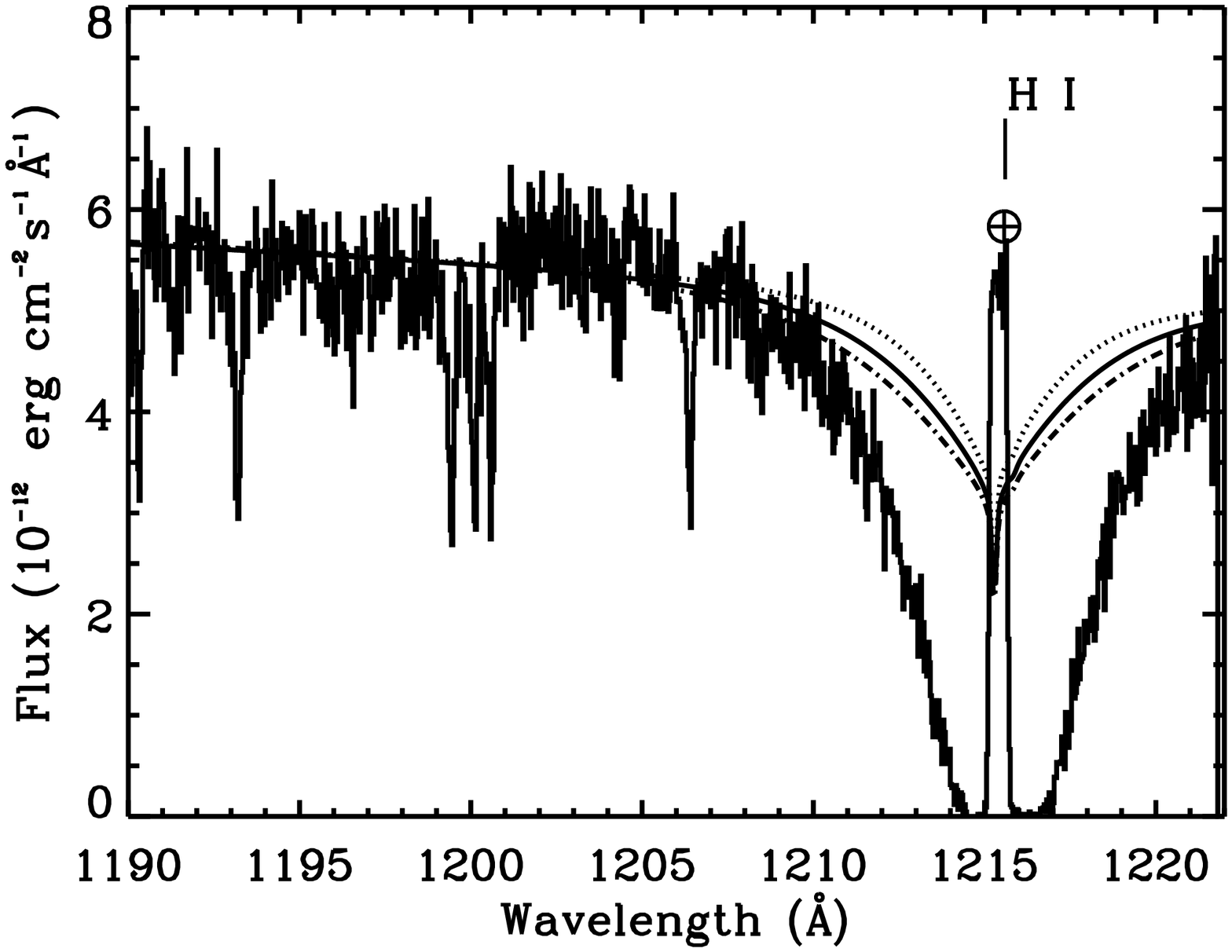}
\caption{\lya~GHRS observations of the WD\,1034$+$001 sightline. Three stellar models are overplotted: \teff~=~100,000 K, log$g$~= 7.5 (best fit model, solid line), \teff~=~115,000 K, log$g$~= 7.2 (dotted line), and \teff~=~90,000 K, log$g$~= 7.8 (dash-dotted line). All the models have log (He/H) = 3.0. The two last models produce the highest and lowest $N$(H\,I), respectively, and are used to determine the uncertainties in $N$(H\,I) associated with the stellar models. Geocoronal emission is annotated with $\oplus$. We derive log $N$(H\,I) = 20.07 $\pm$ 0.07 (1$\sigma$).\label{wd1034himodel}}
\end{center}
\end{figure}

We computed stellar atmosphere models using the atmospheric parameters determined by \citet{1995A&A...298..567W}, i.e., \teff~=~100,000 $\pm~^{15,000}_{10,000}$ K, log $g$ = 7.5 $\pm$ 0.3, and log (He/H) = 3.0. NLTE H+He models were computed with the stellar atmosphere codes TLUSTY and SYNSPEC \citep[see, e.g.,][]{1995ApJ...439..875H}. The line profile of the \ion{He}{2}$\lambda$1215 line uses an approximate Stark broadening treatment developed by \citet{1994A&A...282..151H}, while the hydrogen line profile uses the \citet{1973ApJS...25...37V} theory \citep{1997A&AS..122..285L}. We define as the best fit model the stellar model with \teff~=~100,000 K, log $g$ = 7.5, and log (He/H) = 3.0. The 1 $\sigma$ uncertainties in the stellar parameters determined by \citet{1995A&A...298..567W} are used to compute stellar models that allow us to determine the systematic uncertainties in $N$(H\,I) associated with the uncertainties in the stellar models. Determining the H\,I column density with different stellar models places more credible error bars on $N$(H\,I) than using a single stellar model. The statistical uncertainties are derived by using the best fit stellar model. The systematic uncertainties associated with the stellar models are estimated by considering the most extreme stellar models, i.e., those that yield the strongest and weakest stellar \lya. Figure \ref{wd1034himodel} illustrates such stellar profiles. The figure shows that the best fit stellar model (\teff~=~100,000 K, log$g$ = 7.5; solid line) produces a \lya~line profile that is between the strongest \lya~line (\teff~=~90,000 K, log$g$ = 7.8; dash-dotted line) and the weakest one (\teff~=~115,000 K, log$g$ = 7.2).

\subsubsubsection{\lya~Profile Fitting}

We use the separation between stellar and interstellar absorptions derived by \citet{1985ApJ...292..477S} and \citet{1995A&A...298..567W} to align the data and the stellar models. 
 The different models are then scaled by different factors in order to match the spectrum flux over the region 1190 -- 1195 \AA. The spectrum is then normalized by the different models prior to fitting. 

To determine the H\,I column density we fit three absorption components with independent velocities, one each for H\,I, D\,I, and \hmol~($J$ = 0 -- 3). $N$(D\,I) and $N$(\hmol) are fixed at the adopted values (see Tables \ref{h2} and \ref{natomic}) during the fitting procedure. The $b$-value of the component containing D\,I is fixed at $b$ = 5.8 \kms, derived from the D\,I COG. In order to constrain $b_{\rm H\,I}$~we fit also the higher order H\,I Lyman lines 917 \AA, 918 \AA, 919 \AA, 920 \AA, and 926 \AA~(923~\AA~is blended with stellar N\,IV). The Si\,III line ($\lambda$1206) is also included in the \lya~fit, in a fourth independent absorption component, so we can use a large continuum region in the blue side of the \lya~wing. The normalized data is fitted in the range 1201.15 -- 1221.28 \AA; the continuum is fixed at 1.0 during the fitting procedure. 

\begin{figure}
\begin{center}
\epsscale{1}
\plotone{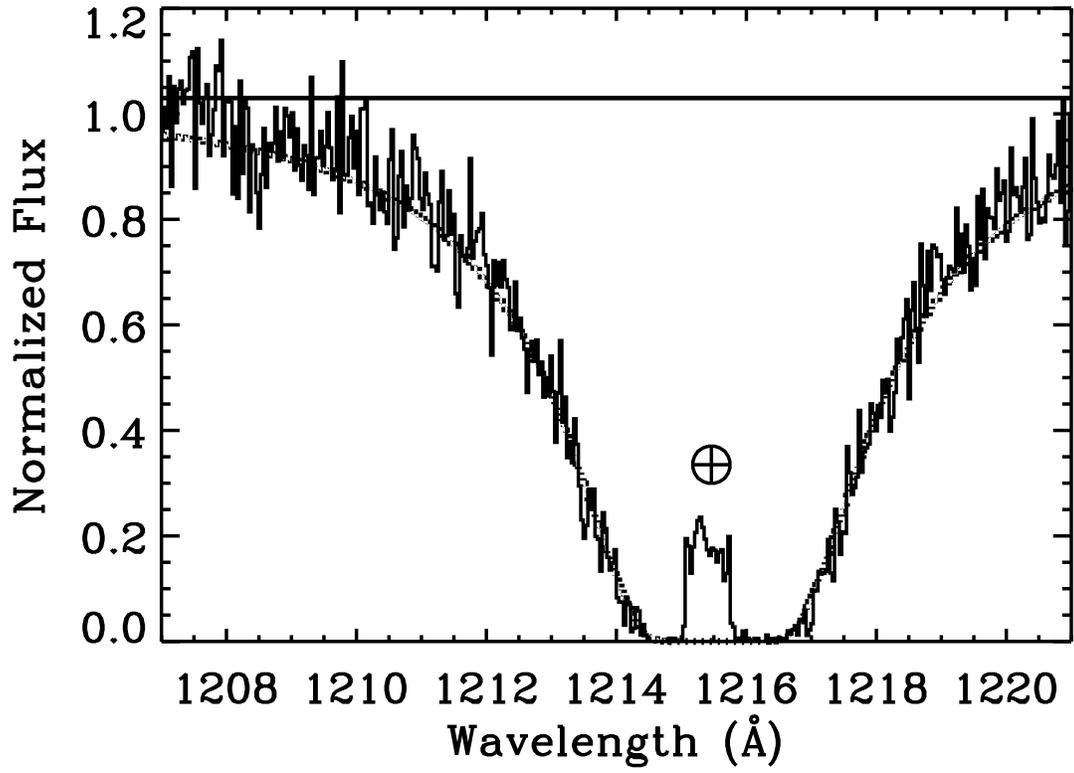}
\caption{Fit to the \lya~interstellar absorption along the WD\,1034$+$001 sightline, using the best fit stellar model to normalize the data. Terrestrial airglow is labeled with $\oplus$. \label{wd1034hifit}}
\end{center}
\end{figure}

The  statistical uncertainties associated with $N$(H\,I) are determined by fitting the data normalized by the best fit model, following the procedure outlined in \citet{2003ApJ...587..235O}. To determine the uncertainties associated with the continuum placement due to the uncertainties in the stellar model we perform two additional fits, in which the continuum level is multiplied by factors of 1.07 and 0.93 during the fitting procedure (corresponding to a 7\% change in the continuum level or to a '$\chi$-by-eye' of 2 $\sigma$). As mentioned above, the effect of systematic uncertainties associated with the stellar models are determined by using the extreme models discussed above to measure the highest and lowest interstellar $N$(H\,I). These models are illustrated in Figure \ref{wd1034himodel}. Figure \ref{wd1034hifit} presents the fit to the \lya~region along this sightline, using the best fit stellar model to normalize the data.

We take a conservative approach to combine the different uncertainties by adding them, rather than adding them in quadrature. Taking into account the different uncertainties in the manner discussed above we determine log $N$(H\,I) = 20.07 $\pm$ 0.07 (1$\sigma$).

\subsubsection{TD1\,32709}
\label{hi-uv0904}

To determine $N$(H\,I) along this sightline we use data obtained with {\it IUE} and {\it FUSE}. High dispersion, large aperture {\it IUE} observations of the \lya~region were obtained in March 1981 (swp13459), March 1994 (swp50226, swp50227), and December 1993 (swp49676, swp49677), for a total exposure time of 53940 s. The calibrated NEWSIPS MXHI data were retrieved from the Multimission archive at the Space Telescope Science Institute. The data were not processed further. 

The different observations of \lya~were coadded (weighted by their uncertainties) to increase the signal-to-noise ratio of the final dataset. A set of stellar models, described below, were computed to take into account the stellar absorption in the vicinity of \lya. Similarly to WD\,1034$+$001, we consider the influence of several stellar models with different temperatures and gravities, on the \lya~profile.

\subsubsubsection{Stellar Model}

We computed NLTE H+He+C+N models using the atmospheric parameters obtained by \citet{1993A&A...273..212D} from fitting the optical spectrum of TD1\,32709. These authors determined \teff~=~46,500 $\pm$ 1,000 K, log $g$ = 5.55 $\pm$ 0.15, and log(He/H) = 2. We define as the best fit model the model corresponding to \teff~=~46,500 K, log $g$ = 5.55, and log(He/H) = 2, and use the uncertainties in the stellar parameters to compute stellar models which are used to determine the uncertainties in $N$(H\,I) associated with the stellar models. We considered also the influence on the adopted He/H by computing another model with \teff~=~45,000 K, log$g$ = 5.7 and log(He/H) = 1. This produced $N$(H\,I) within the uncertainties of the adopted value.

\begin{figure}
\begin{center}
\includegraphics[scale=.75,angle=90]{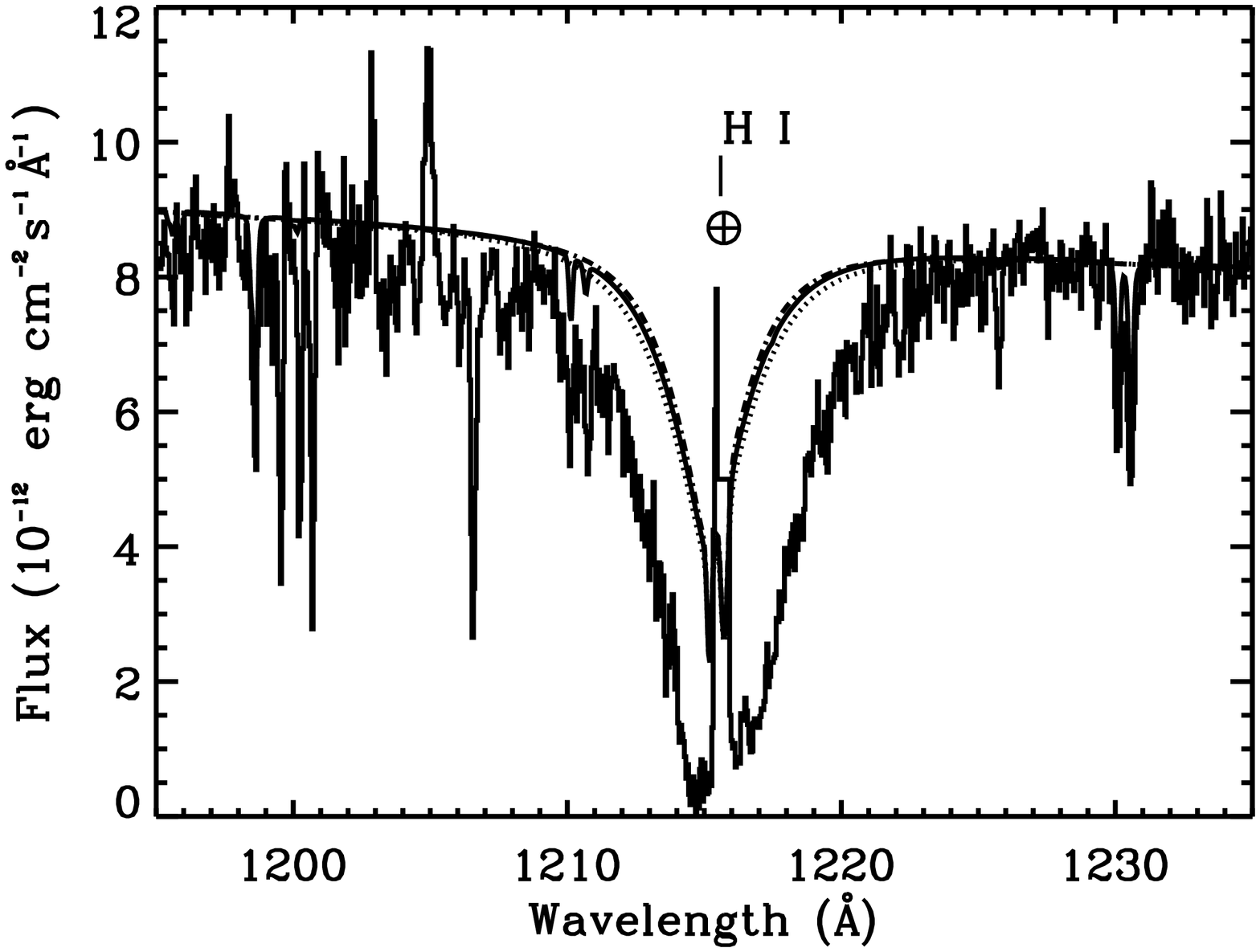}
\caption{\lya~{\it IUE} observations of the TD1\,32709 sightline. Three stellar models are overplotted: \teff~=~46,500 K, log$g$~= 5.55 (best fit model, solid line), \teff~=~45,000 K, log$g$~= 5.7 (dotted line), and \teff~=~47500 K, log$g$~= 5.4 (dash-dotted line). All the models have log (He/H) = 2.0. The two last models produce the lowest and highest $N$(H\,I), respectively, and are used to determine the uncertainties in $N$(H\,I) associated with the stellar models. Geocoronal emission is annotated with $\oplus$. We derive log $N$(H\,I) = 20.03 $\pm$ 0.10 (1$\sigma$).\label{uv0904himodel}}
\end{center}
\end{figure}

Figure \ref{uv0904himodel} presents the best fit stellar model (solid line), corresponding to \teff~=~46,500 K and log$g$ = 5.55, together with the models that produce the strongest (\teff~=~45,500 K and log$g$ = 5.7; dashed line) and weakest (\teff~=~47,500 K and log$g$~=~5.4; dash-dotted line) stellar absorption. These models are used to determine the systematic uncertainties in $N$(H\,I) associated with the stellar models.

\subsubsubsection{\lya~Profile Fitting}

We use the stellar C\,IV doublet at $\lambda$1230 to align the stellar models with the data. The stellar models are then scaled by different factors to match the spectrum flux in the 1200 \AA~and 1225 \AA~regions. Prior to fitting we normalize the data by the different stellar models.

\begin{figure}
\begin{center}
\epsscale{0.75}
\plotone{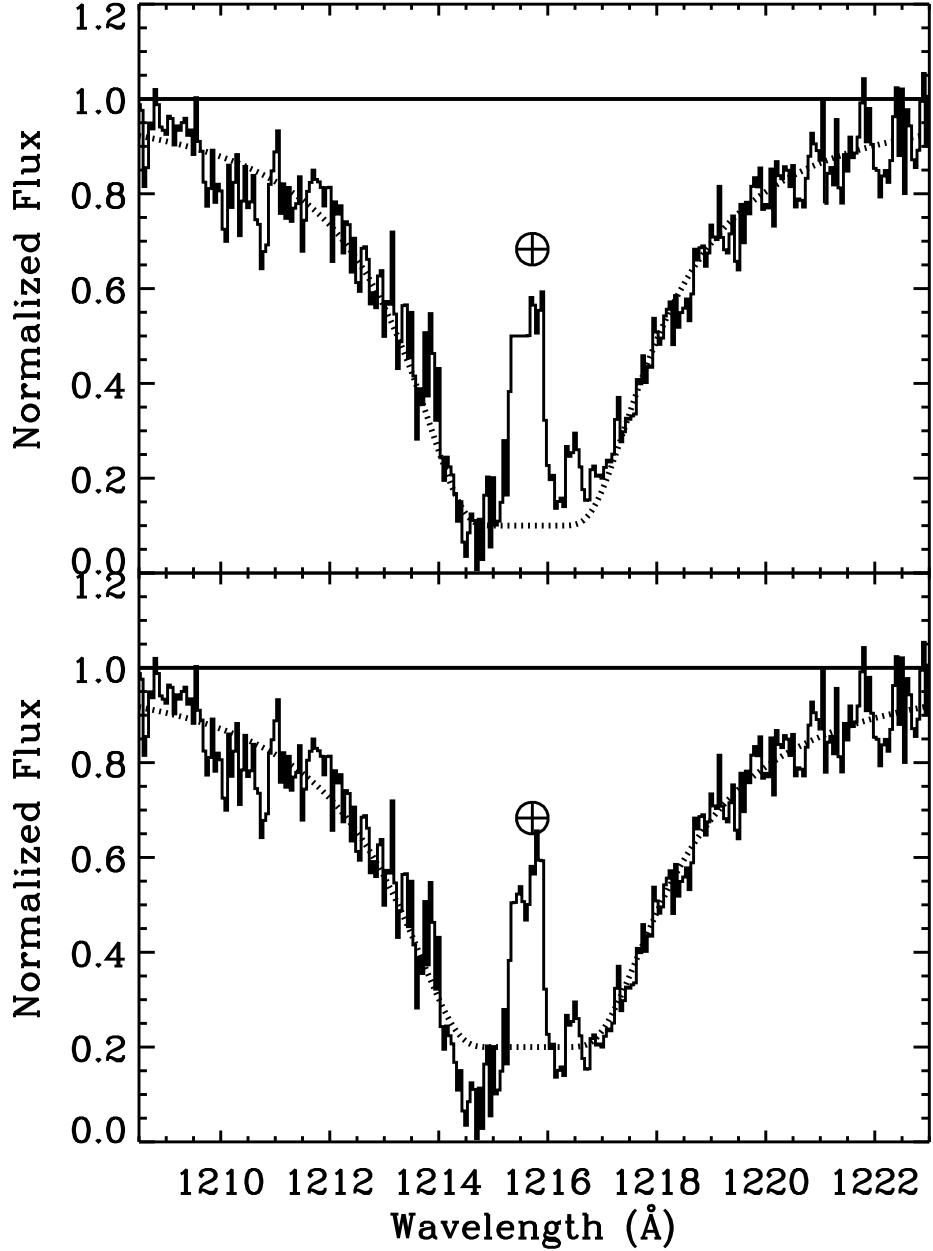}
\caption{Fit to the \lya~interstellar absorption along the TD1\,32709 sightline, using the best fit stellar model to normalize the data. {\it Top panel}: The blue wing of the \lya~line is used to define the zero flux level, yielding log$N$(H\,I) = 20.03. {\it Bottom panel}: The red wing of the \lya~line is used to define the zero flux level, yielding log$N$(H\,I) = 20.12, within the 1 $\sigma$ uncertainty quoted in Table \ref{natomic} for $N$(H\,I) along this sightline. \label{uv0904hifit}}
\end{center}
\end{figure}

To determine $N$(H\,I) we fit three absorption components, one each for H\,I, D\,I + O\,I, and \hmol~(J = 0 -- 2). The column densities of all the species but H\,I, are fixed at the adopted values (see Tables \ref{natomic} and \ref{h2}) during the fitting procedure. The $b$-value of the component containing D\,I and O\,I is fixed at $b$ = 6.0 \kms, derived from the D\,I and O\,I COG analyses. Higher order Ly lines (916 -- 919 \AA) are also included in the fit, to constrain $b_{\rm H\,I}$. These Ly lines were chosen because our models indicate that the interstellar absorption spectrum is not affected by stellar absorption (either from H or He) in this region. Geocoronal emission fills part of the core of the \lya~line and is taken into account during the fitting procedure by introducing a background in the fit. We determine the background level using the blue side of the \lya~core; the airglow emission visible in Figure \ref{uv0904himodel} is not allowed to influence the profile fitting procedure. We estimate the errors associated with this background by performing multiple fits, using the best fit model, where the background is varied between 6 and 12\% of the continuum level. Similarly, errors associated with the continuum level are determined by varying the continuum level, using the best fit model. Figure \ref{uv0904hifit} (top panel) presents the fit to the \lya~interstellar absorption when the best fit stellar model is used to normalize the data and with the zero-flux level defined by the blue side of the \lya~wing, yielding log $N$(H\,I) = 20.03. A similar fit, but with the zero-flux level defined by the red side of the \lya~wing is shown in the bottom panel of this figure. This fit yields log $N$(H\,I) = 20.12. The small emission line on the red side of the airglow emission is likely an artifact of the {\it IUE} data similar to the ones studied by \citet{1996PASP..108..925C}. Taking into account this feature when defining the zero-flux level leads to $N$(H\,I) in the range quoted above (20.3--20.12). The statistical uncertainties associated with $N$(H\,I) are determined by fitting the data normalized by the best fit model, following the procedure outlined in \citet{2003ApJ...587..235O}. Taking all these uncertainties into account, by adding them as in the case of WD\,1034$+$001, we determine log $N$(H\,I) = 20.03 $\pm$ 0.10.

\section{RESULTS}
\label{results}

Table \ref{ratios} presents several ratios, using the column densities summarized in Table \ref{natomic}. The fraction of \hmol~along these sightlines, $f_{\rm H_2}$ = 2$\times$$N$(\hmol)/(2$\times$$N$(\hmol) + $N$(H\,I)), is log $f_{\rm H_2}$ = $-$4.05 $\pm$ 0.58, $-$4.13 $\pm$ 0.46, and $-$5.25 $\pm$ 0.80 for WD\,1034$+$001, BD$+$39\,3226, and TD1\,32709, respectively. Hence, for the three sightlines, $N$(H) = $N$(H\,I) $+$ 2$\times$$N$(\hmol) $\approx$ $N$(H\,I).

\subsection{D/H Ratios}

We derive (D/H)$\times10^5$ = 2.14 $\pm~^{0.53}_{0.45}$, 1.17 $\pm~^{0.31}_{0.25}$, and 1.86 $\pm~^{0.53}_{0.43}$, for WD\,1034$+$001, BD$+$39\,3226, and TD1\,32709, respectively. Only the D/H ratio toward TD1\,32709 is consistent, at the 1$\sigma$ level, with the LB value (D/H = (1.56 $\pm$ 0.04)$\times10^{-5}$) derived by \citet{2004ApJ...609..838W}. The value of D/H toward WD\,1034$+$001 is one of the highest D/H ratios measured in the nearby interstellar medium. The other high values are (1.91 $\pm~^{0.26}_{0.24}$)$\times10^{-5}$ \citep[PG\,0038$+$199,][]{2005astro.ph..1320W}, (2.14 $\pm$ 0.41)$\times10^{-5}$ \citep[Feige\,110,][]{2002ApJS..140...37F}, (2.18 $\pm~^{0.22}_{0.19}$)$\times10^{-5}$ \citep[$\gamma^2$ Vel,][]{2000ApJ...545..277S}, (2.24 $\pm~^{0.64}_{0.52}$)$\times10^{-5}$ \citep[$\alpha$ Cru,][]{1976ApJ...203..378Y} and (2.24 $\pm~^{0.70}_{0.66}$)$\times10^{-5}$ \citep[LSE\,44][]{2005friedman}. 

The D/H values reported above are averages over the entire sightline; however they are dominated by the gas beyond the Local Bubble. WD\,1034$+$001 is the shortest sightline analyzed in this work, $d$ = 155 $\pm~^{58}_{43}$ pc. Even though the uncertainties on its distance put it close to the Local Bubble boundary \citep[$\sim$100 pc,][however, in some directions the boundary can be at 65 or 250 pc]{1999A&A...346..785S}, the bulk of the gas probed by this sightline is outside the Local Bubble (log $N$(H) $<$ 19.2). The Local Bubble contribution to the WD\,1034$+$001 D and H column densities can be estimated by assuming log$N$(H) = 19.2 and using D/H = 1.56$\times10^{-5}$ derived by \citet{2004ApJ...609..838W} for the Local Bubble to determine log$N$(D)$_{\rm LB}$ = 14.39. Subtracting the LB contribution yields then D/H = 2.23$\times10^{-5}$ for the gas outside the LB, an insignificant change from the D/H ratio quoted in Table \ref{ratios} for this sightline and consistent with the quoted uncertainties (D/H$\times10^5$ = 2.14 $\pm~^{0.53}_{0.45}$). For the BD$+$39\,3226 and TD1\,32709 subtracting the LB contribution to the sightline D/H ratios in the manner described above leads to D/H$\times10^5$ = 1.12 and 1.92, respectively, consistent within the uncertainties with the ratios reported in Table \ref{ratios} for these two sightlines.

\subsection{O/H Ratios}

We derive (O/H)$\times10^4$ = 3.39 $\pm~^{1.06}_{0.86}$, 2.09 $\pm~^{0.72}_{0.58}$, and 2.45 $\pm~^{0.90}_{0.71}$, for WD\,1034$+$001, BD$+$39\,3226, and TD1\,32709, respectively. The first and last ratios are consistent, within the quoted uncertainties, with O/H = (3.43 $\pm$ 0.15)$\times10^{-4}$ derived for 13 lines of sight probing gas within 1500 pc (most within 500 pc), with 20.18 $\leq$ log $N$(H\,I) $\leq$ 21.28 \citep[][updated $f$-value]{1998ApJ...493..222M}. O/H along the BD$+$39\,3226 sightline is $\approx$1.86$\sigma$ away from the \citet{1998ApJ...493..222M} ratio. Taking into account the Local Bubble contribution to the O/H ratio along the three sightlines using log$N$(H) = 19.2 and the O/H ratio derived by \citet{2005ApJ...625..232O} for the Local Bubble, O/H$\times10^4$ = 3.45 $\pm$ 0.19, leads to O/H ratios for the gas beyond the LB consistent within the uncertainties with the values quoted in Table \ref{ratios}.

\subsection{D/O Ratios}

We derive (D/O)$\times10^2$ = 6.31 $\pm~^{1.79}_{1.38}$, 5.62 $\pm~^{1.61}_{1.31}$, and 7.59 $\pm~^{2.17}_{1.76}$, for WD\,1034$+$001, BD$+$39\,3226, and TD1\,32709, respectively. \citet{2003ApJ...599..297H} derived D/O = (3.84 $\pm$ 0.16)$\times10^{-2}$ for the Local Bubble from measurements along 14 sightlines. All the D/O ratios reported here are inconsistent, at the 1$\sigma$ level, with the Local Bubble value, and are some of the highest D/O ratios measured in the ISM. The implications of this will be discussed below.

\subsection{N/H, O/N, and D/N Ratios}

The N/H ratios computed for the three sightlines, (N/H)$\times10^5$ = 7.76 $\pm~^{2.82}_{2.20}$, 5.89 $\pm~^{2.04}_{1.63}$, and 8.91 $\pm~^{3.26}_{2.59}$ (for WD\,1034$+$001, BD$+$39\,3226, and TD1\,32709, respectively), are all consistent, within the uncertainties, with N/H = (7.5 $\pm$ 0.4)$\times10^{-5}$ derived by \citet{1997ApJ...490L.103M}.

We derive O/N = 4.37 $\pm~^{1.79}_{1.38}$, 3.55 $\pm~^{1.30}_{1.03}$, and 2.75 $\pm~^{1.01}_{0.80}$. These are consistent, at the 1$\sigma$ level (1.34$\sigma$ for TD1\,32709), with O/N = 4.1 $\pm$ 0.3 derived using the values of O/H and N/H determined by \citet{1998ApJ...493..222M} and \citet{1997ApJ...490L.103M}.

The D/N ratios for the three sightlines are (2.75 $\pm~^{1.00}_{0.78}$)$\times10^{-1}$, (2.00 $\pm~^{0.57}_{0.46}$)$\times10^{-1}$, and (2.09 $\pm~^{0.60}_{0.49}$)$\times10^{-1}$ (for WD\,1034$+$001, BD$+$39\,3226, and TD1\,32709, respectively).

\subsection{$N$ from {\it FUSE} vs. $N$ from {\it ORFEUS} for the BD$+$39\,3226 sightline}

Table \ref{Ncomparison} displays, side by side, the column densities of several species derived in this work from {\it FUSE} data and the column densities derived by \citet{1999A&A...352..287B} using {\it ORFEUS} and {\it IUE} data. The column densities for D\,I, O\,I, Fe\,II, and \hmol~($J$ = 0, 1, 3) are consistent within the quoted uncertainties. For these species, using the {\it FUSE} data to determine their column densities leads in all cases to a significant improvement of the uncertainties. This is particularly important for D\,I, N\,I, and O\,I, for which we want to determine ratios relative to H\,I as accurate as possible. The value of $N$(N\,I) determined by \citet{1999A&A...352..287B} is substantially smaller than the one determined in this work; log $N$(N\,I) = 14.75 $\pm$ 0.25 vs. 15.85 $\pm$ 0.10. We suspect this is because the weakest N\,I line used by \citet{1999A&A...352..287B} to determine $N$(N\,I) ($\lambda$964.6256) suffers from considerable saturation, leading to an underestimation of the value $N$(N\,I). 


\section{DISCUSSION}
\label{discussion}

\subsection{WD\,1034$+$001}
\label{wd1034_disc}

The WD\,1034$+$001 sightline has one of the highest D/H ratios reported in the literature, D/H = (2.14 $\pm~^{0.53}_{0.45}$)$\times10^{-5}$, 1.29$\sigma$ above D/H$_{\rm LB}$ derived by \citet{2004ApJ...609..838W}. However, its O/H and N/H ratios are consistent at the 1$\sigma$ level with previous measurements \citep{1998ApJ...493..222M,1997ApJ...490L.103M}. A D/H ratio consistent with the LB value ($\sim$1.56$\times10^{-5}$), would require log$N$(H\,I) = 20.21. In such case, the O/H and N/H ratios (2.47 $\pm~^{0.80}_{0.63}$)$\times10^{-4}$ and (5.66 $\pm~^{2.06}_{1.60}$)$\times10^{-5}$, respectively) would still be consistent with the values derived by \citet{1998ApJ...493..222M} and \citet{1997ApJ...490L.103M}, while not affecting the high D/O ratio (the same percentage uncertainties were used for the required $N$(H\,I) as the ones derived in this work). In this scenario $\sim$ 34\% of the H\,I would have been missed in our analysis. It seems unlikely that $N$(H\,I) along this sightline could be underestimated by such a large factor. On the other hand, a value of log$N$(D\,I) = 15.26 (38\% smaller than our derived value) together with our derived $N$(H\,I) would also lead to D/H consistent with the LB value, without affecting the O/H and N/H ratios. In this scenario, using the same uncertainties for the required $N$(D\,I) as the ones derived in this work, we would find D/O = (4.57 $\pm~^{1.43}_{1.16}$)$\times10^{-2}$. This is lower than our derived value of (6.31 $\pm~^{1.79}_{1.38}$)$\times10^{-2}$, and consistent with the Local Bubble value derived by \citet{2003ApJ...599..297H}. We derived $N$(D\,I) along this sightline using three different methods, AOD, COG, and PF (see Table \ref{atomicdata}), and it is unlikely that $N$(D\,I) has been overestimated.

Linsky et al. (2005, in prep) looked at correlations between the D/H ratio and other parameters, such as the depletions of Fe and Si, and the kinetic temperature of \hmol~($T_{01}$), in order to try to understand what is the total deuterium abundance in the local Galactic disk. They found that sightlines with higher Fe and Si depletions correspond in general to lower D/H ratios and that sightlines with high  $T_{01}$~have typically higher D/H ratios.

The Fe/H ratio for this sightline, Fe/H = (1.07 $\pm~^{0.33}_{0.27}$)$\times10^{-6}$ (corresponding to log (Fe\,II/H\,I) = $-$5.97), is lower than would be expected for this D/H ratio according to Figure 3 of Linsky et al. (2005, in prep). Using the solar abundance of Fe from \citet{2004astro.ph.10214A}, [Fe/H]$_{\odot}$ = $-$4.55, the depletion of Fe along this sightline is D(Fe) = $-$1.42. This is an unusual sightline, for which non-stellar absorption by C\,IV, N\,V, O\,VI, Si\,IV, S\,III, and S\,IV, has been detected \citep[][and this work]{1985ApJ...292..477S,1995A&A...298..567W}, in addition to the large elliptic halos seen in emission (see $\S$\ref{wd1034_target}).


The presence of highly ionized species, the high D/H and lower than expected Fe/H, and the typical O/H and N/H ratios are consistent with a scenario in which D is released from the grains, but not much Fe. Perhaps the high radiation field of WD\,1034$+$001 plays a special role for this sightline. A similar mechanism has been proposed by Linsky et al. (2005, in prep) to explain the high D/H and low Fe/H ratios along the sightline to $\gamma^2$ Vel.

\subsection{BD$+$39\,3226}
\label{bd39_disc}

For the BD$+$39\,3226 sightline the D/H ratio is only slightly inconsistent with the LB value (1.15$\sigma$ below). For this sightline, D/H and Fe/H ((1.17 $\pm~^{0.34}_{0.28}$)$\times10^{-6}$) follows the correlation derived by Linsky et al. (2005, in prep) for these quantities. Even though interstellar O\,VI is detected along this sightline, we found no evidence of other highly ionized species (such as C\,IV, N\,V, Si\,IV, S\,III, or S\,IV) in either the {\it FUSE} or {\it IUE} data for this star. In this case the separation between stellar and interstellar absorption is high enough ($\sim$255 \kms) to allow for a definitive detection, should an interstellar feature be present. It is possible that along this sightline O\,VI is formed in evaporative interfaces between cool clouds and the hot and diffuse gas in the Local Bubble. This mechanism has been invoked by \citet{2005ApJ...622..377O} to explain the presence of O\,VI along the sightlines to several white dwarfs inside the Local Bubble. D/O along this sightline, (5.62 $\pm~^{1.61}_{1.31}$)$\times10^{-2}$, is not consistent at the 1 $\sigma$ level with the D/O ratio derived for the LB by \citet{2003ApJ...599..297H}, D/O = (3.84 $\pm$ 0.16)$\times10^{-2}$. This is however, in part due to the low O/H ratio derived here for this sightline, O/H = (2.09 $\pm~^{0.72}_{0.58}$)$\times10^{-4}$, which is 1.86$\sigma$ below the O/H ratio derived by \citet{1998ApJ...493..222M}. This O/H ratio is lower than what has been derived in similar studies (see $\S$\ref{litvalues}) and might be indicative of variations of the O/H ratio. Using the O/H ratio derived by the authors above yields D/O $\sim$ 3.4$\times10^{-2}$, in closer agreement with the D/O LB value reported above.

\subsection{TD1\,32709}
\label{td1_disc}

The TD1\,32709 sightline presents a D/H ratio consistent at the 1$\sigma$ level with the LB, while D/O is inconsistent within the quoted uncertainties at the 2$\sigma$ level. O/H is lower than the value derived by \citet{1998ApJ...493..222M} by 1.09$\sigma$. If we assume that the D/H and D/O ratios are high due to $N$(D\,I) being overestimated then we would need log$N$(D\,I) = 15.22 to bring D/H closer to the LB value. Using the same uncertainties for the required D\,I as the ones derived in this work (0.05 dex), the D/O ratio would still be high at (6.31 $\pm~^{1.80}_{1.47}$)$\times10^{-2}$, even with this reduced value for $N$(D\,I). It is unlikely that $N$(H\,I) has been over or underestimated, as a lower $N$(H\,I) would lead to an even larger D/H ratio, while a larger $N$(H\,I) would lead to an even smaller O/H ratio. Increasing $N$(O\,I) so that O/H agrees with the \citet{1998ApJ...493..222M} value still leads to a high D/O ratio, (5.43 $\pm~^{1.55}_{1.26}$) $\times10^{-2}$ (assuming that the increased $N$(O\,I) has the same uncertainties as the one derived in this work). The only possibility to reconcile the D/H and D/O ratios along this sightline with the LB values would be for D and O to be simultaneously over and underestimated by 1.6 $\sigma$ and 1.5 $\sigma$, respectively. However this is unlikely given the number of transitions and methods used to determine $N$(D\,I) and $N$(O\,I).

For this sightline the Fe/H ratio ((0.83 $\pm~^{0.30}_{0.24}$)$\times10^{-6}$) is lower than would be expected for the D/H ratio according to Figure 3 of Linsky et al. (2005, in prep). We searched the {\it IUE} data for this star for absorption by highly ionized species such as C\,IV, N\,V, and Si\,IV. However because the separation between stellar and interstellar absorption is low ($\sim$13 \kms) we cannot determine if the detected absorptions are of stellar or interstellar origin (or both), similarly to what we had concluded for the O\,VI absorption detected in the {\it FUSE} data (see $\S$\ref{uv0904atomic}).

\subsection{Implications for models that explain D/H variations}
\label{implica}

The high D/H ratios derived for the WD\,1034$+$001 and TD1\,32709 sightlines increase the sample of sightlines for which a high D/H ratio has been found. These measurements strengthen the argument that these high ratios can not be due to some systematic problem associated with the different analyses (see also discussion below) or that they are peculiar cases. In particular, it has been argued that the high D/H ratios found in the ISM could be due to systematic effects that affect the $N$(H\,I) measurements \citep{2005guillaume}. Because the D/O ratios derived for these two sightlines are also high and independent of $N$(H\,I), the high D/H ratios are likely due to a high abundance of D and not to a systematic problem with the $N$(H\,I) determination (see also discussion in $\S$\ref{litvalues}). In addition, the high D/O ratios, together with O/H ratios consistent with the \citet{1998ApJ...493..222M} results, contradict the argument by \citet{2006guillaume} against a high present-epoch D/H ratio, on the basis that no correspondingly high D/O ratios have been measured. It is also hard to explain these high D/H ratios on the basis of localized infall of metal poor gas, as one would expect the O/H and N/H ratios for these sightlines to be affected as well. In addition, several studies of the O/H ratio \citep{1998ApJ...493..222M,2003ApJ...591.1000A,2004ApJ...613.1037C,2005ApJ...625..232O} have shown that this ratio is constant across a large range of $N$(O\,I), distance to targets, and average sightline gas density. However, note that in a special case where the infalling gas had a high D/H ratio and metallicity comparable to the local ISM value of \citet{1998ApJ...493..222M}, such a mechanism could not be ruled out. Variable astration in which the amount of D burned in the interior of stars and O produced, varies from sightline to sightline, seems also to be an unlikely mechanism for explaining the high D/H ratios. In the case of WD\,1034$+$001 and TD1\,32709 less deuterium would have to have been destroyed by astration and consequently less O would have been produced by supernovae. This scenario seems implausible because it implies that astration rates would have to be variable on short distance scales in order to explain the nearby sightlines with LB-like D/H ratios and also because it is not supported by our derived O/H and N/H ratios. In addition, in both the infall and variable astration scenarios, these processes would have to occur in a time smaller than the typical mixing time-scale of 350 Myr \citep{2002ApJ...581.1047D}. Our results support the idea that some sightlines in the Milky Way ISM have high D/H ratios. Linsky et al. (2005, in prep) have used the existence of these sightlines together with the deuterium dust-depletion model of \citet{2004oee..symp..320D} to argue for a high present-epoch abundance of deuterium. Linsky et al. (2005, in prep) estimate the total abundance of D (in gas $+$ dust) in the local disk of the Galaxy, D/H $\ge$ (2.19 $\pm$ 0.27)$\times10^{-5}$, by taking the weighted average of the ratios for the five sightlines with the highest values. Our new measurements for WD\,1034$+$001 and TD1\,32709 are in good agreement with this estimate. Below we compare our new measurements with ratios from the literature.

\section{COMPARISON WITH PREVIOUS MEASUREMENTS}

\subsection{D/H, O/H, and D/O vs. $N$(H)}
\label{litvalues}

\begin{figure}
\begin{center}
\epsscale{0.65}
\plotone{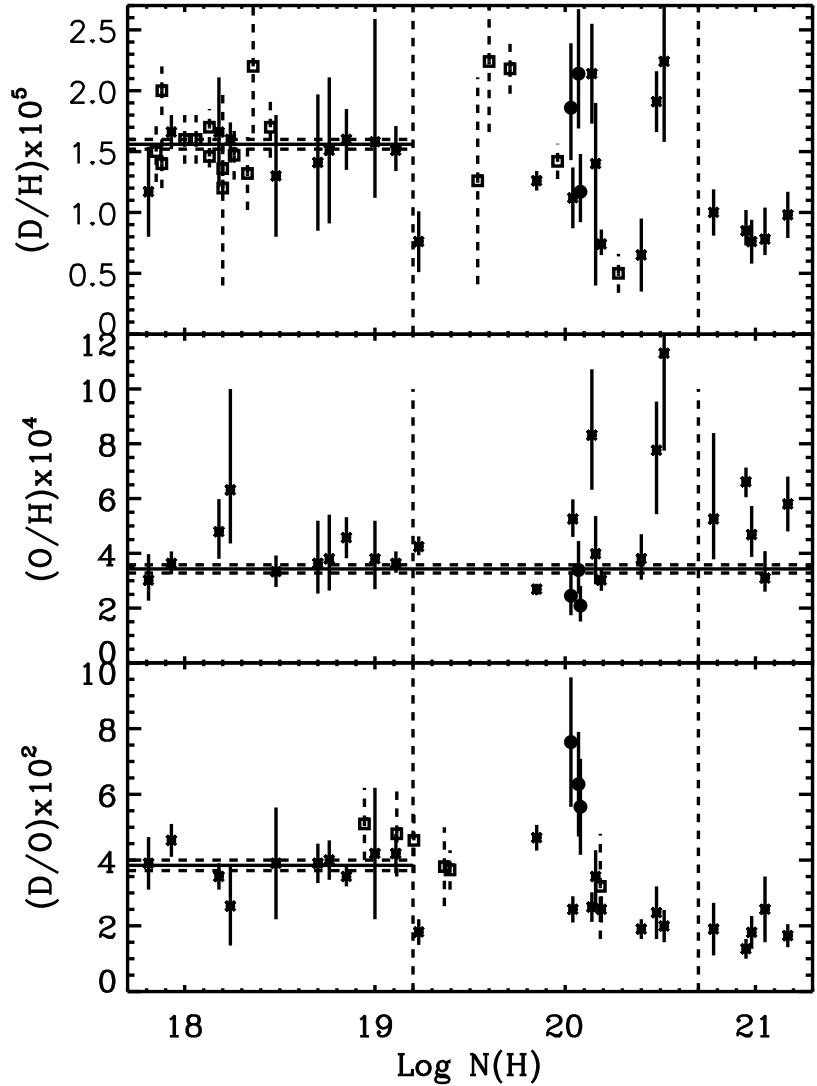}
\caption{D/H, O/H, and D/O ratios as a function of the sightline total hydrogen column density, $N$(H). Sightlines for which all three D/H, O/H, and D/O ratios are available are marked by asterisks (literature values) and filled circles (this work). {\it Top panel:} Sightlines for which no O\,I measurement is available are marked with squares, the error bars are displayed by dotted lines. (D/H)$_{\rm LB}$ = (1.56$\pm$ 0.04)$\times10^{-5}$ \citep[derived by][from a compilation of values in the literature]{2004ApJ...609..838W} is indicated by solid and dashed (1$\sigma$) horizontal lines for log $N$(H) $<$ 19.2, which defines the contour of the Local Bubble (dashed vertical line). For 19.2 $<$ log $N$(H) $<$ 20.7 there is a large scatter in the D/H ratio. For log $N$(H) $>$ 20.7, D/H seems to be constant, but lower than the LB value. {\it Middle panel:} The O/H = (3.43 $\pm$ 0.15)$\times10^{-4}$ derived by \citet{1998ApJ...493..222M} is marked by solid and dashed (1$\sigma$) horizontal lines. {\it Bottom panel:} D/O = (3.84 $\pm$ 0.16)$\times10^{-2}$ derived by \citet{2003ApJ...599..297H} for the LB is marked by solid and dashed (1$\sigma$) horizontal lines. Sightlines for which no $N$(H) is available are indicated by squares, the uncertainties are displayed by dotted lines. For these sightlines we use $N$(O\,I) and the O/H ratio derived by \citet{1998ApJ...493..222M} to estimate $N$(H). \label{ratiosplot}}
\end{center}
\end{figure}

Figure \ref{ratiosplot} presents D/H, O/H and D/O ratios (1$\sigma$ uncertainties, top, middle, and bottom panels, respectively) as a function of the total hydrogen column density, $N$(H), along the sightline. The data used in this figure are presented in Table \ref{alltabledata}. Sightlines for which all the ratios are available are marked by asterisks (literature values) and filled circles (this work). The data in this plot uses the ratios derived in this work together with the compilation by \citet{2004ApJ...609..838W} and the values of \citet{2005friedman} and \citet{2005guillaume}. Sightlines for which no $N$(O) is available are marked with open squares in the top panel (uncertainties displayed as dotted lines for clarity of the plot). Sightlines displayed by open squares with dashed error bars in the bottom panel (corresponding to D/O) do not have $N$(H) measurements available. For these sightlines we estimate $N$(H) using $N$(O\,I) and the O/H ratio of 3.43$\times10^{-4}$ derived by \citet{1998ApJ...493..222M}. Also plotted are the Local Bubble values of D/H = (1.56 $\pm$ 0.04)$\times10^{-5}$ \citep[top panel,][]{2004ApJ...609..838W} and D/O = (3.84 $\pm$ 0.16)$\times10^{-2}$ \citep[bottom panel,][]{2003ApJ...599..297H}. Dashed vertical lines mark the approximate position of the Local Bubble (log $N$(H) $<$ 19.2), and log $N$(H) = 20.7, where a new low and relatively constant D/H ratio seems to emerge. The value of O/H = (3.43 $\pm$ 0.15)$\times10^{-4}$ derived by \citet{1998ApJ...493..222M} for the local ISM is also marked (solid and dashed lines in the middle panel of Figure \ref{ratiosplot}).

For the range of hydrogen column densities 19.2 $<$ log $N$(H) $<$ 20.7, assuming a constant D/H ratio yields the weighted mean (D/H)$\times10^5$ = 1.18 $\pm$ 0.05 (1 $\sigma$ on the mean) and $\chi^2_{\nu}$ = 5.9 for 16 degrees of freedom (where $\chi^2_{\nu}$ is the total $\chi^2$ divided by the number of degrees of freedom, $\nu$). Because some of the ratios in Table \ref{alltabledata} have asymmetric uncertainties we use the average of the uncertainties on each ratio to compute the weighted mean and the uncertainty in the weighted mean for D/H and for the other ratios discussed below. One can consider two possibilities to explain the high value of $\chi^2_{\nu}$. The scatter is real and variations of the D/H ratio do exist, or the uncertainties of the individual measurements have been underestimated. This last possibility can be explored by assuming that the scatter in D/H for 19.2 $<$ log$N$(H) $<$ 20.7 is due to some random systematic error in the determination of D/H, unaccounted for in the quoted uncertainties. One can then estimate by how much the uncertainties would have to be increased to obtain $\chi^2_{\nu}$ = 1. Assuming further that this error can be expressed as a fraction of D/H, and that it is added in quadrature to the quoted uncertainty $\sigma_{\rm D/H}$, $\sigma_{\rm new}$  = ($\sigma_{\rm D/H}^2$ $+$ f$^2\times$(D/H)$^2$)$^{0.5}$, we derive $f$ = 0.40. This implies that the unknown systematic errors would have to be 40\% of D/H. Such a high uncertainty seems unlikely. If we also add the ratios for log$N$(H) $>$ 20.7 to the dataset, we derive the weighted mean D/H = (1.10 $\pm$ 0.04)$\times10^{-5}$ with $\chi^2_{\nu}$ = 5.1  for 21 degrees of freedom, and calculate $f$ = 0.34, still a high and unlikely value. These results show that it is unlikely that the scatter in D/H is due to unknown systematic uncertainties associated with the different measurements, and that it is real.

As mentioned above, the apparent lack of scatter in the D/O ratios, has been used as an argument against the existence of sightlines with high D/H ratios \citep{2005guillaume}. One has however to be careful in analyzing these ratios, as they depend on both the $N$(D) and $N$(O) values. For instance, there are different ways of producing a low D/O. One can have a low D/H ratio, and normal O/H, or a normal D/H and a high O/H \citep[where normal refers to the LB value for D/H and to the][value for O/H]{1998ApJ...493..222M}. This last way must not be common as several studies have concluded that O/H is constant throughout a large range of $N$(H) \citep[see for example][and Oliveira et. al. 2005]{2003ApJ...591.1000A,2004ApJ...613.1037C}. However, as shown in Figure \ref{ratiosplot}, four out of the five sightlines with log $N$(H) $>$ 20.7, also have O/H ratios inconsistent with the \citet{1998ApJ...493..222M} value (solid and dashed horizontal lines in the middle plot). For the mean of these five points we find O/H = (5.33 $\pm$ 0.36)$\times10^{-4}$. However $\chi^2_{\nu}$ = 5.3 for four degrees of freedom, indicating that a single mean may not be appropriate. In other words, for four of these targets D/O is lower than it would have been, if these sightlines had normal O/H ratios. Because of the high average O/H value and considering that the number of measurements at large $N$(H) ($>$20.7 in the log) is statistically small, we think it is necessary when estimating the deuterium abundance at high column densities \citep[which according to the suggestion of][may be the present epoch Milky Way abundance]{2003ApJ...599..297H} to consider both D/H and D/O measurements. In addition, the use of the \citet{1998ApJ...493..222M} value for O/H to convert the average D/O in this region to the average D/H yields an estimate that is low when compared to the direct D/H measurements in this $N$(H) region, not unlike what happens when we do the same exercise for the Local Bubble.

The five points with high H column density presented in Figure \ref{ratiosplot} are near the limit of deuterium column densities that can be measured with an instrument with {\it FUSE}-like properties. The deuterium lines used in these analysis \citep{2003ApJ...586.1094H,2004ApJ...609..838W,2005guillaume} would have been saturated if the D/H ratios along these sightlines were similar or larger than the LB D/H value ($\sim$1.56$\times10^{-5}$). Hence, there is a bias against measuring high D/H ratios at these large H column densities. This could explain in part why only a few high D/H ratios have been reported in the literature.

Additionally, to try to understand the behavior of the D/O ratio in the regime 19.2 $<$ log$N$(H) $<$ 20.7 and the apparent lack of scatter in D/O compared to D/H, we can consider the O/H ratios for the sightlines for which D/H $>$ 1.9$\times10^{-5}$ in the top panel of Figure \ref{ratiosplot} ($\alpha$ Cru, $\gamma^2$ Vel, WD\,1034$+$001, Feige\,110, PG\,0038$+$199 and LSE\,44; as a function of increasing $N$(H)). For these sightlines with high D/H, one would expect also a high D/O, assuming that O/H is similar to the \citet{1998ApJ...493..222M} value, hence producing more scatter in the D/O measurements. However, two of these sightlines do not have O/H measurements ($\alpha$ Cru and $\gamma^2$ Vel), while Feige\,110, PG\,0038$+$199, and LSE\,44 have abnormally high O/H ratios (see Table \ref{alltabledata}). The remaining sightline (WD\,1034$+$001) has O/H consistent with the \citet{1998ApJ...493..222M} value. Hence, the apparent lack of scatter in the D/O ratio, which has been used as an argument against a high present-epoch D/H ratio, is the result of small number statistics, and cannot be used to draw definitive conclusions about the behavior of the D/O ratio in this regime of $N$(H).

\subsection{D/O vs. Fe/O}
\label{iron}

\begin{figure}
\begin{center}
\includegraphics[scale=.5,angle=90]{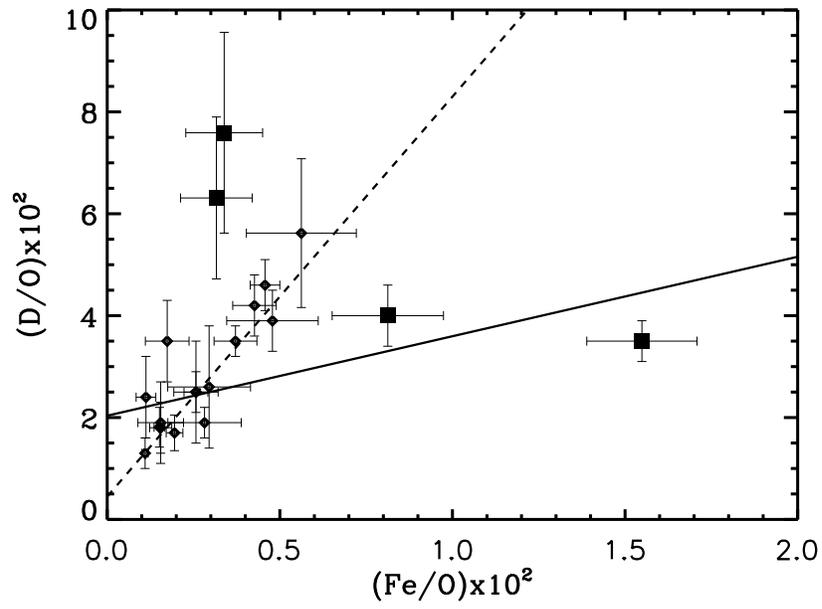}
\caption{D/O as a function of Fe/O. The solid line represents the fit to all the data points presented. The dashed line represents the fit to the data when the four data points displayed by squares are removed from the fit. See $\S$\ref{iron} for discussion. \label{ironplot}}
\end{center}
\end{figure}

Linsky et al. (2005, in prep) have looked at the correlation between the deuterium abundance and the depletion of Fe and found that in sightlines with low D/H ratios Fe is typically more depleted than in sightlines with higher D/H ratios. A different way of considering the relationship between the deuterium and iron abundances is to look at D/O versus Fe/O. By taking the ratio of these abundances relatively to oxygen rather than to hydrogen one can avoid systematic problems that might affect the determination of $N$(H\,I) \citep[as could be the case for the Feige\,110 sightline, as argued by][]{2005guillaume}. Figure \ref{ironplot} presents D/O as a function of Fe/O. The solid line represents the fit to all data points in this plot, yielding D/O = (2.04 $\pm$ 0.15)$\times10^{-2}$ + (1.56 $\pm$ 0.30)$\times10^{-2}\times$Fe/O, with $\chi^2$ = 76 for 18 degrees of freedom (only uncertainties in D/O taken into account). This fit is mostly influenced by the sightline with the largest Fe/O ratio and small D/O uncertainties, G191$-$B2B, represented by a square in the plot. Removing this sightline from the fit yields D/O = (0.99 $\pm$ 0.23)$\times10^{-2}$ $+$ (5.64 $\pm$ 0.73)$\times10^{-2}$$\times$D/O, with $\chi^2$ = 39 for 17 degrees of freedom. The other three sightlines represented by squares also have Fe/O ratios which are inconsistent with the trend displayed by the majority of the data plotted in Figure \ref{ironplot}. Excluding all the sightlines displayed by squares in the plot from our fit increases the slope slightly more yielding D/O = (0.47 $\pm$ 0.28)$\times10^{-2}$ + (7.85 $\pm$ 0.99)$\times10^{-2}\times$Fe/O, with $\chi^2$ = 16 for 14 degrees of freedom. The probability of $\chi^2$ being larger than 16 for 14 degrees of freedom is $\sim$30\%. Thus, the trend of decreasing Fe/H with decreasing D/H observed by Linsky et al. (2005, in prep) is also observed when one considers D/O versus Fe/O. We note however that there is no particular reason for not taking into account the points displayed in Figure \ref{ironplot} by squares or that it is theoretically expected that $N$(D) depends linearly on $N$(Fe).

\subsection{D/H and $\langle n_{\rm H}\rangle$}
\label{nh_disc}

Another way of trying to understand the behavior of D/H in the ISM, besides looking for correlations with the abundances of other species \citep[e.g.,][Linsky et al. 2005, in prep]{2005ApJ...620L..39P}, is to study D/H as a function of the average sightline gas volume density, $\langle n_{\rm H}\rangle$ = $N$(H)/$d$.

Elemental abundances measured in the interstellar gas in the solar vicinity are generally depleted with respect to the solar values \citep{1974ApJ...193L..35M,1986ApJ...301..355J}. This depletion effect has been commonly interpreted as the result of a fraction of the elements being locked up in dust grains \citep{2004oee..symp..339J}. Furthermore, several studies have shown that the strength of the depletions increases as the average gas density increases \citep[see e. g.,][]{1979ApJ...229..136S,1984ApJ...284..157H,1985ApJ...290L..21S,1986ApJ...301..355J,1987ip...symp..533J,1994ApJ...424..748C}. The idea that D might be depleted in grains was first proposed by \citet{1982auva.nasa...54J} and updated recently by \citet{2004oee..symp..320D} \citep[see also][]{2003ARA&A..41..241D}. \citet{1984ApJ...284..157H} looked for a correlation between the deuterium abundance and the sightline density. Using $Copernicus$~data for 14 sightlines and a limited range of densities, they found no evidence of significant depletion of interstellar deuterium. However, since $Copernicus$, several space-born observatories ($IMAPS$, $HST$, $FUSE$) have been used to measure the interstellar abundance of deuterium and there are now D/H measurements for more than forty sightlines. It is then appropriate to revisit the relationship between D/H and the average sightline gas density.

We use the simple model developed by \citet{1985ApJ...290L..21S} \citep[see also][]{1986ApJ...301..355J} to explain the correlation between the level of depletion and the mean line of sight density, to explore the relationship between D/H and $\langle n_{\rm H}\rangle$. In this model, the interstellar medium is composed of many warm diffuse clouds, higher density cold diffuse clouds which occur with a lower spatial frequency and cold dense clouds with even lower spatial frequency. At low densities, contributions from the diffuse warm gas dominates that of the other two types of clouds. Cold diffuse clouds start to prevail with increasing densities, and at even higher densities, contributions from cold dense clouds outweigh everything else. Therefore, any correlation of a physical property with $\langle n_{\rm H}\rangle$ can be attributed partly to differences of the properties of the types of clouds. 

\citet{1986ApJ...301..355J} used the model developed by \citet{1985ApJ...290L..21S} to derive Equation \ref{depletionequation}, which predicts the abundance of element $X_i$, $A(X_i)$ = log $N$($X_i$) $-$ log $N$(H) ($N$(H) = $N$(H\,I) +2$N$(\hmol)), as a function of the sightline properties:

\begin{equation}
{A(X_i) = A_c(X_i) + {\rm log} [1 +\frac{n_{w}}{\langle n_{\rm H} \rangle}(\frac{\delta_w}{\delta_c}-1)]}
\label{depletionequation}
\end{equation}
where $\delta_w$($X_i$) is the depletion of element $X_i$~in the warm gas and $\delta_c$($X_i$) in the cold clouds, and $n_w$~is the mean density of the warm gas along the line of sight

Equation \ref{depletionequation} can be simplified to
\begin{equation}
{A(X_i) = A_c(X_i) + {\rm log} [1 +\frac{n_{w}}{\langle n_{\rm H} \rangle}(10^{(A_w(X_i) - A_c(X_i))}-1)]}
\label{depletionequation2}
\end{equation}
where $A_w(X_i)$ and $A_c(X_i)$ are the abundances of $X_i$ in the warm diffuse medium and cold clouds, respectively. This equation assumes that diffuse cold clouds and dense cold clouds have the same level of depletion and that for $\langle n_{\rm H}\rangle$/$n_{\rm w} <$ 1, all the neutral hydrogen along the line of sight is in warm gas \citep[note that for the gas densities considered here, cold dense clouds are not expected to contribute significantly to the sightline properties,][]{1985ApJ...290L..21S}. Hence, for $\langle n_{\rm H}\rangle< n_{\rm w}$, $A(X_i)$ has the constant value $A_w(X_i)$, characteristic of warm gas.

\begin{figure}
\begin{center}
\epsscale{0.7}
\plotone{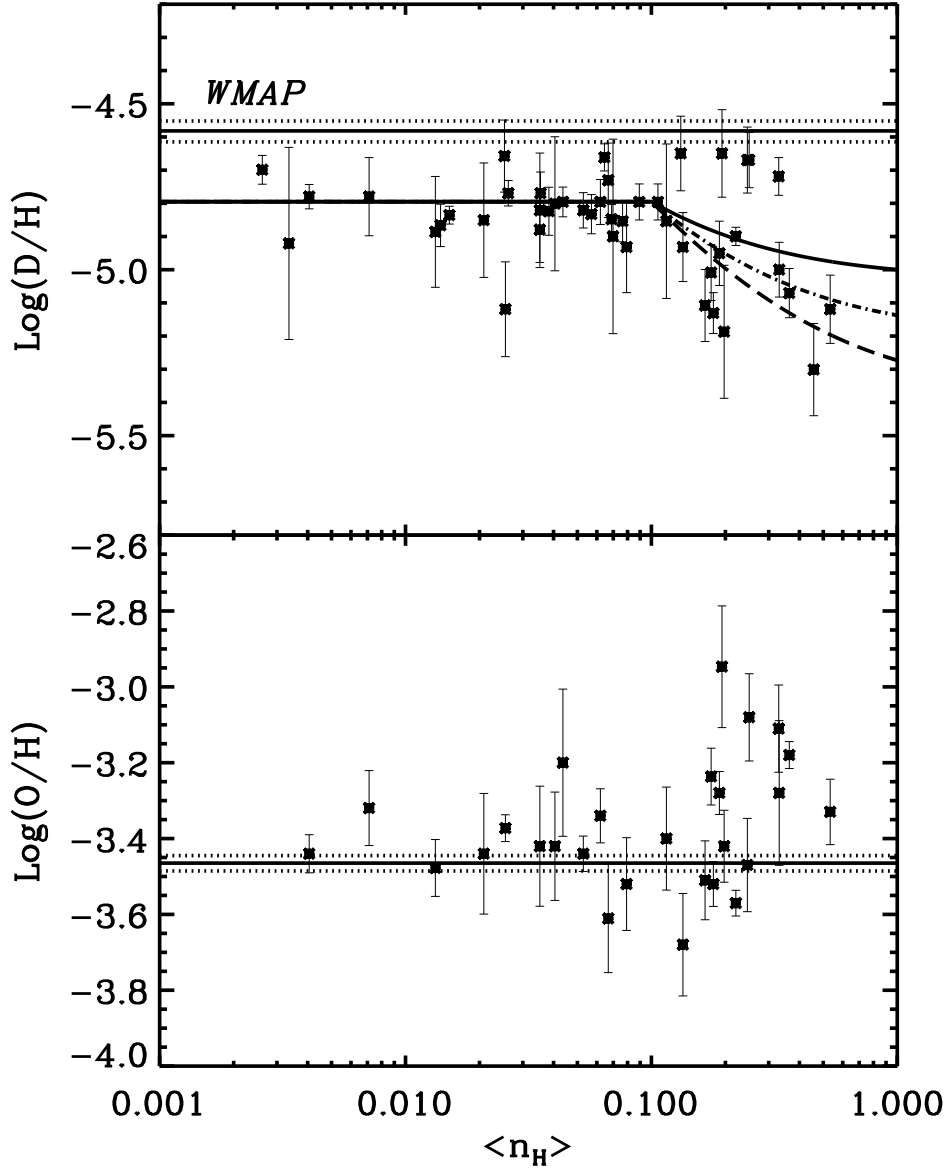}
\caption{{\it Top panel}: Log(D/H) as a function of the average sightline density, $\langle n_{\rm H}\rangle$ (cm$^{-3}$). The {\it WMAP}-based result of (D/H)$_{\rm prim}$ = (2.62 $\pm~^{0.18}_{0.20}$)$\times10^{-5}$ \citep{2003ApJS..148..175S} is also plotted ({\it solid} and {\it dotted} lines at log(D/H) = $-$4.57). The {\it solid} line represents the fit of Equation \ref{depletionequation2} to all data points, yielding  $A_w(D)$ = $-$4.80 (D/H = 1.58$\times10^{-5}$) and $A_c(D)$ = $-$5.03 (D/H = 0.93$\times10^{-5}$). The {\it dash-dotted} and {\it dashed} lines correspond to $A_c(D)$ = $-$5.20 (D/H = 0.63$\times10^{-5}$) and $A_c(D)$ = $-$5.38 (D/H = 0.42$\times10^{-5}$), respectively. See $\S$\ref{nh_disc} for discussion. {\it Bottom panel}: Log(O/H) as a function of $\langle n_{\rm H}\rangle$. The solid and dashed lines represent O/H = (3.43 $\pm$ 0.15)$\times10^{-4}$ derived by \citet{1998ApJ...493..222M}.\label{dh_density}}
\end{center}
\end{figure}

Figure \ref{dh_density} (top panel) presents log(D/H) as a function of the average sightline density, $\langle n_{\rm H}\rangle$. For comparison, we plot also log(O/H) (bottom panel). All the data have been presented previously in the top and middle panels of Figure \ref{ratiosplot} and are also listed in Table \ref{alltabledata}. Also plotted is (D/H)$_{\rm prim}$ = (2.62 $\pm~^{0.18}_{0.20}$)$\times10^{-5}$, determined from analysis of the {\it Wilkinson Microwave Anisotropy Probe (WMAP)} satellite data \citep[top panel,][]{2003ApJS..148..175S} and O/H = (3.43 $\pm$ 0.15)$\times10^{-4}$ from \citet{1998ApJ...493..222M} (bottom panel). Like \citet{1986ApJ...301..355J} we adopt $n_{\rm w}$ = 0.10 cm$^{-3}$ below, to determine $A_w(D)$ and $A_c(D)$. As it can be seen in Figure \ref{dh_density}, $\langle n_{\rm H}\rangle$ = 0.10 cm$^{-3}$ seems to be a natural point, after which the scatter in D/H increases. However choosing a value in the range 0.08 $< \langle n_{\rm H}\rangle <$ 0.12 does not affect $A_w(D)$ determined below. A non-linear least squares fit ($\chi^2$ minimization) is used to fit $A(D)$ from Equation \ref{depletionequation2} above to the data plotted in the top panel of Figure \ref{dh_density}.

We derive $A_w(D)$ = $-$4.80 $\pm$ 0.01 ($\langle n_{\rm H}\rangle <$0.10 cm$^{-3}$) corresponding to D/H = 1.58$\times10^{-5}$ and $\chi^2_{\nu}$ = 1.2 (26 degrees of freedom). This D/H ratio is consistent with the one derived for the LB by \citet{2004ApJ...609..838W}, D/H = (1.56 $\pm$ 0.04)$\times10^{-5}$. We note however, that not all data points with $\langle n_{\rm H}\rangle <$0.10 cm$^{-3}$ correspond to targets located inside the LB, although all sightlines inside the LB have $\langle n_{\rm H}\rangle \leq$0.10 cm$^{-3}$. There are five sightlines of this type, outside the LB; $\beta$\,Cen, $\lambda$\,Sco, $\gamma^2$\,Vel, $\zeta$\,Pup, and TD1\,32709. Three of these have D/H values that are consistent with the low density average but $\lambda$~Sco (D/H$\times10^5$ = 0.76 $\pm$ 0.25) and $\gamma^2$~Vel (D/H$\times10^5$ = 2.18 $\pm~^{0.22}_{0.19}$) are exceptional. Thus, for low densities, as well as for higher densities, which we discuss below, some sightlines apparently defy the trends by statistically significant margins.


For $\langle n_{\rm H}\rangle >$0.10 cm$^{-3}$, $A_c(D)$ is mostly affected by the five points with a high D/H ratio, i.e., log(D/H) $> -$4.8. We derive $A_c(D)$ = $-$5.03 (D/H = 0.93$\times10^{-5}$) with $\chi^2_{\nu}$ = 4.7 (16 degrees of freedom), using all ratios corresponding to sightlines with $\langle n_{\rm H}\rangle >$0.10 cm$^{-3}$ (solid line in Figure \ref{dh_density}). If we assume that these five ratios are exceptional, corresponding to sightlines with different conditions, and remove them from our fit, we would derive $A_c(D)$ = $-$5.20 (D/H = 0.63$\times10^{-5}$) with $\chi^2_{\nu}$ = 2.4 (11 degrees of freedom). The corresponding fit is represented by a dashed-dot line in the top panel of Figure \ref{dh_density}. This fit is now heavily dominated by a single ratio corresponding to the BD$+$28\,4211 sightline (with a small uncertainty compared to the other sightlines). If in addition we exclude this ratio so that the fit is not dominated by one point we would derive, $A_c(D)$ = $-$5.38 (D/H = 0.42$\times10^{-5}$) with $\chi^2_{\nu}$ = 1.5 (10 degrees of freedom), represented by a dashed line in the top panel of Figure \ref{dh_density}.

For O/H in the bottom panel of Figure \ref{dh_density} we derive the weighted mean of O/H = (3.78 $\pm$ 0.18)$\times10^{-4}$ (1 $\sigma$ in the mean) for $\langle n_{\rm H}\rangle <$ 0.1 cm$^{-3}$, with $\chi^2_{\nu}$ = 0.8 for 11 degrees of freedom. The square-root of the weighted average variance (standard deviation) is 0.56$\times10^{-4}$. Considering only the sightlines with $\langle n_{\rm H}\rangle >$ 0.1 cm$^{-3}$ we derive the weighted mean O/H = (3.41 $\pm$ 0.15)$\times10^{-4}$ with a standard deviation of 1.40$\times10^{-4}$ and $\chi^2_{\nu}$ = 5.8 for 14 degrees of freedom. The O/H data does not show the same trend with $\langle n_{\rm H}\rangle$ as the one displayed by D/H. Recent studies have shown that oxygen is not expected to be depleted for $\langle n_{\rm H}\rangle$ $<$ 1.5 cm$^{-3}$ \citep{2004ApJ...613.1037C}. However, the bottom panel of Figure \ref{dh_density} does seem to indicate that for some of the densest sightlines there is an anti-correlation between D/H and O/H, which could be an indication of astration. 

Another interesting comparison is the depletion of Fe compared to that implied by Figure \ref{dh_density} for D. \citet{1986ApJ...301..355J} found that the abundance of Fe decreases by a factor of three from the warm gas value, by $\langle n_{\rm H}\rangle$ = 1.0. Figure \ref{dh_density} indicates a factor of two or more decrease over the same density range for D, supporting the correlation between the values of D/H and the Fe depletion reported by Linsky et al. (2005, in prep).

Analyzing D/H as a function of the average sightline density removes the need for the three regimes displayed in Figure \ref{ratiosplot}. In this picture, D/H is constant and has the same value as in the LB for $\langle n_{\rm H}\rangle <$0.10 cm$^{-3}$, i. e. for sightlines dominated by warm diffuse gas, implying that D/H is constant in a broader sense than has been considered before. For $\langle n_{\rm H}\rangle >$0.10 cm$^{-3}$, D/H decreases with increasing density, as more of the gas is associated with cold diffuse clouds. Considering that for other elements such as Si and Fe this has been considered as evidence of depletion of these elements into dust grains \citep[see for e.g.][]{1986ApJ...301..355J}, then our study strongly suggests that D might also be depleted into dust grains as suggested by \citet{2004oee..symp..320D} (but see Linsky et al. 2005, in prep, for discussions on other possible explanations, such as variable astration and localized infall of metal-poor gas). However, as noted above, in both regimes there are a few sightlines which are exceptional, with high D/H ratios. It is possible that along these sightlines the conditions are such that D has either been released from the grains, does not deplete onto grains, or is only mildly depleted. Also, while there are no accurate quantitative estimates for significant production of D by local sources, the possibility of such sources cannot be entirely dismissed \citep[see for example][]{1999ApJ...511..502M,2003ApJ...597...48P}. The true value of D/H in cold diffuse clouds, $A_c(D)$, is difficult to derive at this point given the small number of sightlines with high gas density. Nevertheless, it seems reasonable to estimate that in cold diffuse clouds D/H $<$ 0.93$\times10^{-5}$, which is the result of our fit when all the sightlines with $\langle n_{\rm H}\rangle >$0.10 cm$^{-3}$ are included in the analysis.

\section{SUMMARY}
\label{summary}

We have used data obtained with {\it FUSE} together with archival data from {\it IUE} and {\it HST} to derive column densities and ratios of column densities along the lines of sight to WD\,1034$+$001, BD$+$39\,3226, and TD1\,32709. The D/H derived here for two of these sightlines are not consistent at the 1 $\sigma$ level with the Local Bubble value; none of the D/O ratios are consistent, at the 1 $\sigma$ level with the Local Bubble value, and present some of the highest values in the literature. Considered along with the ratios published for other sightlines, our results reinforce the scatter in the D/H measurements and indicate that there is also scatter in the D/O ratio, implying that the high D/H ratios derived here and along other sightlines are unlikely to be due to problems with the $N$(H\,I) determinations. We estimate that the additional, presently unknown, systematic errors on the determination of $N$(H\,I) would have to be of the order of 40\% of each D/H ratio, in order to bring all the D/H ratios into agreement. Thus, our results support the idea that some sightlines in the Milky Way ISM have high D/H ratios. For sightlines with log$N$(H) $>$ 20.7 the O/H ratios are not consistent with a single value, and are inconsistent with the local ISM value derived by \citet{1998ApJ...493..222M}. The hint of anti-correlation between D/H and O/H for these sightlines could be an indication of astration.

The trend between decreasing D/H with decreasing Fe/H observed by Linsky et al. (2005, in prep) is also observed when D/O versus Fe/O are considered. These ratios are not subject to possible systematic uncertainties in the determination of $N$(H).

In order to try to understand the behavior of the deuterium abundance we performed a  study of D/H versus the sightline average gas density. Our results indicate that as long as the gas probed by the sightlines is in warm diffuse clouds with $\langle n_{\rm H}\rangle <$0.10 cm$^{-3}$ D/H seems to be constant and have the Local Bubble value. In addition, our study shows that D/H seems to decrease with increasing sightline gas density, similarly to what has been observed for other elements such as Fe and Si, which strongly supports the idea that D might be depleted into dust grains. Finally, a few sightlines do not follow the trend, but show exceptionally high D/H ratios.

\acknowledgments

This work is based on data obtained for the Guaranteed Time Team by the NASA-CNES-CSA \fuse~mission operated by The Johns Hopkins University. Financial support to U. S. participants has been provided in part by NASA contract NAS5-32985 to Johns Hopkins University. Based on observations made with the NASA/ESA Hubble Space Telescope and the International Ultraviolet Explorer, obtained from the Data Archive at the Space Telescope Science Institute, which is operated by the Association of Universities for Research in Astronomy, Inc. under NASA contract NAS5-26555. Support for MAST for non-HST data is provided by the NASA Office of Space Science via grant NAG5-7584 and by other grants and contracts. The profile fitting procedure, Owens.f, used in this work was developed by M. Lemoine and the French \fuse~Team. We thank Todd Tripp and Scott Friedman for their help in determining $N$(H\,I) from {\it IUE} data and W. V. Dixon for useful discussions regarding the scattered light background in the {\it FUSE} data. We thank also Guillaume H{\' e}brard, Jeff Linsky, and Paule Sonnentrucker for comments that helped to improve the paper.

\bibliography{ms}
\bibliographystyle{apj}

\clearpage

\begin{deluxetable}{lccccccc}
\tablewidth{0pc}
\tablecaption{Stellar Properties \label{star_properties}}
\tablehead{ 
\colhead{Star} & \colhead{T$_{\rm eff}$} & \colhead{log~$g$} &\colhead{$d$} &\colhead{$l$} &\colhead{$b$} & \colhead{Sp. Type}&\colhead{$v_{\rm PH}-v_{\rm ISM}$}\\
\colhead{}  &  \colhead{(K)}             &\colhead{(cm$^{-2}$)} &\colhead{(pc)} &\colhead{($^\circ$)}   &\colhead{($^\circ$)} &\colhead{} &\colhead{(km s$^{-1}$)}}
\startdata
WD\,1034$+$001\tablenotemark{a} & 100,000 $\pm~^{15,000}_{10,000}$ & 7.5 $\pm$ 0.3 & 155 $\pm~^{58}_{43}$ & 247.55 & $+$47.75 & DO & $\sim +$50 \\
BD$+$39\,3226\tablenotemark{b}  & 45,000 & 5.5 & 290 $\pm~^{140}_{70}$ & 65.00 & $+$28.77& sdO& $\sim -$255\\ 
TD1\,32709\tablenotemark{c} & 46,500 $\pm$ 1,000 & 5.55 $\pm$ 0.15 & 520 $\pm$ 90 & 232.98 & $+$28.12 & sdO & $\sim -$13\tablenotemark{d}\\
\enddata
\tablenotetext{a}{From \citet{1995A&A...298..567W}.}
\tablenotetext{b}{From \citet {1999A&A...352..287B}.}
\tablenotetext{c}{From \citet{1993A&A...273..212D}.}
\tablenotetext{d}{This work.}
\end{deluxetable}

\begin{deluxetable}{lcccccc}
\tablewidth{0pc}
\tablecaption{Log of \fuse~observations \label{fuse_obs}}
\tablehead{ 
\colhead{Star} & \colhead{Program ID} & \colhead{Aperture}& \colhead{Mode} &\colhead{Time (Ks)} & \colhead{Date} &\colhead{CalFUSE}} 
\startdata
WD\,1034$+$001 & P1042003 & MDRS & TTAG & 6.7 & 2004 April 04 & 2.4.2 \\
BD$+$39\,3226 & P3020501 & MDRS & HIST & 5.2 & 2004 May 21 & 2.4.2 \\
TD1\,32709 & P2051301 & MDRS & HIST & 1.8 & 2004 March 12 & 2.4.1 \\
	   & P2051302 & MDRS & HIST & 3.2 & 2004 May 4    & 2.4.1 \\
	   & P2051303 & MDRS & HIST & 2.4 & 2004 May 5    & 2.4.1 \\
\enddata
\end{deluxetable}

\begin{deluxetable}{lccccc}
\tablewidth{0pc}
\tablecaption{Atomic data and analysis methods for the lines used in the analyses$^{a}$ \label{atomicdata}}
\tablehead{ 
\colhead{Species} & \colhead{Wavelength (\AA)} & \colhead{Log $f\lambda$} & \colhead{WD\,1034$+$001} &\colhead{BD$+$39\,3226}& \colhead{TD1\,32709}}
\startdata
H\,I	& 1215.670	& 2.70	& P	&\ldots	& P  \\
D\,I	& 916.1785	&$-$0.28& P	&\ldots & P \\
\ldots	& 917.8797	&$-$0.07& \ldots&\ldots	&A, P \\	
\ldots	& 919.1013	& 0.04  & A, C, P&A, C, P& A, C, P \\
\ldots	& 920.713	& 0.17	& A, C, P  & C, P	&A, C, P \\
\ldots	& 922.899	& 0.31	& A, C, P	&\ldots & C \\
\ldots	& 925.974	& 0.47	& C, P	& C, P	& C, P \\
\ldots	& 930.495	& 0.65	&\ldots	& \ldots& C, P \\
\ldots	& 937.548	& 0.86	& C	& C	& C, P \\
\ldots	& 949.485	& 1.12	&\ldots & C	& C\\
\ldots	& 972.272	& 1.45	& C	& C	& C\\
%
C\,II	& 1036.3367	& 2.09	& B/S & A	& A \\
C\,II*	& 1037.018	& 2.11	& A	& A	& A\\
C\,III	& 977.0200	& 2.87	& A	& A	& B/S\\
N\,I	& 951.2948	&$-$1.66& \ldots&\ldots	& P\\
\ldots	& 952.303	& 0.25	& C	& C	& C \\
\ldots  & 952.415	& 0.21	&\ldots & C	& C \\
\ldots  & 952.523	&$-$0.24& A, C, P&A, C, P&A, C, P \\
\ldots  & 953.415	& 1.10	& C	& C	& C \\
\ldots	& 953.655	& 1.38	& C	& C	& C \\
\ldots	& 953.9699	& 1.52  & C	& C	& \ldots \\
\ldots  & 954.1042	& 0.81	& C	& C	& C \\
\ldots	& 955.8819	&$-$1.54&\ldots	& \ldots& P \\
\ldots	& 959.4937	&$-$1.30&\ldots & \ldots & P\\
\ldots	& 963.990	& 1.54	& C	& C	& C \\
\ldots	& 964.626	& 0.96	& C	& C	& C \\
\ldots  & 1134.1653	& 1.69	& C	& C	& C \\
\ldots  & 1134.4149	& 1.53  & C	& C	& C \\
\ldots  & 1134.9803	& 1.24  &\ldots &\ldots	& C \\
N\,II	& 1083.994	& 2.10	& A & A	& A\\
O\,I	& 916.8150	& $-$0.36 & C	&\ldots & \ldots\\
\ldots	& 919.658	& $-$0.06 & C	& C	& C \\
\ldots	& 919.917	& $-$0.79 & A, P & A, C, P & A, C, P \\
\ldots	& 921.875	& 0.04    & C	&\ldots & \ldots \\
\ldots	& 922.200	& $-$0.65 & P	&\ldots & P \\
\ldots	& 924.950	& 0.15	  & C	& C	& \ldots \\
\ldots	& 925.446	& $-$0.49 & A, C& A, C	& C \\
\ldots  & 929.5168      & 0.32    &\ldots&\ldots& C \\
\ldots	& 930.257	& $-$0.30 & C 	& C	& C \\
\ldots	& 936.6295	& 0.53	  &\ldots& C	& C \\
\ldots	& 948.6855	& 0.77	  &\ldots&\ldots& C \\
\ldots	& 950.885	& 0.18	  & C	 & C	& \ldots \\
\ldots	& 971.738	&1.13	  &\ldots& C	& C \\
\ldots	& 974.070	& $-$1.82 & P, C &\ldots& P, C \\
\ldots	& 976.4481	& 0.51	  &\ldots&\ldots& C \\
\ldots  & 1039.2301	&0.98	  &\ldots&\ldots& C \\
O\,VI	& 1031.9261	& 2.13	& P	& P	& B/S\\
\ldots  & 1037.6167	& 1.83	& P	&\ldots	& B/S\\
Si\,II	& 1020.699	& 1.22	& A	& A	& A \\
P\,II	& 1152.8180	& 2.45	& A	& A	& A \\
S\,III	& 1012.4950	& 1.65	& A	&\ldots	& \ldots\\
S\,IV	& 1062.6780	& 1.64	& A	&\ldots	& \ldots\\
Ar\,I	& 1048.220	& 2.44	& A	& A	&A \\
Fe\,II	& 1055.2617	& 0.90	&\ldots	&A, C, P& \ldots \\
\ldots	& 1063.176	& 1.76	& C	& C	& C, P \\
\ldots	& 1081.9352	&$-$0.90& C, P	& P	& C \\ 
\ldots	& 1096.8770	& 1.55	& C, P	& C, P	& P \\
\ldots	& 1112.0480	& 0.84	& C, P	& P	& C \\
\ldots	& 1121.975	& 1.36	& C, P	& C, P	& \ldots \\
\ldots	& 1125.448	& 1.26	& \ldots& C, P	& \ldots \\
\ldots	& 1133.6654	& 0.80	&A, C, P& C, P	& A, C, P\\
\ldots	& 1142.3656	& 0.68	& P	&\ldots	& C \\
\ldots	& 1143.2260	& 1.31	& C, P	& C, P	&A, C, P \\
\ldots	& 1144.938	& 2.08	& C, P	& C	& \ldots \\
\enddata
\tablenotetext{a}{A, C, and P, denote lines that are analyzed with apparent optical depth, curve of growth, and profile fitting techniques, respectively. B/S means that no column densities were derived from these lines due to blending with stellar features.}
\end{deluxetable}

\begin{deluxetable}{lcccc}
\tablewidth{0pc}
\tablecaption{Equivalent widths of the atomic lines used with the COG method (m\AA)\label{eqw}}
\tablehead{ 
\colhead{Species} & \colhead{Wavelength (\AA)}  & \colhead{WD\,1034$+$001} &\colhead{BD$+$39\,3226}& \colhead{TD1\,32709}}
\startdata
D\,I	& 919.1013	 & 16.96 $\pm$ 2.46 & 12.85 $\pm$ 2.72 & 6.46 $\pm$ 4.23 \\
\ldots	& 920.713		& 18.68 $\pm$ 2.44  & 14.68 $\pm$ 3.77	& 18.59 $\pm$ 2.23 \\
\ldots	& 922.899		& 27.49 $\pm$ 3.47	&\ldots & 26.93 $\pm$ 5.13 \\
\ldots	& 925.974		& 39.78 $\pm$ 3.00	&  26.78 $\pm$ 2.60	& 26.29 $\pm$ 4.33 \\
\ldots	& 930.495		&\ldots	& \ldots& 41.78 $\pm$ 3.10 \\
\ldots	& 937.548		& 46.06 $\pm$ 4.42	& 34.77 $\pm$ 3.06	& 48.31 $\pm$ 6.60 \\
\ldots	& 949.485		&\ldots & 53.68 $\pm$ 2.88	& 53.61 $\pm$ 5.80\\
\ldots	& 972.272		& 67.64 $\pm$ 8.91	& 71.00 $\pm$ 4.91	& 74.34 $\pm$ 6.52\\
N\,I	& 952.303		& 62.29 $\pm$ 3.43	& 37.71 $\pm$ 1.68	& 61.98 $\pm$ 3.17 \\
\ldots  & 952.415		&\ldots & 44.11 $\pm$ 1.59	& 49.23 $\pm$ 2.52 \\
\ldots  & 952.523	& 19.40 $\pm$ 2.71 & 21.96 $\pm$ 1.52 &29.63 $\pm$ 2.82 \\
\ldots  & 953.415		& 67.19 $\pm$ 3.58	& 54.64 $\pm$ 2.05	& 70.61 $\pm$ 3.48 \\
\ldots	& 953.655		& 75.78 $\pm$ 3.84 & 63.32 $\pm$ 2.51	& 81.25 $\pm$ 3.53 \\
\ldots	& 953.9699	  & 78.27 $\pm$ 3.54	& 65.91 $\pm$ 2.14	& \ldots \\
\ldots  & 954.1042		& 68.76 $\pm$ 3.49	& 51.17 $\pm$ 1.86	& 57.25 $\pm$ 3.59 \\
\ldots	& 963.990		& 67.68 $\pm$ 4.73	& 77.37 $\pm$ 2.54	& 70.22 $\pm$ 3.19 \\
\ldots	& 964.626		& 66.84 $\pm$ 4.10	& 64.60 $\pm$ 2.50	& 59.59 $\pm$ 3.37 \\
\ldots  & 1134.1653		& 84.50 $\pm$ 2.76	& 73.40 $\pm$ 1.31	& 101.82 $\pm$ 2.35 \\
\ldots  & 1134.4149	  & 86.95 $\pm$ 2.48	& 81.27 $\pm$ 1.41	& 107.53 $\pm$ 2.57 \\
\ldots  & 1134.9803	  &\ldots &\ldots	& 97.89 $\pm$ 2.01 \\
O\,I	& 916.8150	 & 48.33 $\pm$ 3.92	&\ldots & \ldots\\
\ldots	& 919.658	 & 56.28 $\pm$ 3.32	& 44.98 $\pm$ 2.57	& 44.90 $\pm$ 3.31 \\
\ldots	& 919.917	 & \ldots & 18.63 $\pm$ 1.47 & 27.62 $\pm$ 2.91 \\
\ldots	& 921.875	    & 59.46 $\pm$ 3.33	&\ldots & \ldots \\
\ldots	& 924.950		  & 66.17 $\pm$ 3.59	& 57.68 $\pm$ 2.53	& \ldots \\
\ldots	& 925.446	 & 44.46 $\pm$ 3.37 & 39.13 $\pm$ 2.14	& 41.62 $\pm$ 2.88 \\
\ldots  & 929.5168          &\ldots&\ldots& 72.33 $\pm$ 3.22 \\
\ldots	& 930.257	 & 54.78 $\pm$ 3.54	& 41.69 $\pm$ 2.11	& 50.04 $\pm$ 3.72 \\
\ldots	& 936.6295		  &\ldots& 70.93 $\pm$ 2.69	& 85.04 $\pm$ 3.10 \\
\ldots	& 948.6855		  &\ldots&\ldots& 88.06 $\pm$ 3.87 \\
\ldots	& 950.885		  & 64.30 $\pm$ 3.76	 & 54.69 $\pm$ 1.75	& \ldots \\
\ldots	& 971.738	& \ldots & 90.73 $\pm$ 3.24	& 84.34 $\pm$ 5.02 \\
\ldots	& 974.070	& 4.86 $\pm$ 2.06 & \ldots & 6.91 $\pm$ 1.91	\\
\ldots	& 976.4481		  &\ldots&\ldots& 74.83 $\pm$ 3.26 \\
\ldots  & 1039.2301		  &\ldots&\ldots& 101.94 $\pm$ 1.88 \\
Fe\,II	& 1055.2617		&\ldots	&8.29 $\pm$ 0.94& \ldots \\
\ldots	& 1063.176		& 54.53 $\pm$ 2.81	& 41.14 $\pm$ 1.28	& 26.09 $\pm$ 1.23 \\
\ldots	& 1081.9352	& 21.03 $\pm$ 3.90 	& \ldots	& 8.00 $\pm$ 1.64 \\ 
\ldots	& 1096.8770		& 21.03 $\pm$ 1.75	& 34.52 $\pm$ 1.43	& \ldots \\
\ldots	& 1112.0480		& 11.19 $\pm$ 1.52	& \ldots	& 5.87 $\pm$ 1.05 \\
\ldots	& 1121.975		& 22.58 $\pm$ 2.07	& 14.14 $\pm$ 0.91	& \ldots \\
\ldots	& 1125.448		& \ldots & 22.41 $\pm$ 1.09	& \ldots \\
\ldots	& 1133.6654		& 5.70 $\pm$ 1.90 & 6.48 $\pm$ 0.64	& 5.28 $\pm$ 1.01\\
\ldots	& 1142.3656		& \ldots &\ldots	& 5.84 $\pm$ 0.94 \\
\ldots	& 1143.2260		& 16.01 $\pm$ 1.92	& 22.98 $\pm$ 1.06	&16.43 $\pm$ 1.47 \\
\ldots	& 1144.938		& 55.82 $\pm$ 2.48	& 57.39 $\pm$ 1.34	& \ldots \\
\enddata
\end{deluxetable}

\begin{deluxetable}{lccc}
\tablewidth{0pc}
\tablecaption{\hmol~Column Densities (Log)\label{h2}}
\tablehead{ 
\colhead{\hmol} & \colhead{WD\,1034$+$001} & \colhead{BD$+$39\,3226}& \colhead{TD1\,32709}} 
\startdata
$J$~= 0	& 14.55 $\pm$ 0.10 & 15.17 $\pm$ 0.15	& 13.50 $\pm$ 0.15	\\
$J$~= 1	& 15.63 $\pm$ 0.15 & 15.41 $\pm~^{0.06}_{0.10}$	& 14.35 $\pm$ 0.15	\\
$J$~= 2	& 14.60 $\pm$ 0.10 & 14.47 $\pm$ 0.10	& 13.44 $\pm$ 0.15	\\
$J$~= 3	& 14.30 $\pm$ 0.10 & 13.95 $\pm$ 0.10	& 13.33 $\pm$ 0.15	\\
$J$~= 4	& \ldots & 13.38 $\pm$ 0.10	& \ldots	\\
$J$~= 5	& \ldots & 13.49 $\pm$ 0.07	& \ldots	\\
Total	& 15.72 $\pm~^{0.13}_{0.12}$ & 15.65 $\pm~^{0.06}_{0.07}$ & 14.48 $\pm~^{0.12}_{0.11}$	\\
\enddata
\end{deluxetable}

\begin{deluxetable}{lccc}
\tablewidth{0pc}
\tablecaption{Column Densities (Log) Along the WD\,1034$+$001 Sightline\label{nwd1034}}
\tablehead{ 
\colhead{Species} & \colhead{AOD} & \colhead{COG}& \colhead{PF}} 
\startdata
D\,I	& 15.27 $\pm~^{0.11}_{0.14}$ ($\lambda$919) & 15.40 $\pm~^{0.07}_{0.06}$ & 15.40 $\pm~^{0.05}_{0.03}$\\
N\,I	& 15.89 $\pm~^{0.08}_{0.09}$ ($\lambda$952.5) & 16.07 $\pm~^{0.22}_{0.14}$ & 15.84 $\pm~^{0.12}_{0.13}$\\
O\,I	& 16.57 $\pm$ 0.07 ($\lambda$919.9)	& 16.62 $\pm$ 0.05  & 16.57 $\pm~^{0.07}_{0.08}$\\	
Fe\,II	& 14.19 $\pm~^{0.11}_{0.15}$ ($\lambda$1133) & 14.14 $\pm~^{0.11}_{0.10}$ & 14.05 $\pm~^{0.05}_{0.03}$\\
\enddata
\tablecomments{The transitions associated with the AOD results quoted in the table are given in (). Uncertainties are 1$\sigma$.}
\end{deluxetable}

\begin{deluxetable}{lccc}
\tablewidth{0pc}
\tablecaption{Column Densities (Log) Along the BD$+$39\,3226 Sightline\label{nbd39}}
\tablehead{ 
\colhead{Species} & \colhead{AOD} & \colhead{COG}& \colhead{PF}} 
\startdata
D\,I	& 15.27 $\pm~^{0.08}_{0.11}$ ($\lambda$919) & 15.15 $\pm~^{0.08}_{0.07}$ & 15.15 $\pm~^{0.04}_{0.03}$\\
N\,I	& 15.93 $\pm~^{0.12}_{0.17}$ ($\lambda$952.5) & 15.88 $\pm~^{0.11}_{0.09}$ & 15.80 $\pm~^{0.06}_{0.05}$ \\
O\,I	& 16.33 $\pm$ 0.08 ($\lambda$919.9) & 16.31 $\pm~^{0.07}_{0.06}$ & 16.51 $\pm$ 0.04 \\
Fe\,II	& 14.05 $\pm$ 0.05 ($\lambda$1055) & 14.13 $\pm~^{0.06}_{0.05}$ & 14.21 $\pm$ 0.02 \\
\enddata
\tablecomments{The transitions associated with the AOD results quoted in the table are given in (). Uncertainties are 1$\sigma$.}
\end{deluxetable}

\begin{deluxetable}{lccc}
\tablewidth{0pc}
\tablecaption{Column Densities (Log) Along the TD1\,32709 Sightline\label{ntd1}}
\tablehead{ 
\colhead{Species} & \colhead{AOD} & \colhead{COG}& \colhead{PF}} 
\startdata
D\,I	& 15.33 $\pm~^{0.08}_{0.09}$ ($\lambda$919) & 15.29 $\pm~^{0.05}_{0.04}$ & 15.26 $\pm$ 0.03\\
N\,I	& 15.93 $\pm$ 0.04 ($\lambda$952.5) & 16.02 $\pm~^{0.16}_{0.12}$ & 15.97 $\pm~^{0.09}_{0.11}$ \\
O\,I	& 16.36 $\pm~^{0.07}_{0.08}$ ($\lambda$919.9) & 16.45 $\pm~^{0.09}_{0.03}$ & 16.48 $\pm~^{0.06}_{0.05}$\\
Fe\,II	& 13.98 $\pm~^{0.08}_{0.10}$ ($\lambda$1133) & 14.03 $\pm$ 0.05 & 13.87 $\pm~^{0.03}_{0.02}$\\ 
\enddata
\tablecomments{The transitions associated with the AOD results quoted in the table are given in (). Uncertainties are 1$\sigma$.}
\end{deluxetable}



\begin{deluxetable}{lccc}
\tablewidth{0pc}
\tablecaption{Atomic Column Densities (Log) \label{natomic}}
\tablehead{ 
\colhead{Species} & \colhead{WD\,1034$+$001} & \colhead{BD$+$39\,3226}& \colhead{TD1\,32709}} 
\startdata
H\,I & 20.07 $\pm$ 0.07 & 20.08 $\pm$ 0.09\tablenotemark{a} & 20.03 $\pm$ 0.10 \\
D\,I  & 15.40 $\pm$ 0.07 & 15.15 $\pm$ 0.05  & 15.30 $\pm$ 0.05 \\
C\,II  & blend  & $\ge$ 14.24  &$\ge$ 14.65 \\
C\,II* & $\ge$ 13.88 & $\ge$ 13.63   & $\ge$ 13.94\\
C\,III & $\ge$ 13.66  & $\ge$ 13.40  & blend\\
N\,I  & 15.96 $\pm$ 0.12 & 15.85 $\pm$ 0.10  & 15.98 $\pm$ 0.10 \\
N\,II & $\ge$ 14.15 & $\ge$ 14.25	& $\ge$ 14.35\tablenotemark{b}\\
O\,I  & 16.60 $\pm$ 0.10 & 16.40 $\pm$ 0.10   & 16.42 $\pm$ 0.10\\
O\,VI  & 13.08 $\pm$ 0.08\tablenotemark{c} & 13.14 $\pm$ 0.05  & \ldots \\
Si\,II	 & $\ge$ 14.68  & $\ge$ 14.58  & $\ge$ 14.66 \\
P\,II	 & $\ge$ 13.16 & $\ge$ 13.01  &$\ge$ 13.19 \\
S\,III  & $\ge$ 14.17\tablenotemark{c} & \ldots  &  \ldots \\
S\,IV	& $\ge$ 14.08\tablenotemark{c} & \ldots  &  \ldots \\
Ar\,I	 &  $\ge$ 13.65 & $\ge$ 13.45   & $\ge$ 13.64 \\
Fe\,II & 14.10 $\pm$ 0.10 & 14.15 $\pm$ 0.07  & 13.95 $\pm$ 0.10\\
\enddata
\tablenotetext{a}{From \citet{1999A&A...352..287B}.}
\tablenotetext{b}{Possibly contaminated by stellar N\,II absorption.}
\tablenotetext{c}{Circumstellar contribution possible, see $\S$\ref{wd1034atomicdis}.}
\end{deluxetable}

\begin{deluxetable}{lccc}
\tablewidth{0pc}
\tablecaption{Ratios of Column Densities\label{ratios}}
\tablehead{ 
\colhead{Ratio} & \colhead{WD\,1034$+$001} & \colhead{BD$+$39\,3226}& \colhead{TD1\,32709}} 
\startdata
(D\,I/H\,I)$\times10^{5}$ & 2.14 $\pm~^{0.53}_{0.45}$	& 1.17 $\pm~^{0.31}_{0.25}$ & 1.86 $\pm~^{0.53}_{0.43}$ \\
(N\,I/H\,I)$\times10^{5}$ & 7.76 $\pm~^{2.82}_{2.20}$	& 5.89 $\pm~^{2.04}_{1.63}$ & 8.91 $\pm~^{3.26}_{2.59}$ \\
(O\,I/H\,I)$\times10^{4}$ & 3.39 $\pm~^{1.06}_{0.86}$	& 2.09 $\pm~^{0.72}_{0.58}$ & 2.45 $\pm~^{0.90}_{0.71}$ \\
(D\,I/O\,I)$\times10^{2}$ & 6.31 $\pm~^{1.79}_{1.38}$ & 5.62 $\pm~^{1.61}_{1.31}$ & 7.59 $\pm~^{2.17}_{1.76}$ \\
(D\,I/N\,I)$\times10^{1}$ & 2.75 $\pm~^{1.00}_{0.78}$ & 2.00 $\pm~^{0.57}_{0.46}$ & 2.09 $\pm~^{0.60}_{0.49}$  \\
O\,I/N\,I & 4.37 $\pm~^{1.79}_{1.38}$	& 3.55 $\pm~^{1.30}_{1.03}$	& 2.75 $\pm~^{1.01}_{0.80}$ \\
\enddata
\end{deluxetable}


\begin{deluxetable}{lcc}
\tablewidth{0pc}
\tablecaption{BD$+$39\,3226: $N_{\rm {\it FUSE}}$ vs. $N_{\rm {\it ORFEUS} + {\it IUE}}$ \label{Ncomparison}}
\tablehead{ 
\colhead{Species}&\colhead{log~$N_{\it FUSE}$\tablenotemark{a}} & \colhead{log~$N_{\it ORFEUS + IUE}$\tablenotemark{b}} } 
\startdata
D\,I	& 15.15 $\pm$ 0.05 & 15.16 $\pm$ 0.13 \\
C	& $\geq$ 14.38	& 16.40 $\pm$ 0.75 \\
N\,I	& 15.85 $\pm$ 0.10	& 14.75 $\pm$ 0.25 \\
O\,I	& 16.40 $\pm$ 0.10 & 16.40 $\pm~^{0.75}_{0.50}$ \\
Si	& $\geq$ 14.58	& 14.80 $\pm$ 0.20 \\ 
Fe\,II	& 14.15 $\pm$ 0.07 & 14.10 $\pm$ 0.15 \\
\hmol~($J$~= 0)	& 15.17 $\pm$ 0.15 & 15.10 $\pm$ 0.20		\\
\ldots~($J$~= 1) & 15.41 $\pm~^{0.06}_{0.10}$ & 15.50 $\pm~^{0.20}_{0.30}$	\\
\ldots~($J$~= 2) & 14.47 $\pm$ 0.10 & 14.10 $\pm~^{0.15}_{0.20}$	\\
\ldots~($J$~= 3) & 13.95 $\pm$ 0.10 & 13.95 $\pm~^{0.15}_{0.10}$	\\
\ldots~($J$~= 4) & 13.38 $\pm$ 0.10 & 13.80 $\pm~^{0.30}_{0.20}$	\\
\ldots~($J$~= 5) & 13.49 $\pm$ 0.07 & $<$ 13.40	\\
\enddata
\tablenotetext{a}{This work.}
\tablenotetext{b}{\citet{1999A&A...352..287B}. No $f$-value corrections were applied to $N$.}
\end{deluxetable}


\begin{deluxetable}{lccccccc}
\rotate
\setlength{\tabcolsep}{0.04in}
\tablewidth{0pc}
\tablecaption{Compilation of Sightline Ratios\label{alltabledata}}
\tablehead{ 
\colhead{Star} & \colhead{$d$ (pc)} & \colhead{log~$N$(H)} &\colhead{(D/H)$\times10^5$}&\colhead{(D/O)$\times10^2$}&\colhead{(O/H)$\times10^4$}&\colhead{$\langle n_{\rm H}\rangle$ (cm$^{-3}$)}&\colhead{Ref.}}
\startdata
Sirius B & 2.64& 17.81 $\pm$ 0.11 & 1.17 $\pm$ 0.37 & 3.9 $\pm$ 0.8	& 3.02 $\pm~^{0.95}_{0.75}$ & 0.079  &2, 2, 1\\
$\epsilon$ Eri & 3.22& 17.88 $\pm$ 0.04 & 1.40 $\pm$ 0.20 & 	\ldots	& \ldots & 0.076 & 3, 3, -\\	
Procyon	& 3.50& 18.06 $\pm$ 0.05 & 1.60 $\pm$ 0.20 & 	\ldots	& \ldots & 0.106 &4\\
$\epsilon$ Ind	& 3.63& 18.00 $\pm$ 0.05 & 1.60 $\pm$ 0.20 & 	\ldots	& \ldots &  0.089  &5\\
36 Oph	& 5.99& 17.85 $\pm$ 0.08 & 1.50 $\pm$ 0.25 & 	\ldots	& \ldots &  0.038 &6\\
$\beta$ Gem & 10.34& 18.26 $\pm$ 0.04 & 1.47 $\pm$ 0.20 &  \ldots	& \ldots &  0.057 &3, 3, -\\
Capella	 & 12.9& 18.24 $\pm$ 0.04 & 1.60 $\pm~^{0.14}_{0.19}$ & 2.6 $\pm$ 1.2 & 6.31 $\pm~^{3.69}_{1.95}$ & 0.044&4, 4, 7\\
$\beta$ Cas & 16.7& 18.13 $\pm$ 0.03 & 1.70 $\pm$ 0.15 & 	\ldots	& \ldots &  0.026 &3\\	
$\alpha$ Tri & 19.7& 18.33 $\pm$ 0.04 & 1.32 $\pm$ 0.30 & 	\ldots	& \ldots &  0.035 &3\\ 
$\lambda$ And & 25.8& 18.45 $\pm$ 0.08 & 1.70 $\pm$ 0.25 & 	\ldots	&  \ldots &  0.035 &7\\
$\beta$ Cet & 29.4& 18.36 $\pm$ 0.05 & 2.20 $\pm$ 0.55 & 	\ldots	&  \ldots & 0.025  &8\\
HR\,1099 & 29 & 18.13 $\pm$ 0.02 & 1.46 $\pm$ 0.09 & 	\ldots	& \ldots & 0.015   &8\\	
$\sigma$ Gem & 37 & 18.20 $\pm$ 0.04 & 1.36 $\pm$ 0.20 & 	\ldots	& \ldots &  0.014  &3\\	
WD\,1634$-$573& 37 & 18.85 $\pm$ 0.06 & 1.60 $\pm$ 0.25 & 3.5 $\pm$ 0.3 & 4.57 $\pm$ 0.75& 0.062 &9\\
WD\,2211$-$495& 53 & 18.76 $\pm$ 0.15 & 1.51 $\pm$ 0.60 & 4.0 $\pm$ 0.6 & 3.80 $\pm~^{1.61}_{1.16}$ & 0.035 &10\\
HZ\,43	& 68 & 17.93 $\pm$ 0.03 & 1.66 $\pm$ 0.14 & 4.6 $\pm$ 0.5 & 3.63 $\pm~^{0.44}_{0.70}$ & 0.004 &11 \\
G191$-$B2B & 69 & 18.18 $\pm$ 0.09 & 1.66 $\pm$ 0.45 & 3.5 $\pm$ 0.4 & 4.79 $\pm~^{1.19}_{0.99}$ & 0.007 &12 \\
Feige\,24 & 74 & 18.47 $\pm$ 0.03 & 1.30 $\pm$ 0.50 &  3.9 $\pm$ 1.7 & 3.33 $\pm~^{0.59}_{0.56}$ & 0.013 & 13, 13, 14\\	
WD\,0621$-$376& 78 & 18.70 $\pm$ 0.15 & 1.41 $\pm$ 0.56 & 3.9 $\pm$ 0.6 & 3.63 $\pm~^{1.56}_{1.10}$ & 0.021  & 15\\
GD\,246	& 79 & 19.11 $\pm$ 0.03 & 1.51 $\pm~^{0.20}_{0.17}$ & 4.2 $\pm$ 0.6 & 3.63 $\pm~^{0.44}_{0.34}$ & 0.053  &16\\	
$\alpha$ Vir &80&19.00 $\pm$ 0.10&1.58 $\pm~^{1.01}_{0.46}$ & 4.2 $\pm$ 2.0 & 3.80 $\pm~^{1.39}_{1.11}$&0.041 &17, 17, 18\\
31\,Com		& 94 & 17.88 $\pm$ 0.03 & 2.00 $\pm$ 0.20 &  	\ldots 	&  \ldots & 0.003  &3\\	
$\alpha$ Cru 	& 98 & 19.60 $\pm$ 0.10 & 2.24 $\pm~^{0.64}_{0.52}$ & 	\ldots	&  \ldots & 0.132  &17, 17, -\\
BD$+$28\,4211	& 104 & 19.85 $\pm$ 0.02 & 1.26 $\pm$ 0.08 &  4.7 $\pm$ 0.4 	& 2.69 $\pm~^{0.23}_{0.22}$ & 0.221 &19, 1, 1 \\
$\theta$ Car 	& 135 & 20.28 $\pm$ 0.10 & 0.50 $\pm$ 0.16 & 	\ldots	&  \ldots & 0.457 &20\\
$\beta$ Cma 	& 153 & 18.20 $\pm$ 0.16 & 1.20 $\pm~^{1.10}_{0.50}$ & 	\ldots	&  \ldots &0.003  &21 \\	
WD\,1034$+$001	& 155 & 20.07 $\pm$ 0.07 & 2.14 $\pm~^{0.53}_{0.45}$	& 6.31 $\pm~^{1.79}_{1.38}$& 3.39 $\pm~^{1.06}_{0.86}$& 0.246  &22 \\
$\beta$ Cen 	& 161 & 19.54 $\pm$ 0.05 & 1.26 $\pm~^{1.25}_{0.45}$ & 	\ldots	& \ldots &  0.070  & 17\\	
Feige\,110	& 179 & 20.14 $\pm$ 0.09 & 2.14 $\pm$ 0.41 & 2.6 $\pm$ 0.5 & 8.32 $\pm~^{2.41}_{1.99}$ &  0.245 & 23, 23, 35\\
$\gamma$ Cas	& 188 & 20.04 $\pm$ 0.04 & 1.12 $\pm$ 0.25 & 2.5 $\pm$ 0.4 & 5.25 $\pm~^{0.72}_{0.35}$ &   0.189  &24, 24, 25 \\
$\lambda$ Sco &216 & 19.23 $\pm$ 0.03 & 0.76 $\pm$ 0.25&1.8 $\pm~^{0.4}_{0.3}$&4.24 $\pm~^{0.38}_{0.31}$ &0.026  & 26\\	
$\gamma^2$ Vel 	& 258 & 19.71 $\pm$ 0.03 & 2.18 $\pm~^{0.22}_{0.19}$ & 	\ldots	&  \ldots &0.064  &27, 27, -\\
$\delta$ Ori	& 281 & 20.19 $\pm$ 0.03 & 0.74 $\pm~^{0.12}_{0.09}$ & 2.5 $\pm$ 0.4 & 2.80 $\pm$ 0.40	& 0.179 &28, 28, 25 \\
BD$+$39\,3226 & 290 & 20.08 $\pm$ 0.09 &1.17 $\pm~^{0.31}_{0.25}$ &5.62 $\pm~^{1.61}_{1.31}$&2.09 $\pm~^{0.72}_{0.58}$&0.134&29, 22, 22\\
PG\,0038$+$199&297& 20.48 $\pm$ 0.04 & 1.91 $\pm~^{0.26}_{0.24}$ & 2.4 $\pm~^{1.0}_{0.5}$ & 7.76 $\pm~^{1.78}_{2.33}$ & 0.330 &30\\
$\iota$ Ori	& 407 & 20.16 $\pm$ 0.10 & 1.40 $\pm~^{0.50}_{1.00}$ &  3.5 $\pm$ 0.8 & 3.98 $\pm~^{1.38}_{1.11}$ & 0.115 &31, 31, 25\\
$\epsilon$ Ori	& 412 & 20.40 $\pm$ 0.08 & 0.65 $\pm$ 0.30 & 1.9 $\pm$ 0.3 & 3.80 $\pm~^{0.90}_{0.76}$ &0.198  &31, 31, 25\\
$\zeta$ Pup	& 429 & 19.96 $\pm$ 0.03 & 1.42 $\pm~^{0.15}_{0.14}$ & 	\ldots	&  \ldots &  0.069 & 27\\	
TD1\,32709	& 520 & 20.03 $\pm$ 0.10 & 1.86 $\pm~^{0.53}_{0.43}$ & 7.59 $\pm~^{2.17}_{1.76}$ & 2.45 $\pm~^{0.90}_{0.71}$ &0.067&22\\
LSE\,44		& 554 & 20.52 $\pm~^{0.11}_{0.14}$ & 2.24 $\pm~^{0.70}_{0.67}$ & 1.99 $\pm~^{0.65}_{0.34}$ & 11.3 $\pm~^{4.8}_{3.6}$ & 0.194 & 32 \\	
LS\,1274	& 580 & 20.98 $\pm$ 0.04 & 0.76 $\pm$ 0.18 & 1.8 $\pm$ 0.5 & 4.68 $\pm~^{1.05}_{0.81}$	& 0.534 &33, 33, 1\\
JL\,9		& 590 & 20.78 $\pm$ 0.05 & 1.00 $\pm$ 0.19 & 1.9 $\pm$ 0.8 & 5.25 $\pm~^{3.14}_{1.47}$ & 0.331  &33\\
HD\,195965	& 794	& 20.95 $\pm$ 0.03 & 0.85 $\pm~^{0.17}_{0.12}$ & 1.3 $\pm$ 0.3 & 6.61 $\pm~^{0.52}_{0.56}$ & 0.364 &34\\
HD\,191877	& 2200	& 21.05 $\pm$ 0.05 & 0.78 $\pm~^{0.26}_{0.13}$ & 2.5 $\pm$ 1.0 & 3.09 $\pm~^{0.99}_{0.49}$ & 0.165 &34\\
HD\,90087	& 2740	& 21.17 $\pm$ 0.05 & 0.98 $\pm$ 0.19	& 1.7 $\pm$ 0.4	& 5.8 $\pm$ 1.0	& 0.175	& 35\\
\hline
WD\,2004$-$605& 52 & 19.11 & \ldots &	4.8 $\pm$ 1.3	&  \ldots &   \ldots & 1 \\
WD\,1631$+$781& 67 & 19.36 & \ldots &	3.8 $\pm$ 1.2	&  \ldots &   \ldots & 1 \\
WD\,2331$-$475& 82 & 18.94 & \ldots &	5.1 $\pm$ 1.1	& \ldots &   \ldots & 16 \\
HZ\,21		& 115 & 19.20 & \ldots &	4.6 $\pm$ 1.0	&  \ldots &   \ldots & 16 \\
Lan\,23		& 122 & 20.18 & \ldots &	3.2 $\pm$ 1.6	&  \ldots &  \ldots & 16\\
CPD$-$31\,1701	& 131 & 19.39 & \ldots &	3.7 $\pm$ 0.6	&   \ldots &  \ldots & 1\\
\enddata
\tablecomments{The reference numbers refer to $N$(H), $N$(D), and $N$(O), respectively. When only one reference is given it refers to all the $N$ quoted for that particular sightline. The second part of the table contains sightlines for which only D/O is available. These ratios are displayed in Figure \ref{ratiosplot} for comparison purposes only. For these sightlines the quoted $N$(H) are estimated by using $N$(O) and O/H = (3.43 $\pm$ 0.15)$\times10^{-4}$ from \citet{1998ApJ...493..222M}.}
\tablerefs{(1) \citet{2003ApJ...599..297H}, (2) \citet{1999A&A...350..643H}, (3) \citet{1997ApJ...488..760D}, (4) \citet{1995ApJ...451..335L}, (5) \citet{1996ApJ...470.1157W}, (6) \citet{2000ApJ...537..304W}, (7) \citet{2002ApJ...581.1168W}, (8) \citet{1997ApJ...474..315P}, (9) \citet{2002ApJS..140...91W}, (10) \citet{2002ApJS..140..103H}, (11) \citet{2002ApJS..140...19K}, (12) \citet{2002ApJS..140...67L}, (13) \citet{2000ApJ...544..423V}, (14) \citet{2005ApJ...625..232O}, (15) \citet{2002ApJS..140...81L}, (16) \citet{2003ApJ...587..235O}, (17) \citet{1976ApJ...203..378Y}, (18) \citet{1979ApJ...228..127Y}, (19) \citet{2002ApJS..140...51S}, (20) \citet{1992ApJS...83..261A}, (21) \citet{1985ApJ...296..593G}, (22) This work, (23) \citet{2002ApJS..140...37F}, (24) \citet{1980ApJ...242..576F}, (25) \citet{1998ApJ...493..222M}, (26) \citet{1983ApJ...264..172Y}, (27) \citet{2000ApJ...545..277S}, (28) \citet{1999ApJ...520..182J}, (29) \citet{1999A&A...352..287B}, (30) \citet{2005astro.ph..1320W}, (31) \citet{1979ApJ...229..923L}, (32) \citet{2005friedman}, (33) \citet{2004ApJ...609..838W}, (34) \citet{2003ApJ...586.1094H}, (35) \citet{2005guillaume}.}
\end{deluxetable}
\clearpage


\clearpage

\end{document}